%% file: Main.tex
\documentclass[seceqn]{elsart}
\usepackage[pagebackref]{hyperref}
\usepackage{amsmath,amssymb,epsfig,natbib,bm}

\journal{Physics Reports}

\begin{document}

\newcommand{\eprint}[1]{\href{http://arxiv.org/abs/#1}{#1}}
\newcommand{\ISBN}[1]{\href{http://cosmologist.info/ISBN/#1}{ISBN: #1}}
\newcommand{\adsurl}[1]{\href{#1}{ADS}}
\providecommand{\url}[1]{\href{#1}{#1}}
\providecommand{\newblock}{}

\newcommand{\tTheta}{{\tilde{\Theta}}}
\newcommand{\tC}{{\tilde{C}}}
\newcommand{\tX}{{\tilde{X}}}
\newcommand{\tY}{{\tilde{Y}}}
\newcommand{\tQ}{{\tilde{Q}}}
\newcommand{\tU}{{\tilde{U}}}

\newcommand{\GeV}{\text{GeV}}

\newcommand{\dFT}[1]{\frac{\ud^2 #1}{2\pi}\,}
\newcommand{\arcmin}{\text{arcmin}}

\newcommand{\grad}{\nabla}
\newcommand{\vort}{\varpi}
\newcommand\ba{\begin{eqnarray}}
\newcommand\ea{\end{eqnarray}}
\newcommand\be{\begin{equation}}
\newcommand\ee{\end{equation}}
\newcommand\lagrange{{\cal L}}
\newcommand{\uD}{{\mathrm{D}}}
\newcommand{\curl}{\,\mbox{curl}\,}
\newcommand\del{\nabla}
\newcommand\Tr{{\rm Tr}}
\newcommand\fourth{{1\over 8}}
\newcommand\bibi{\bibitem}
\newcommand{\kf}{\beta}
\newcommand{\kfprod}{\alpha}
\renewcommand\H{{\cal H}}
\newcommand\K{{\cal K}}
\newcommand\opacity{\tau_c^{-1}}
\newcommand{\tot}{{\text{tot}}}
\newcommand{\hatkappa}{{\kappa}}
\newcommand{\hatlambda}{\hat{\lambda}}

\newcommand{\vzero}{\mathbf{0}}
\newcommand{\Psil}{\Psi_l}
\newcommand{\bsigma}{{\bar{\sigma}}}
\newcommand{\bI}{\bar{I}}
\newcommand{\bq}{\bar{q}}
\newcommand{\bv}{\bar{v}}
\renewcommand\P{{\cal P}}

\newcommand{\la}{\langle}
\newcommand{\ra}{\rangle}

\newcommand\xx{\mbox{\boldmath $x$}}
\newcommand{\phpr} {\phi'}
\newcommand{\gam}{\gamma_{ij}}
\newcommand{\sqgam}{\sqrt{\gamma}}
\newcommand{\delk}{\Delta+3{\K}}
\newcommand{\dph}{\delta\phi}
\newcommand{\om} {\Omega}
\newcommand{\dom}{\delta^{(3)}\left(\Omega\right)}
\newcommand{\rar}{\rightarrow}
\newcommand{\Rar}{\Rightarrow}
\newcommand\bigdot[1] {\stackrel{\mbox{{\huge .}}}{#1}}
\newcommand\bigddot[1] {\stackrel{\mbox{{\huge ..}}}{#1}}
\newcommand{\Mpc}{\text{Mpc}}

\newcommand{\ud}{{\text{d}}}
\newcommand{\vnhat}{{\hat{\mathbf{n}}}}
\newcommand{\valpha}{{\boldsymbol{\alpha}}}
\newcommand{\vdelta}{{\boldsymbol{\delta}}}
\newcommand{\vxi}{{\boldsymbol{\xi}}}
\newcommand{\vgrad}{{\boldsymbol{\nabla}}}
\newcommand{\vk}{{\mathbf{k}}}
\newcommand{\vl}{{\mathbf{l}}}
\newcommand{\vlhat}{{\hat{\mathbf{l}}}}
\newcommand{\vr}{{\mathbf{r}}}
\newcommand{\vL}{{\mathbf{L}}}
\newcommand{\vq}{{\mathbf{q}}}
\newcommand{\vqp}{{{\mathbf{q}}_+}}
\newcommand{\vqm}{{{\mathbf{q}}_-}}
\newcommand{\Cgl}{C_{\text{gl}}}
\newcommand{\Cgltwo}{C_{\text{gl},2}}

\newcommand{\primvar}{\mathcal{R}}

\newcommand{\begm}{\begin{pmatrix}}
\newcommand{\enm}{\end{pmatrix}}

\newcommand{\threej}[6]{{\begm #1 & #2 & #3 \\ #4 & #5 & #6 \enm}}
\newcommand{\fsky}{f_{\text{sky}}}

\newcommand{\cla}{\mathcal{A}}
\newcommand{\clb}{\mathcal{B}}
\newcommand{\clc}{\mathcal{C}}
\newcommand{\cld}{\mathcal{D}}
\newcommand{\cle}{\mathcal{E}}
\newcommand{\clf}{\mathcal{F}}
\newcommand{\clg}{\mathcal{G}}
\newcommand{\clh}{\mathcal{H}}
\newcommand{\cli}{\mathcal{I}}
\newcommand{\clj}{\mathcal{J}}
\newcommand{\clk}{\mathcal{K}}
\newcommand{\cll}{\mathcal{L}}
\newcommand{\clm}{\mathcal{M}}
\newcommand{\cln}{\mathcal{N}}
\newcommand{\clo}{\mathcal{O}}
\newcommand{\clp}{\mathcal{P}}
\newcommand{\clq}{\mathcal{Q}}
\newcommand{\clr}{\mathcal{R}}
\newcommand{\cls}{\mathcal{S}}
\newcommand{\clt}{\mathcal{T}}
\newcommand{\clu}{\mathcal{U}}
\newcommand{\clv}{\mathcal{V}}
\newcommand{\clw}{\mathcal{W}}
\newcommand{\clx}{\mathcal{X}}
\newcommand{\cly}{\mathcal{Y}}
\newcommand{\clz}{\mathcal{Z}}
\newcommand{\CMBFAST}{\textsc{cmbfast}}
\newcommand{\CMBEASY}{\textsc{cmbeasy}}
\newcommand{\CAMB}{\textsc{camb}}
\newcommand{\RECFAST}{\textsc{recfast}}
\newcommand{\COSMOMC}{\textsc{CosmoMC}}
\newcommand{\Healpix}{\textsc{healpix}}
\newcommand{\HALOFIT}{\textsc{halofit}}

\newcommand{\Ctot}{{\tilde{C}^\tot}}

\newcommand{\Omtot}{\Omega_{\mathrm{tot}}}
\newcommand{\Omb}{\Omega_{\mathrm{b}}}
\newcommand{\Omc}{\Omega_{\mathrm{c}}}
\newcommand{\Omm}{\Omega_{\mathrm{m}}}
\newcommand{\omb}{\omega_{\mathrm{b}}}
\newcommand{\omc}{\omega_{\mathrm{c}}}
\newcommand{\omm}{\omega_{\mathrm{m}}}
\newcommand{\Omdm}{\Omega_{\mathrm{DM}}}
\newcommand{\Omnu}{\Omega_{\nu}}

\newcommand{\Oml}{\Omega_\Lambda}
\newcommand{\OmK}{\Omega_K}

\newcommand{\Hunit}{~\text{km}~\text{s}^{-1} \Mpc^{-1}}
\newcommand{\Gyr}{{\rm Gyr}}

\newcommand{\nrun}{n_{\text{run}}}

\newcommand{\lmax}{l_{\text{max}}}

\newcommand{\zre}{z_{\text{re}}}
\newcommand{\mpl}{m_{\text{Pl}}}

\newcommand{\vpsi}{\mathbf{\psi}}
\newcommand{\vTheta}{\mathbf{\Theta}}

\newcommand{\vv}{\mathbf{v}}
\newcommand{\vvhat}{\hat{\mathbf{v}}}

\newcommand{\vd}{\mathbf{d}}
\newcommand{\vC}{\mathbf{C}}
\newcommand{\vX}{\mathbf{X}}
\newcommand{\vn}{\mathbf{n}}
\newcommand{\vy}{\mathbf{y}}
\newcommand{\tP}{\tilde{P}}
\newcommand{\tvC}{\tilde{\mathbf{C}}}
\newcommand{\tvT}{\tilde{\mathbf{T}}}
\newcommand{\tvX}{\tilde{\mathbf{X}}}
\newcommand{\tvn}{\tilde{\mathbf{n}}}
\newcommand{\tvy}{\tilde{\mathbf{y}}}

\newcommand{\nvC}{\bar{\mathbf{C}}}
\newcommand{\nvT}{\bar{\mathbf{T}}}
\newcommand{\nvX}{{\bar{\mathbf{X}}}}
\newcommand{\nvn}{\bar{\mathbf{n}}}
\newcommand{\nvy}{\bar{\mathbf{y}}}

\newcommand{\mN}{\bm{N}}
\newcommand{\eV}{\,\text{eV}}
\newcommand{\vtheta}{\bm{\theta}}
\newcommand{\tT}{\tilde{T}}
\newcommand{\tE}{\tilde{E}}
\newcommand{\tB}{\tilde{B}}

\newcommand{\mCh}{\hat{\bm{C}}}
\newcommand{\Ch}{\hat{C}}

\newcommand{\Bt}{\tilde{B}}
\newcommand{\Et}{\tilde{E}}
\newcommand{\bld}[1]{\mathrm{#1}}
\newcommand{\mLambda}{\bm{\Lambda}}
\newcommand{\mA}{\bm{A}}
\newcommand{\mC}{\bm{C}}
\newcommand{\mQ}{\bm{Q}}
\newcommand{\mU}{\bm{U}}
\newcommand{\mX}{\bm{X}}
\newcommand{\mV}{\bm{V}}
\newcommand{\mP}{\bm{P}}
\newcommand{\mR}{\bm{R}}
\newcommand{\mW}{\bm{W}}
\newcommand{\mD}{\bm{D}}
\newcommand{\mI}{\bm{I}}
\newcommand{\mH}{\bm{H}}
\newcommand{\mM}{\bm{M}}
\newcommand{\mS}{\bm{S}}
\newcommand{\mzero}{\bm{0}}
\newcommand{\mL}{\bm{L}}

\newcommand{\btheta}{\bm{\theta}}
\newcommand{\bphi}{\bm{\psi}}

\newcommand{\vb}{\mathbf{b}}
\newcommand{\vA}{\mathbf{A}}
\newcommand{\vAt}{\tilde{\mathbf{A}}}
\newcommand{\ve}{\mathbf{e}}
\newcommand{\vE}{\mathbf{E}}
\newcommand{\vB}{\mathbf{B}}
\newcommand{\vo}{\mathbf{o}}
\newcommand{\vEt}{\tilde{\mathbf{E}}}
\newcommand{\vBt}{\tilde{\mathbf{B}}}
\newcommand{\vEw}{\mathbf{E}_W}
\newcommand{\vBw}{\mathbf{B}_W}
\newcommand{\vx}{\mathbf{x}}
\newcommand{\vXt}{\tilde{\vX}}
\newcommand{\vXb}{\bar{\vX}}
\newcommand{\vTb}{\bar{\vT}}
\newcommand{\vTt}{\tilde{\vT}}
\newcommand{\vY}{\mathbf{Y}}
\newcommand{\vBwr}{{\vBw^{(R)}}}
\newcommand{\RW}{{W^{(R)}}}

\newcommand{\mUt}{\tilde{\mU}}
\newcommand{\mVt}{\tilde{\mV}}
\newcommand{\mDt}{\tilde{\mD}}

\newcommand{\Rot}{\begm \mzero &\mI \\ -\mI & \mzero \enm}
\newcommand{\Pt}{\begm \vEt \\ \vBt \enm}

\newcommand{\edth}{\,\eth\,}
\renewcommand{\beth}{\,\overline{\eth}\,}

\newcommand{\xil}{\tilde{\xi}}

\newcommand{\PsiN}{\Psi_{\text{N}}}
\newcommand{\PhiN}{\Phi_{\text{N}}}

\newcommand{\sE}{{}_{|s|}E}
\newcommand{\sB}{{}_{|s|}B}
\newcommand{\sElm}{\sE_{lm}}
\newcommand{\sBlm}{\sB_{lm}}
\newcommand{\angarg}{}

\newcommand{\muK}{\mu\rm{K}}

\let\Oldsection\section
\newcommand{\bsection}[1]{\Oldsection{\bf{#1}}}

\newcommand{\alt}{\lesssim}
\newcommand{\agt}{\gtrsim}

\begin{frontmatter}


\title{Weak Gravitational Lensing of the CMB}


\author{Antony Lewis\thanksref{al}}
\address{Institute of Astronomy, Madingley Road, Cambridge, CB3 0HA, UK.\thanksref{cita}}
\thanks[al]{Latest contact details at \url{http://cosmologist.info}}
\thanks[cita]{Formerly: CITA, 60 St. George St, Toronto M5S 3H8, ON, Canada.}

\author{Anthony Challinor}
\address{Astrophysics Group, Cavendish Laboratory, J.J.\ Thomson Avenue,
Cambridge CB3 0HE, U.K.}

\begin{abstract}

Weak gravitational lensing has several important effects on the cosmic microwave background (CMB): it changes the CMB power spectra, induces non-Gaussianities, and generates a $B$-mode polarization signal that is an important source of confusion for the signal from primordial gravitational waves. The lensing signal can also be used to help constrain cosmological parameters and lensing mass distributions. We review the origin and calculation of these effects.
Topics include: lensing in General Relativity, the lensing potential, lensed temperature and polarization power spectra, implications for constraining inflation, non-Gaussian structure, reconstruction of the lensing potential, delensing, sky curvature corrections, simulations, cosmological parameter estimation, cluster mass reconstruction, and moving lenses/dipole lensing.


\end{abstract}

\begin{keyword}
Cosmic Microwave Background; Gravitational Lensing
\PACS 98.80.Es, 98.70.Vc, 98.62.Sb, 8.80.Hw
\end{keyword}

\end{frontmatter}
\newpage
\tableofcontents

\newpage


\bsection{Introduction}
\label{sec:intro}

\input{intro}

\bsection{Lensing in General Relativity}
\label{sec:GR}
\input{GR}

\bsection{The lensing potential}
\label{sec:potential}
\input{potential}

\bsection{The lensed CMB temperature power spectrum}
\label{sec:temp}
\input{TempCl}

\bsection{Lensing of CMB polarization}
\label{sec:pol}
\input{pol}

\bsection{Non-Gaussian structure}
\label{sec:nongauss}
\input{nongauss.tex}

\bsection{Reconstructing the lensing potential}
\label{sec:recon}
\input{recon.tex}

\bsection{Delensing the lensed sky}

\label{sec:unlens}
\input{unlens.tex}

\bsection{Sky curvature}

\label{sec:curv}
\input{curv.tex}

\bsection{Simulations}

\label{sec:sim}
\input{sim.tex}

\bsection{Status and applications of CMB lensing}

\label{sec:apps}
\input{apps.tex}

\bsection{Summary and outlook}

It is a general prediction of metric theories of gravity that gravitational lensing must occur. It is therefore a robust prediction that the CMB must be lensed by potential gradients along the line of sight. As we have seen, this weak lensing of the CMB has several important observable effects, most of which are now well understood theoretically in standard models. In summary the lensing:
\begin{itemize}
\item
Perturbs the observed position of the points on the last scattering surface without changing the frequency dependence so the spectrum remains blackbody.
\item
Can be modelled accurately using an integrated lensing potential that is only weakly sensitive to non-linear structures on scales above tens of arcminutes.
\item
Smooths out the acoustic oscillations in the temperature, $E$-polarization and cross-power spectra by several percent.
\item
Generates $B$-mode polarization at the $\mu \text{K}$ level by lensing of $E$-polarization. This is a potential source of confusion for detecting primordial gravitational waves. A significant fraction of the lens-induced $B$-modes come
from small-scale lenses and the non-linear density field introduces
$\sim 10\%$ corrections to the $B$-mode power on all scales.
\item
Produces a distinctive small-scale non-Gaussian trispectrum (four-point function), with smaller signals in the higher even $n$-point functions.
\item
Introduces no non-Gaussian signals in the small-scale odd $n$-point functions if small-scale correlations between the lensing potential and e.g. kinetic-SZ are neglected.
\item
Introduces a large-scale non-Gaussian bispectrum (three-point function) due to correlation between the lensing potential and large-scale temperature.
\item
Can in principle be delensed by using high-resolution polarization observations to reconstruct the lensing potential and re-mapping points to the last scattering surface.
\item
Can be used to improve constraints on some cosmological parameters that are poorly constrained by the CMB power spectra alone, such as the dark energy model
and sub-$eV$ neutrino masses.
\item
Can be simulated accurately in the Gaussian limit, and also from $N$-body simulation for more realistic small-scale analyses.
\item
May be able to constrain the mass distribution of high-redshift clusters
from their lensing signature observed at high-resolution and sensitivity.
\end{itemize}

At the time of writing there is no detection of the lensing signal, though this is expected to change fairly soon. CMB observations are already at the level of precision where the lensing effect on the temperature power spectra must shortly be included for an accurate analysis. This will certainly be the
case once data from the future Planck satellite\footnote{\url{http://www.rssd.esa.int/index.php?project=Planck}}, currently scheduled for launch in early
2008, becomes available~\cite{Lewis:2005tp}.
Polarization experiments already taking data, most notably QUAD\footnote{\url{http://www.astro.cf.ac.uk/groups/instrumentation/projects/quad/}}, have the sensitivity to detect the $B$-modes due to lensing after $\sim$ one year of integration.
The next generation of polarization experiments targeting $B$-modes from
gravitational waves, such as Clover\footnote{\url{http://www.astro.cf.ac.uk/groups/instrumentation/projects/clover/}} and
QUIET\footnote{\url{http://quiet.uchicago.edu/}}, are being designed so that
lens-induced $B$-modes are the limiting source of confusion for detecting
primary $B$-modes from gravity waves (assuming foreground
subtraction is efficient). For such instruments, lensing reconstruction
and delensing will start to become beneficial. Proposed future missions such as the Inflation Probe\footnote{\url{http://universe.nasa.gov/program/inflation.html}} (CMBpol) would be able to measure the lensed $B$-mode signal at high precision, and delensing (or a full likelihood analysis) will become essential to place good constraints on primordial gravity waves.

Although CMB lensing is generally well understood, there remain areas for further work. In particular the delensing methods for subtracting out the $B$-mode lensing signal have so far only been developed in idealized cases (flat sky without boundaries, simple noise and beam properties, no non-linear structures, etc.). There are also no known general methods for performing a full accurate parameter likelihood analysis on lensed data once simple approximations are no longer good enough. Such methods may be required to obtain accurate parameter constraints from future high-resolution measurements of the $E$ (and $B$) polarization. In principle a near-optimal likelihood analysis would additionally be able to constrain the gravity waves without a separate explicit delensing step.
Analysis of the observed temperature field on small scales will require accurate modelling of kinetic-SZ and other non-linear signals, including correlations with the lensing signal. It may also be important to account for lensing in non-standard models, for example when trying to find matched circles from non-trivial topologies. If the primordial fluctuations turn out to be significantly non-Gaussian or anisotropic this will also significantly complicate the full lensed analysis and require much further work.

\subsection*{Acknowledgements}

We thank Chris Hirata, Wayne Hu, and Oliver Zahn for figures and data.
AL acknowledges a PPARC Advanced fellowship and
AC a Royal Society University Research Fellowship. The Beowulf cluster used for some calculations was funded by the Canada Foundation for Innovation and the Ontario Innovation Trust.

\providecommand{\apj}{Astrophys. J. }\providecommand{\apjl}{Astrophys. J.
  }\providecommand{\mnras}{MNRAS}

\end{document}

%% file: intro.tex
Observations of the cosmic microwave background (CMB) temperature and polarization are a powerful probe of the early universe and cosmological models. There are now convincing measurements from space, ground and balloons, and many future observations are planned at ever greater resolution and/or sensitivity\footnote{\url{http://lambda.gsfc.nasa.gov/product/}}. The basic theory of the CMB is well understood, with a robust prediction for acoustic oscillations in the primordial plasma giving rise to a characteristic pattern of peaks in the anisotropy power spectra (for reviews see e.g. Refs.~\cite{LL,Hu:2001bc,Hu:2002aa,DodelsonBook,HuSciAm,Challinor:2004bd,Scott:2006xy,Padmanabhan:2006kz}). As ever better data becomes available, observations can resolve smaller effects including  various non-linear signals that are important on small scales. One of the most important of these is weak lensing: the deflection of CMB photons coming from the last scattering surface by potential gradients along our line of sight~\cite{Blanchard87,Cole89,Linder90,Seljak:1996ve,Metcalf:1997ih}. The lensing has a quantitatively important effect on the temperature power spectrum, as well as introducing qualitatively new non-Gaussian and polarization signals. The lensing signal has the same frequency spectrum as the unlensed CMB, and hence cannot easily be distinguished. Understanding and modelling lensing will therefore be essential for the correct interpretation of near-future CMB data.

This review is aimed at readers familiar with the basic theory of the unlensed CMB; however we shall start with a very brief overview of the relevant properties. We then describe why the effect of lensing is important, give order of magnitude estimates, and present a heuristic physical derivation of the deflection angle. We shall use the weak lensing approximation, so we discuss what this means and whether it is a valid approximation. We give a rigorous derivation of the deflection using General Relativity (GR) and discuss technical issues in Section~\ref{sec:GR}. In Section~\ref{sec:potential} we describe how the main effects of lensing can be encapsulated into a single lensing potential, how this affects the CMB, and how to relate it to the matter power spectrum. In Section~\ref{sec:temp} we derive the important several-percent effect on the CMB temperature power spectrum due to lensing, and in Section~\ref{sec:pol} extend this to the polarization. We explain the importance of the `$B$-mode' polarization signal from lensing, in particular how it is a source of confusion with any primordial signal from gravitational waves. In addition to the $B$-mode signal, lensing also introduces a characteristic non-Gaussian signal in the CMB, as discussed in Section~\ref{sec:nongauss}.

The extra information in the non-Gaussian structure of the lensed CMB sky actually provides enough information to learn something about the distribution of the matter causing the lensing. In Section~\ref{sec:recon} we discuss how the lensing potential can be reconstructed from observations of the lensed sky alone. If we know something about the lensing potential, this in principle allows us to `delens' the sky to recover what the CMB would have looked like without lensing. We discuss methods for doing this in Section~\ref{sec:unlens}, and in particular how the $B$-mode lensing signal can be subtracted to get at any residual interesting $B$-mode signal from primordial gravitational waves.

For most of this review we shall use the flat-sky approximation, as it greatly simplifies many of the derivations. In Section~\ref{sec:curv} we describe the small but potentially important corrections due to the fact that in reality observation directions have spherical geometry about our observation point. In Section~\ref{sec:sim} we describe the simulation of lensed CMB skies, essential for testing pipelines and assessing the importance of non-linear physics and other real-world complications. Finally in Section~\ref{sec:apps} we describe the current observational status and various possible applications of CMB lensing. These include using lensing to constrain dark energy, neutrino mass and other cosmological parameters, as well as learning about the properties of lensing clusters and galaxies. We also discuss the equivalent effects of moving lenses and dipole lensing.

\subsection{The unlensed CMB}

Observations show that the CMB temperature is approximately uniform at a temperature of $2.725$K, with anisotropies at the $\clo(10^{-5})$ level~\cite{COBE}. The statistics of the observed CMB anisotropies can be explained as arising from acoustic oscillations in the primordial photon-baryon fluid, where the initial perturbations had a Gaussian, purely adiabatic, nearly scale-invariant spectrum~\cite{Spergel:2003cb}. The small size of the perturbations means that linear perturbation theory is very accurate, and the acoustic oscillations can be modelled accurately to give robust predictions. What we observe is the epoch at which the universe became neutral at a redshift $z \sim 10^3$, after which CMB photons can propagate more or less freely towards us. The perturbations on this high-redshift last-scattering surface are the main contribution to the CMB anisotropy observed, with additional large-scale anisotropies from the integrated Sachs-Wolfe~\cite{Sachs:67} (ISW) effect from evolving potentials along the line of sight. The process of recombination is not instantaneous as it takes a while for most of the protons and electrons to combine into neutral hydrogen, though viewed today it is extremely thin compared to the distance to the last scattering surface. It is however slow enough for a photon quadrupole to develop, which then gives rise to polarization through Thomson scattering~\cite{Rees:1968,Polnarev:1985,Hu:1997hv}. Thus the last scattering surface is expected to have an interesting polarization signal as well as an anisotropy in the CMB temperature.

The late-time evolution of potentials gives an additional large-scale temperature anisotropy through the (late-)ISW effect, which means that the large scale CMB temperature is to some extent correlated with the large scale structure having the evolving potentials. Another important late time effect is that of reionization: the epoch at which energetic photons from stars and quasars can reionize the neutral hydrogen gas. This is thought to have happened at redshifts somewhere between $6 \alt z \alt 20$. The effect is important because the free electrons after reionization scatter some of the CMB photons, damping the small scale anisotropies. Furthermore the CMB quadrupole from last scattering will scatter from the free electrons giving rise to a polarization signal that can dominate the expected signal from the last scattering surface on large scales. Angular scales smaller than a couple of degrees are however only damped by the reionization: the small-scale CMB signal of most concern for CMB lensing is just proportional to the signal coming from the last scattering surface. The proportionality constant $e^{-2\tau}$ depends on the optical depth $\tau$ to the last scattering surface.

There are important second-order signals other than lensing, for example the Sunyaev-Zel'dovich (SZ) effect~\cite{SZ} (for reviews see e.g. Refs.~\cite{Hu:2001bc,Cooray:2004cd}). These are generally generated at low redshift, and are almost unaffected by lensing~\cite{Wu:2003au}. We shall therefore concentrate on the process of lensing of the linear CMB signal from the last scattering surface. Since the width of the last scattering surface is small ($\sim 100 \Mpc$ comoving) compared to its distance from us ($\sim 14 000 \Mpc$ comoving), it is a good approximation to consider it as a single lens source plane. The effect of lensing on the reionization and ISW signals is generally very small as they are only important on large scales.

\subsection{CMB lensing order of magnitude}

Gravitational lensing of light by matter is expected even in Newtonian physics. If a particle passes a point mass at a safe distance, a simple Newtonian argument gives a deflection by an angle $2\Psi(R)/c^2$ (assuming the deflection is small), where $\Psi(R)$ is the potential at the point of closest approach on the undeflected path. As discussed below, a  relativistic argument using General Relativity in fact gives twice this result. For weak lensing of the CMB we are interested in the deflection of a photon's direction as it travels from the last scattering surface to our observation point. On its way the photon will encounter various under- and over-densities, where in matter domination the potentials due to these perturbations are constant in the linear regime. The depth of the potentials is $\sim 2 \times 10^{-5}$, so we might expect each potential encountered to give a deflection $\delta\beta \sim 10^{-4}$. The characteristic size of potential wells given by the scale of the peak of the matter power spectrum is $\sim 300 \Mpc$ (comoving), and the distance to last scattering is about $14000\Mpc$, so the number passed through is $\sim 50$. If the potentials are  uncorrelated this would give an r.m.s. total deflection $\sim 50^{1/2}\times 10^{-4} \sim 7\times 10^{-4}$, corresponding to about $\sim 2$ arcminutes. We might therefore expect the lensing to become an order unity effect on the CMB at $l \agt 3000$. In fact the unlensed CMB has very little power on these scales due to damping, so lensing can dominate the observed power at $l \agt 3000$ in the absence of any other secondaries. Of course we have neglected several important effects here, including the correlation of the lensing clumps and the correct conversion factors between deflection angles and observed angles; however these order-of-magnitude results are about right.

The deflection angles will be correlated over the sky by an angle given by the angular size of a characteristic potential, which is $\sim 300/7000 \sim 2^\circ$ for a potential mid-way to last scattering. So although the deflection angles are much smaller than the size of the degree-scale primary CMB acoustic peaks, they are correlated over a comparable scale. This means that the lensing can also have an important effect on the scale of the primary acoustic peaks. If a degree sized hot spot at last scattering is lensed, we expect the spot to appear larger or smaller by $\sim 2'$, corresponding to a fractional change in size of $\sim 2/60 \sim 3\%$. The lensing won't make a spot larger or smaller on average, but it will change the statistics of the size distribution. The distribution will be $3\%$ wider, corresponding to a $\sim 3\%$ broadening on the acoustic peak: after lensing the size of the spot is less well defined on average. We therefore expect lensing to have percent level effects on the CMB spectrum on the scale of the primary acoustic peaks, and to dominate the spectrum on arcminute scales.

\subsection{Weak-lensing deflection angle}
\label{subsec:gravity}

The idea that gravity can bend light dates back over two hundred years~\cite{Perlick04}.
In this subsection we give some simple non-rigorous derivations to get the correct lowest order result for the deflection angle.
In Section~\ref{sec:GR} we use General Relativity to give a more rigorous derivation and discuss higher-order corrections.

Consider weak lensing of a photon with velocity $\vv$ by a point of mass $M$ within Newtonian theory. By weak lensing we mean that deflections due to the lensing are a small perturbations, so deflection angles etc. are small and may be treated accurately using first-order results. The acceleration due to the mass is given by the gradient of the potential $\Psi$, causing a small transverse acceleration $\dot{v}_\perp = -\grad_\perp\Psi = GM\cos\theta/r^2$, where $\theta$ is the angle of the photon from the mass relative to its angle at distance of closest approach $R_0$. Integrating over the photon path for constant speed $|\vv|=c$ gives a total deflection angle $v_\perp/|\vv| = 2 G M / c^2 R_0$, which is the standard Newtonian result. From now on we shall use natural units with $c=1$.

What changes in General Relativity? The idea that acceleration is due to a force holds in GR, with $ D_\chi \vvhat = -\grad_\perp\Psi$ in the Newtonian gauge, where $D_\chi$ represents the covariant derivative along the photon world line. However to relate this local acceleration to a change in observed coordinate we also have to account for the curvature of space. The total effect is a local deflection angle $\delta\beta = -2\delta\chi \grad_\perp \Psi$, where $\delta\chi$ is a small distance along the photon path.
Essentially the GR result is the inertial Newtonian result plus an equal term from the effect of spacetime curvature.

\begin{figure}
\begin{center}
\psfig{figure=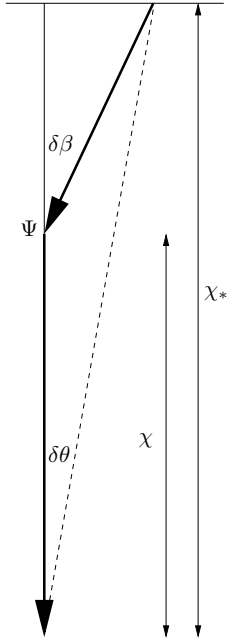,width=3cm}
\caption{Weak lensing geometry for a source (the CMB) at comoving distance $\chi_*$ lensed by a potential $\Psi$ at distance $\chi$, assuming a flat universe. The lensing deflection by an angle $\delta\beta$ changes the observed angle of the source by an angle $\delta\theta$.
\label{geom}}
\end{center}
\end{figure}

Since potentials are generally small ($|\Psi| \alt 10^{-3}$; smaller on linear scales) the deflection angles are small, consistent with our small angle assumption.
Now let's consider how this deflection affects the observed angle $\theta$ of an object at comoving distance $\chi_*$. Comoving distances are related to angles via the angular diameter distance $f_K(\chi)$, where
\begin{equation}
f_K(\chi)=\begin{cases}
 K^{-1/2} \sin (K^{1/2} \chi) & \text{for $K>0$, closed}  , \\
\chi & \text{for $K=0$, flat}  , \\
|K|^{-1/2} \sinh (|K|^{1/2} \chi) & \text{for $K<0$, open}. \\
\end{cases}
\label{f_K_def}
\end{equation}
The comoving distance that the source appears to have moved due to the lensing is, in the small angle approximation, $f_K(\chi_*-\chi)\delta\beta = f_K(\chi_*)\delta\theta$ (see Fig.~\ref{geom}). Solving for $\delta\theta$, the deflection due to the source at $\chi$ is
\begin{equation}
\delta\theta_\chi = \frac{f_K(\chi_*-\chi)\delta\beta}{f_K(\chi_*)} = - \frac{f_K(\chi_*-\chi)}{f_K(\chi_*)} 2\delta\chi\grad_\perp\Psi
\end{equation}
in the direction of $\vgrad_\perp\Psi$. Adding up the deflections from all the potential gradients between us and the source, we have a total deflection
\begin{equation}
\valpha = -2 \int_0^{\chi_*} \d\chi \frac{f_K(\chi_*-\chi)}{f_K(\chi_*)}\vgrad_\perp\Psi(\chi\vnhat; \eta_0 -\chi).
\label{delta_theta}
\end{equation}
This is the main result that tells us the deflection angle in terms of the potential gradients along the line of sight.
The quantity $\eta_0 -\chi$ is the conformal time at which the photon was at position $\chi \vnhat$.
It is only valid for weak lensing (small angles), and is only valid to lowest order in the potential. For a flat universe $f_K(\chi_*-\chi)/f_K(\chi_*) = 1- \chi/\chi_*$.

The lensed CMB temperature in a direction $\vnhat$ is given by the unlensed temperature in a deflected direction $\vnhat'$, $\tT(\vnhat) = T(\vnhat') = T(\vnhat + \valpha)$. The derivative of the deflection angle defines a magnification matrix (see e.g. Ref.~\cite{Bartelmann:1999yn})
\begin{equation}
A_{ij}\equiv \delta_{ij} + \frac{\partial}{\partial\theta_i} \alpha_j = \begm 1-\kappa-\gamma_1\, && -\gamma_2 + \omega \\ -\gamma_2 -\omega\, && 1-\kappa+\gamma_1 \enm.
\label{mag_def}
\end{equation}
An infinitesimal source with surface brightness $I(\vnhat + \vdelta \xi)$ at position $\vdelta\xi$ about $\vnhat$ before lensing, becomes, after lensing, $I(\vnhat' + \mA \vdelta\xi)$.  At lowest order the magnification of the intensity $\mu \equiv |\mA|^{-1} = 1/[(1-\kappa)^2+\omega^2 - |\gamma|^2]\approx 1+2\kappa$ is determined by the convergence, $\kappa= -\half \grad\cdot\valpha$. The shear $\gamma_1 + i\gamma_2$ determines the area-preserving distortion, and the antisymmetric piece $\omega$ determines the rotation. Since Eq.~\eqref{delta_theta} is purely a derivative, the antisymmetric rotation $\omega$ vanishes at lowest order.

Shear is the most important effect for galaxy lensing~\cite{Bartelmann:1999yn} as it makes the galaxy ellipticities correlated in a direction determined by the local shear. It can be used without knowing anything about the distribution of the source galaxies. However for the CMB we actually know a lot about the statistical distribution of the unlensed CMB, so shearing is not the only relevant effect: we can also learn information from perturbation-scale (i.e. non-infinitesimal) magnifications. Including shear and convergence effects on finite scales is most easily described as the remapping of points by the deflection angle. For this reason we shall usually discuss the deflection angle directly, though the shearing effect on the CMB is real and may be observable through changes to the hot and cold spot ellipticity distribution~\cite{Bernardeau:1998mw,vanWaerbeke:1999jd}.

\subsection{Photon distribution}
\label{subsubsec:brightness}

The observed angle of a particular photon from the last scattering surface will be altered due to lensing by potential gradients along the line of sight. In general there will also be a perturbation to the frequency due to time-varying potentials. For linear perturbations this is just the integrated Sachs-Wolfe effect, in general it is the Rees-Sciama effect. By discussing only lensing we are explicitly not considering these frequency shifts, so they  must be accounted for separately along with other secondary signals like SZ.

Lensing by transverse gradients does not change the frequency of the photons, and hence does not change the frequency distribution in a given direction except via the change in the angle of the source. Hence the lensed CMB has the same blackbody spectrum as the unlensed CMB, and multi-frequency observations cannot be used to separate the lensing signal.

The number of photons arriving from a particular sized patch of the last scattering surface will be altered due to lensing: a magnifying lens along the line of sight will cause more photons to reach us from the patch than without lensing. However the angular size subtended by the patch is also increased proportionately, so the \emph{number of photons per unit solid angle} remains the same. This is the rule that lensing conserves surface brightness. Since the frequency is also not altered by lensing, it means that lensing of an isotropic last scattering surface is completely unobservable: photons are simply moved around but arrive in the same distribution as before. It is only the presence of anisotropies on the last scattering surface that make the lensing effect interesting.

\subsection{Is weak lensing a good approximation?}
\label{subsubsec:stronglensing}

There are several different aspects of the weak lensing approximation, and `weak' may be used to mean slightly different things in different contexts. The main criterion here is that deflection angles should be small, which is indeed the case with typical deflections being of the order of arcminutes. Deflections from non-linear structure are also small, with clusters typically giving angles of the order of an arcminute or less, and smaller structures such as galaxies giving only arcsecond deflections. Only a minuscule fraction of lines of sight will come close to black holes or other dense bodies that violate the small-angle assumption.

Small deflections are essential for the validity of the Born approximation: calculating the lensing effect using the potential gradients along the \emph{undeflected} path. If deflection angles were large compared to the scale of lensing perturbations this would not be a good approximation and we would have to tackle the much harder problem of ray-tracing along the full deflected photon path. As it is, the Born approximation is excellent for lensing on scales larger than a few arcminutes (see e.g. Refs.~\cite{Hirata:2003ka,Shapiro:2006em}). Lensing from a single cluster is also accurately modelled: the deflection occurs only in a very thin section of the photon path over which the deflected and undeflected paths are the same. Lensing from multiple clusters is more complicated (see Section~\ref{sec:higherorder}), but rare and irrelevant for a CMB calculation on larger scales. On very small scales numerical simulations can account for the important additional effects (Section~\ref{sec:sim}).

Another use of the term weak lensing is to exclude cases of `strong' lensing where lines of sight cross, giving caustics, multiple images and points of infinite magnification and shear. This can happen even for tiny deflection angles if the source is small. For example massive clusters will have a CMB Einstein ring at a radius of $\sim 1$ arcminute from the centre, corresponding to the radius at which deflected rays meet at a point on the last scattering surface. Most of our analysis will in fact apply equally well to `strong' lensing cases, all that we shall require is that deflection angles are small.

Strong lensing and significant magnifications are in fact very common: a large fraction of lines of sight will have intersected other lines of sight by last scattering~\cite{Holz:1997ic,Ellis:1998ha}, and many ray bundles will have experienced significant (several percent) magnification due to encounters with small scale structures along the line of sight. Much of this effect comes from encounters with very small sub-cluster scale structures, and implies that the light cone has a very complex structure on sub-arcminute scales. For the CMB this is fortunately not a problem. On arcminute scales the CMB is very smooth due to diffusion damping. Magnification of a smooth surface does not change what you see at all: if there is a galactic-scale area of large magnification of the CMB this will be essentially unobservable. To have an interesting effect we need arcminute-scale coherent magnification, corresponding to
arcminute scale deflection angles. This is much larger than the amount of deflection produced by most of the structures giving rise to large magnifications. We can therefore proceed to calculate the effect on the CMB down to arcminute scales without worrying at all about small-scale lenses.\footnote{The contrary claim in Ref.~\cite{Lieu:2004vp} is erroneous.} What matters most for the CMB is the lensing effect of large-scale nearly-linear structures.

Another possible worry is whether non-linear clumping in the late universe violates calculations based on angular diameter distances and a low-order expansion about a homogeneous Friedmann-Robertson-Walker (FRW) universe. For example if two lines of sight happen to travel exclusively through voids, they will have diverged relative to the propagation expected in a flat FRW universe~\cite{Dyer:1972}. However it turns out that on average the effect on angular diameter distances is cancelled by convergence caused by propagation through denser regions~\cite{Weinberg:76,Peacock:1986,Holz:1997ic,Kibble:2004tm}. Whether the variance is important has historically been the matter of much debate (see e.g. Refs.~\cite{Cole89,Fukushige:1994jv,Seljak:1996ve}). As discussed above the total magnification variance may be large, but this variation comes largely from very small structures and is largely irrelevant for the CMB\footnote{Ref.~\cite{Lieu:2004tn} claims a several-percent effect on the main CMB acoustic peaks; this is incorrect because the scale dependence and correlation of the magnification were not properly accounted for.}: only a convergence field correlated on scales of many arcminutes can affect CMB acoustic scales. Also note that potentials are everywhere small, ($\alt 10^{-3}$ even including non-linear features), so we might expect that a systematic analysis based on linear theory that we present in later sections should be accurate to within fairly small non-linear corrections~\cite{Seljak:1996ve} (see also Refs.~\cite{Seitz:1994xf,Holz:1997ic,Ellis:1998ha,Hamana:2004ih}).

We discuss higher-order corrections to the first-order lensing result in more detail in Section~\ref{sec:higherorder}.

%% file: GR.tex

To give a rigorous analysis of the lensing effect we must consider directly the propagation
of light rays through the perturbed metric of spacetime. We do this in
linearized form using a standard metric-based approach in Section~\ref{sec:metric}. An alternative treatment based on the dynamics of infinitesimally-separated rays is given in
Section~\ref{sec:bundle}, which allows us to define the lensing deflection entirely in terms of directly observable quantities (in linear theory). We employ and assume familiarity with standard General Relativity with the $(+---)$ signature and units such that $c=1$.

This section on General Relativity is mathematically more sophisticated than the rest of the review, and tackles some interesting conceptual and technical issues. However readers who are happy with Eq.~\eqref{delta_theta} can skip this section and proceed directly to Section~\ref{sec:potential} as the material in this section is not required for the remainder of the review. For more general reviews of gravitational lensing see Refs.~\cite{Bartelmann:1999yn,SchneiderBook,Petters:2001,Mollerach:2002}.

\subsection{Perturbed photon paths}
\label{sec:metric}

We aim to compute the perturbed paths of light rays (null geodesics)
in the real clumpy universe.
In general, solving
for the null geodesics requires numerical ray-tracing methods through
large simulation volumes~\cite{Jain:1999ir}. However, for CMB lensing
on scales $\agt 1\,\arcmin$ only weak lensing is expected to be important
(see Section~\ref{subsubsec:stronglensing}) in which case the perturbations of
rays are small and a linearized treatment is appropriate.

We only consider lensing by density perturbations [for lensing by gravitational
waves, see Refs.~\cite{Kaiser:1996wk,Dodelson:2003bv}],
and for convenience we work in the
conformal Newtonian gauge (for a discussion of gauge issues see Section~\ref{sec:gauges}). The line element is then
\begin{equation}
\ud s^2 = a^2(\eta)[(1+2\PsiN)\ud \eta^2 - (1+2\PhiN)\gamma_{ij} \ud x^i
\ud x^j] ,
\label{eq:metric}
\end{equation}
where $\eta$ is conformal time and the unperturbed spatial metric $\gamma_{ij}$
is such that
\begin{equation}
\gamma_{ij} \ud x^i \ud x^j = \ud \chi^2 + f_K^2(\chi)(\ud \theta^2 +
\sin^2\theta \ud\phi^2).
\end{equation}
We shall work to first order in the scalar potentials $\PhiN$ and $\PsiN$, so equalities below should be taken to apply at first order only.
Since for lensing we are only calculating the path of null geodesics with $\ud s^2 = 0$, we can equally well divide through by $a^2(\eta)(1+2\PhiN)$ and use the simpler conformally-related metric
\begin{equation}
\ud \hat{s}{}^2{} = (1+4\Psi)\ud \eta^2 - \gamma_{ij} \ud x^i
\ud x^j.
\label{eq:confmetric}
\end{equation}
Here we defined the `Weyl potential' $\Psi \equiv (\PsiN - \PhiN)/2$, so called because the general linear scalar-mode Weyl tensor can be expressed in terms of derivatives of $\Psi$ only. The conformal invariance of null geodesics means that they can only depend on the conformally-invariant Weyl part of the Riemann tensor.

Affinely-parameterized null geodesics are solutions of the geodesic
equation,
\begin{equation}
\frac{\ud^2 x^\mu}{\ud \hat{\lambda}^2} + \Gamma^\mu_{\nu\rho}
\frac{\ud x^\nu}{\ud \hat{\lambda}}\frac{\ud x^\rho}{\ud \hat{\lambda}} = 0,
\label{eq:nullgoedesic}
\end{equation}
with $\hat{g}_{\mu \nu} (\ud x^\mu / \ud \hat{\lambda}) (\ud x^\nu / \ud \hat{\lambda}) = 0$ and $\ud \hat{s}^2=0$. Note that we have denoted the affine
parameter in the conformal metric with an overhat since it necessarily
differs from that in the original frame.
In an unperturbed model, there are incoming radial solutions
(with a focus at the origin at conformal time $\eta_0$) with
$\chi = \eta_0 - \eta$.

The $0$-component of the geodesic equation gives
\begin{equation}
\frac{\ud^2 \eta}{\ud \hat{\lambda}^2} +2 \left(\frac{\ud \eta}{\ud \hat{\lambda}}\right)^2
\frac{\ud \Psi}{\ud\eta}
+ 2\frac{\ud \eta}{\ud \hat{\lambda}} \frac{\ud x^i}{\ud \hat{\lambda}} \frac{\partial
\Psi}{\partial x^i} = 0,
\label{eq:etadotdot}
\end{equation}
where the derivative $\ud \Psi / \ud \eta = \partial_\eta \Psi +
(\ud x^i / \ud \eta)\partial_i \Psi$ is along the (perturbed) ray. Equation~\eqref{eq:etadotdot} is the only equation in this section that does not also apply in the original metric of Eq.~\eqref{eq:metric} since we cannot simply replace
$\hat{\lambda}$ by $\lambda$.
With Eq.~\eqref{eq:etadotdot}, we can
eliminate the affine parameter $\hat{\lambda}$ from~\eqref{eq:nullgoedesic}
in favour of $\eta$, giving
\begin{equation}
\frac{\ud^2 x^i}{\ud \eta^2} - 2\frac{\ud x^i}{\ud \eta} \left(\frac{\ud \Psi}
{\ud \eta} + \frac{\ud x^j}{\ud \eta}
\frac{\partial \Psi}{\partial x^j}\right) + 2 \gamma^{ij} \frac{\partial \Psi}
{\partial x^j} + {}^{(3)}\bar{\Gamma}^i_{jk}
\frac{\ud x^j}{\ud \eta}\frac{\ud x^k}{\ud \eta}=0,
\label{eq:dxidetasq}
\end{equation}
where ${}^{(3)}\bar{\Gamma}^i_{jk}$ are the connection coefficients of the
unperturbed three geometry $\gamma_{ij}$.
It is very convenient to consider an observer located at the origin
of the spatial coordinates, in which case we are interested in rays that
focus at $x^i = 0$. For such rays, $\ud \chi / \ud \eta = -1 + O(\Psi)$
and $\ud \theta / \ud \eta = O(\Psi)$ with an equivalent result for $\phi$.
Making use of these results in Eq.~(\ref{eq:dxidetasq}), and evaluating
the background connection coefficients, we find
\begin{eqnarray}
\frac{\ud^2 \chi}{\ud \eta^2} + 2 \frac{\ud \Psi}{\ud \eta} &=& 0 ,
\label{eq:dchidetasq} \\
\frac{\ud^2 \theta}{\ud \eta^2} - 2 \frac{\ud \ln f_K(\chi)}
{\ud \chi} \frac{\ud \theta}{\ud \eta}
 + \frac{2}{f_K^2(\chi)} \frac{\partial \Psi}{\partial \theta}
&=& 0 , \label{eq:dthetadetasq} \\
\frac{\ud^2 \phi}{\ud \eta^2} - 2 \frac{\ud \ln f_K(\chi)}
{\ud \chi} \frac{\ud \phi}{\ud \eta}
+ \frac{2}{f_K^2(\chi)}\frac{1}{\sin^2\theta}
\frac{\partial \Psi}{\partial \phi} &=& 0 , \label{eq:dphidetasq}
\end{eqnarray}
%
which determine the perturbed rays up to first order in $\Psi$.

Equation~(\ref{eq:dchidetasq}) has $\ud \chi / \ud \eta + 2\Psi$ as a first
integral, and this must equal $-1$ by the null condition for the perturbed
ray. Integrating again, we find
\begin{equation}
\chi = \eta_0 - \eta - 2 \int_{\eta_0}^\eta \Psi \, \ud \eta',
\label{eq:radialdelay}
\end{equation}
where the integral is along the ray. Since we are only working to
first-order in $\Psi$, we can evaluate the integral along the unperturbed path
$\theta=\text{const.}$, $\phi=\text{const.}$ and $\chi = \eta_0 - \eta$.
(This is the Born approximation.) Integrating to a fixed $\eta$,
Equation~(\ref{eq:radialdelay}) implies a radial displacement. Alternatively,
integrating to a fixed $\chi$ implies a variation in the conformal time
at emission; this is a time delay if the potential is negative on-average
along the ray, i.e.\ it passes mostly through overdense regions.
At the last-scattering surface, the r.m.s.\ radial delay is $\sim
1\, \text{Mpc}$ and is coherent over very large angular scales~\cite{Hu:2001yq}.
Despite the delay being relatively large, the
effect on the CMB is only small due to geometric suppression effects
and the large coherence scale.
If we think of plane-wave fluctuations, at a given angular scale $l^{-1}$
the dominant plane waves to contribute to the CMB fluctuations have their
wavevector perpendicular to the line of sight (and have magnitude $k \sim
l / f_K(\chi_*)$ where $f_K(\chi_*)$ is the comoving angular-diameter
distance to last scattering). A radial displacement has no effect on such
plane waves since the displacement then lies in the wavefront.
The largest effect of the radial displacement in the CMB power spectra is
actually felt through its cross-correlation with the transverse lensing
displacement, but is still well below $10^{-4}$ of the primary power for
the temperature and polarization at $l\sim 1000$ (where its contribution
peaks), and below
$10^{-3}$ for the cross-power between temperature and polarization~\cite{Hu:2001yq}.
Since these effects are much smaller than those due to the transverse
deflections, we shall not consider radial delays any further.

Equations~\eqref{eq:dthetadetasq} and~\eqref{eq:dphidetasq} can be integrated twice back to a conformal time $\eta_0-\eta_*= \chi_*$ using the zero-order result $\chi=
\eta_0 - \eta$ since it only appears in the argument of a function multiplying
first-order terms. Then, changing integration order, the integral over $\int f_K^{-2}(\eta_0-\eta')\ud\eta'$ can be done using the explicit form in Eq.~\eqref{f_K_def}. The result is~\cite{Kaiser:1996tp}
%
\begin{eqnarray}
\theta(\eta_0-\chi_*) &=& \theta_0 -
\int_0^{\chi_*} \ud \chi \, \frac{f_K(\chi_* - \chi)}{f_K(\chi_*)f_K(\chi)}
2 \frac{\partial}{\partial \theta}\Psi(\chi \vnhat ; \eta_0-\chi) , \label{eq:thetadeflect} \\
\phi(\eta_0-\chi_*) &=& \phi_0 -  \int_0^{\chi_*} \ud \chi \,
\frac{f_K(\chi_* - \chi)}{f_K(\chi_*)f_K(\chi)}
\frac{2}{\sin^2\theta}
\frac{\partial}{\partial \phi}\Psi(\chi \vnhat ; \eta_0-\chi) , \label{eq:phideflect}
\end{eqnarray}
where $\theta_0$ and $\phi_0$ label the line of sight $\vnhat$.
The small angular displacements $\theta-\theta_0$ and $\phi-\phi_0$
form the components of the displacement vector $\valpha$ on the sphere.
Noting that $\alpha_\theta = \theta-\theta_0$, but
$\alpha_\phi = \sin^2\theta (\phi-\phi_0)$, we find
\begin{equation}
\valpha = -2  \int_0^{\chi_*} \ud \chi\,
\frac{f_K(\chi_* - \chi)}{f_K(\chi_*)f_K(\chi)}
\nabla_\vnhat\Psi(\chi \vnhat ; \eta_0-\chi)
\label{eq:geo1}
\end{equation}
where $\vnhat$ is the covariant derivative on the sphere.
This derivation of the deflection agrees with the heuristic derivation
given earlier since we can identify the transverse derivative
in Eq.~\eqref{delta_theta} with $[1/f_K(\chi)]\nabla_\vnhat$.

This result actually holds for any theory described in differential geometry with a metric conformally equivalent to Eq.~\eqref{eq:metric}. General Relativity specifically only enters through the relation between the Weyl potential and the stress-energy tensor, as determined by the Einstein equation. The result is the Poisson equation
\begin{equation}
(\Delta+3K/a^2)\Psi = 4\pi G \left(\bar{\delta\rho} + \Pi\right),
\label{Poisson}
\end{equation}
where $\bar{\delta\rho}$ is the comoving total density perturbation (i.e. evaluated in the rest-frame of the total energy), $\Pi$ is the anisotropic stress, and $\Delta$ is the three-dimensional Laplacian. In matter (and dark energy) domination where lensing is important the anisotropic stress is small and the Weyl potential is then directly related to the comoving matter perturbations.

\subsection{Geodesic deviation}
\label{sec:bundle}

An alternative way to describe gravitational lensing concentrates
on the gravitational distortion of the invariant cross-section of
infinitesimal bundles of light rays~\cite{Blandford:1991,Wald:GR,Seitz:1994xf,Bartelmann:1999yn,VanWaerbeke:2003uq,Perlick04,Sachs61,SchneiderBook}.
The geodesic deviation equation can be used to follow the evolution in the
separation of nearby rays. The advantage of this approach is that it deals
exclusively with physically-observable quantities; the disadvantage
is that by not following rays at finite separation it is not
straightforward to characterize the global properties of the light cone.
This means that some important properties
of strong-lensing situations (i.e.\ where there is multiple covering of
the wave front)
such as the number of images a source
forms, are not readily calculable. However, here we are largely
interested in weak lensing and in this limit the approach is essentially
equivalent to that given above based on the perturbed geodesics.

We shall express the propagation of the cross-section of the bundle of rays
from the 1+3-covariant viewpoint of a field of observers with 4-velocity $u^a$.
Working in the geometric-optics limit, we consider two null geodesics
$x^a(\lambda)$ and $x^a(\lambda) + \delta x^a(\lambda)$
that lie in the past light cone of some event $A$ (so all generators
of the light cone focus at $A$). Here $\lambda$ is an affine parameter,
which we assume takes the same value at $A$ for all rays there.
In addition, we chose $\lambda$ such that $u_a \ud x^a / \ud \lambda$ varies
smoothly across the geodesics at $A$.
For rays with infinitesimal separation, the connecting vector
$\xi^a(\lambda) \equiv \delta x^a(\lambda)$ is then also infinitesimal
and lies in the null surface. This means that
$\xi^a k_a = 0$ where $k^a = \ud x^a / \ud\lambda$ is the wavevector.
The equation of motion of the connecting vector is $\uD \xi^a / \uD \lambda
\equiv k^b \nabla_b \xi^a = \xi^b \nabla_b k^a$ where the
right-hand side involves the (covariant) difference of the wavevectors at
$\lambda$ on the two geodesics. This follows since, in some arbitrary
coordinate chart,
\begin{eqnarray}
\frac{\uD \xi^\mu}{\uD \lambda} &=& \frac{\ud \xi^\mu}{\ud \lambda}
+ \Gamma^\mu_{\nu\rho} \xi^\nu k^\rho \nonumber \\
&=& k^\mu(x+\delta x) - k^\mu(x) + \Gamma^\mu_{\nu\rho} \xi^\nu k^\rho
\nonumber \\
&=& \xi^\nu \partial_\nu k^\mu + \Gamma^\mu_{\nu\rho} \xi^\nu k^\rho
\nonumber \\
&=& \xi^\nu \nabla_\nu k^\mu.
\end{eqnarray}

We now introduce the projected connecting vector $\xi^a_\perp$ that connects
the two geodesics in the local rest space of an observer with velocity
$u^a$, i.e.\ $\xi^a_\perp u_a = 0$. In general, $\xi^a_\perp$ will
differ from $\xi^a$ by a vector proportional to $k^a$ (a displacement along
the geodesic):
\begin{equation}
\xi^a_\perp = \xi^a - \frac{\xi \cdot u}{k\cdot u} k^a .
\label{eq:sachs_1}
\end{equation}
This can be written in terms of the screen projection tensor,
\begin{equation}
\clh_{ab} \equiv g_{ab} - u_a u_b + e_a e_b = g_{ab} +
\frac{k_a k_b}{(k\cdot u)^2} - 2 \frac{u_{(a}k_{b)}}{k\cdot u},
\label{eq:sachs_2}
\end{equation}
where $k^a = (k\cdot u)(u^a + e^a)$ with the (unit) direction $e^a$
perpendicular to $u^a$, as $\xi^a_\perp = \clh^a{}_b \xi^b$.
The screen-projection tensor projects perpendicular to both $u^a$ and
$e^a$, so it projects into the two-dimensional rest-space of
$u^a$ that is perpendicular to the spatial direction of the ray.
Under a (local) change of
velocity $u^a \mapsto \tilde{u}^a$, the new projected connecting vector
is obtained from that with respect to $u^a$ by applying the new
screen projection tensor: $\xi^a_\perp \mapsto \tilde{\xi}^a_\perp
= \tilde{\clh}^a{}_b \xi^b = \tilde{\clh}^a{}_b \xi_{\perp}^b$. Since
$\xi^a_\perp$ and $\tilde{\xi}^a_{\perp}$ differ only by a (null) vector
parallel to $k^a$, dot products between projected connected vectors for
neighbouring rays are independent of the 4-velocity $u^a$ and so the
shape and size of the cross-section of a ray bundle are observer-independent.

The propagation equation for $\xi^a_\perp$ follows from that for
$\xi^a$; we find
\begin{equation}
\clh_{ab} \frac{\uD\xi^b_\perp}{\uD\lambda} = \clh_{ab} \xi_\perp^c \nabla_c
k^b,
\label{eq:sachs_3}
\end{equation}
where we have used $k^b \nabla_b k^a = 0$.
The right-hand side involves the fully projected (2-dimensional) tensor
$\clh_a{}^c \clh_b{}^d \nabla_c k_d$. This is symmetric since, in the
geometric optics approximation, $k_a = \nabla_a S$ where $S$ is the phase
of the radiation field. The symmetric tensor can be decomposed
irreducibly into a trace part and a trace-free part,
\begin{equation}
\clh_a{}^c \clh_b{}^d \nabla_{(c} k_{d)} = \theta \clh_{ab} - \sigma_{ab},
\label{eq:sachs_4}
\end{equation}
where $\theta \equiv \nabla^a k_a$ defines the rate of expansion of the beam
cross-section, and the trace-free shear $\sigma_{ab}$ defines the
area-preserving distortion. These identifications follow from
substituting Eq.~(\ref{eq:sachs_4}) into Eq.~(\ref{eq:sachs_3}):
\begin{equation}
\clh_{ab} \frac{\uD\xi^b_\perp}{\uD\lambda} = \theta \xi_{\perp\, a}
- \sigma_{ab} \xi_\perp^b.
\label{eq:sachs_5}
\end{equation}
The expansion $\theta$ is thus half the fractional rate of change of
invariant area of the ray bundle.
If we differentiate Eq.~(\ref{eq:sachs_3}) again
and use the Ricci identity, $2\nabla_{[a} \nabla_{b]} k_c = R_{abc}{}^d k_d$,
we obtain the geodesic deviation equation for the projected connected
vector in 1+3-covariant form:
\begin{equation}
\clh_a{}^c \frac{\uD}{\uD\lambda}\left( \clh_{cb} \frac{\uD \xi_\perp^b}
{\uD\lambda} \right) = \clh_a{}^c \xi_\perp^b k^d k^e R_{dbce} =
\clh_a{}^c \xi_\perp^b k^d k^e C_{dbce} - \frac{1}{2}
\xi_{\perp a} R_{bc} k^b k^c.
\label{eq:sachs_5b}
\end{equation}
In the last equality, we have decomposed the Riemann tensor in the source
term into a part that depends on the Weyl tensor, $C_{abcd}$,
and a part that depends on the Ricci tensor, $R_{ab}$.
The former describes the tidal gravitational action of nearby matter, while
the latter is only operative where there is matter (including
dark energy) present by the Einstein field equations. The ten degrees
of freedom in the Weyl tensor can be represented by two projected
(perpendicular to $u^a$) symmetric trace-free (PSTF)
tensors $E_{ab}$ (`electric' part) and $H_{ab}$ (`magnetic' part):
\begin{equation}
E_{ab} \equiv u^c u^d C_{acbd} \quad , \quad
H_{ab} \equiv \frac{1}{2} \epsilon_{acd} C_{be}{}^{cd} u^e ,
\label{eq:sachs_8b}
\end{equation}
where $\epsilon_{abc} \equiv \eta_{abcd} u^d$ is the projected alternating
tensor. Writing the Weyl term in Eq.~(\ref{eq:sachs_5b}) in terms of
the electric and magnetic parts, we find
\begin{equation}
\clh_a{}^c \frac{\uD}{\uD\lambda}\left( \clh_{cb} \frac{\uD \xi_\perp^b}
{\uD\lambda} \right) =
-2 (k\cdot u)^2 \xi_\perp^b [\clh_{\langle a}{}^c \clh_{b\rangle}{}^d
E_{cd} - \clh_{(a}{}^c \epsilon_{b)}{}^d H_{cd}]
- \frac{1}{2} \xi_{\perp a} R_{bc} k^b k^c,
\label{eq:sachs_8c}
\end{equation}
where $\epsilon_{ab} \equiv \epsilon_{abc} e^c$ and the angle brackets
denote the PSTF part taken in the 2-dimensional space defined by the
projection $\clh_{ab}$.

It is convenient to express $\xi_{\perp}^a$ in terms of a two-dimensional
orthonormal spacelike basis along the ray, $E_1^a$ and $E_2^a$,
where $\clh^a{}_{b} E_I^b = E_I^a$, $I=1,2$. If we transport these vectors according
to $\clh_{ab} \uD E_I^b / \uD\lambda = 0$ (i.e.\ as parallel as the projection
properties allow), Eq.~(\ref{eq:sachs_5b}) can be written as
\begin{equation}
\frac{\ud^2 \xi_I}{\ud \lambda^2} = \mathcal{T}_{IJ} \xi_J ,
\label{eq:sachs_8d}
\end{equation}
where $\xi_I \equiv -E_I^a \xi_{\perp a}$
and the (symmetric) optical tidal matrix
is $\mathcal{T}_{IJ} \equiv - E_I^b E_J^c k^a k^d R_{abcd}$. Given the
initial condition that $\xi_\perp^a = 0$ at the focus $A$, the solution
of Eq.~(\ref{eq:sachs_8d}) can be written as~\cite{Seitz:1994xf}
\begin{equation}
\xi_I(\lambda) = -(k\cdot u)_A^{-1} \cld_{IJ}(\lambda)
\left. \frac{\ud \xi_J}{\ud \lambda}\right|_A = \cld_{IJ}(\lambda)
\delta \theta_J ,
\label{eq:sachs_8e}
\end{equation}
where the Jacobi map satisfies
\begin{equation}
\frac{\ud^2 \cld_{IJ}}{\ud \lambda^2} = \mathcal{T}_{IK} \cld_{KJ} ,
\label{eq:sachs_jacobi}
\end{equation}
with the initial condition $\cld_{IJ}|_A = 0$
and $\ud \cld_{IJ} / \ud \lambda |_A = -(k\cdot u)_A \delta_{IJ}$.
In Eq.~(\ref{eq:sachs_8e}), we have introduced the (screen) components
$\delta\theta_I$ of the
infinitesimal angular separation of the rays at $A$; these are equal
to $-(k\cdot u)^{-1}_A \ud \xi_I / \ud \lambda |_A$ where the proportionality
constant arises from converting $\ud \lambda$ to a proper-time interval.
The optical expansion and shear can be determined
from the Jacobi map (except at caustics where $\cld_{IJ}$ is not invertible)
by
\begin{equation}
\frac{\ud \cld_{IK}}{\ud \lambda} \cld_{KJ}^{-1} =
\theta \delta_{IJ} + \sigma_{IJ} ,
\label{eq:sachs_8f}
\end{equation}
where we have used Eq.~\eqref{eq:sachs_5}. We shall
see shortly that for weak lensing in a linearly-perturbed Robertson-Walker
geometry we can relate the symmetric trace-free (STF) part of the Jacobi map
to the lensing shear, and the trace to the convergence.

We end this subsection by considering the effect of a change of
4-velocity $u^a$ on the Jacobi map.
If we change $u^a \mapsto \tilde{u}^a$, then, given
the basis $E_I^a$ (appropriate to $u^a$) along the ray, the basis
$\tilde{E}_I^a \equiv \tilde{\clh}^a{}_{b} E_I^b$ is appropriate to
$\tilde{u}^a$ since it
satisfies $\tilde{\clh}_{ab} \uD \tilde{E}_I^b / \uD \lambda = 0$. Under both
changes, the components of $\xi_\perp^a$ are invariant but the components
of the angular separation of rays at $A$ transform to $\tilde{\delta \theta}_I
= (k\cdot u/k\cdot \tilde{u})_A \delta \theta_I$. It follows that the
Jacobi map must scale so as to leave $\cld_{IJ} / (k\cdot u)_A$ independent
of $u^a$. This behaviour is seen to be consistent with its differential
equation on noting that $\clt_{IJ}$ is invariant but the initial
condition on $\ud \cld_{IJ} / \ud\lambda$ scales with $k\cdot u |_A$.
Finally, we note that the components of the optical expansion and shear
are invariant; in 1+3-covariant form the shear transforms as
$\sigma_{ab} \mapsto \tilde{\sigma}_{ab} =
\tilde{\clh}_a{}^c \tilde{\clh}_b{}^d \sigma_{cd}$, which leaves the
square $\sigma_{ab} \sigma^{ab}$ invariant.

\subsubsection{Jacobi map in linearly-perturbed cosmology}

The symmetry of the background Robertson-Walker model demands that the
Jacobi map must be $\propto \delta_{IJ}$. The determinant of the map is
the (invariant) area of the ray bundle per unit solid angle at the focus $A$
and so
\begin{equation}
\cld_{IJ}(\lambda) = a(\lambda) f_K(\chi)\delta_{IJ} ,
\label{eq:sachs_jacobi_a}
\end{equation}
where $a(\lambda)$ is the scale factor at affine parameter $\lambda$ and
$\chi$ is the comoving distance the ray then travels to $A$. The trace-free
part of $\cld_{IJ}$ thus vanishes in the background and so provides a
gauge-invariant description of gravitational lensing in perturbed
Robertson-Walker cosmologies. For weak lensing, the STF part describes the
shear of images and can be related to $\nabla_{\langle a} \alpha_{b\rangle}$,
where $\alpha_a$ is the deflection angle, as we show shortly. The antisymmetric
part of $\cld_{IJ}$ describes rotation of images; it vanishes to
linear order in the perturbations by virtue of Eq.~\eqref{eq:sachs_jacobi},
the initial conditions, and the fact that the optical tidal matrix
is symmetric and, at zero-order, proportional to the identity matrix.

The STF part, $\cld_{\la IJ\ra}$, of the Jacobi map evolves
according to the STF part of Eq.~\eqref{eq:sachs_jacobi}.
To linear order in the perturbations, we find
\begin{equation}
\frac{\ud^2 \cld_{\la IJ\ra}}{\ud\lambda^2} -
\frac{1}{a f_K(\chi)} \frac{\ud^2 [af_K(\chi)]}{\ud \lambda^2}
\cld_{\la IJ\ra} = af_K(\chi) \mathcal{T}_{\la IJ\ra},
\label{eq:sachs_jacobi_b}
\end{equation}
where we have used Eq.~(\ref{eq:sachs_jacobi_a}) and the fact that
the STF part of the optical tidal matrix, $\mathcal{T}_{\la IJ\ra}$ vanishes
in the background (it depends only on the Weyl curvature).
For scalar perturbations, the magnetic part
of the Weyl tensor vanishes and the electric part can be written as
a gradient term: $E_{ab} = - \uD_{\langle a} \uD_{b \rangle} \Psi$, where
the derivatives here are projected perpendicular to $u^a$. The Weyl tensor
is gauge-invariant and is generated from the sum of the
metric perturbations in the conformal Newtonian gauge: $\Psi =
(\PsiN - \PhiN)/2$. Expressing the Weyl tensor in terms of
$\Psi$, we find
\begin{equation}
\mathcal{T}_{\la IJ\ra} = 2 (k\cdot u)^2 E_{\langle I}^a E_{J\rangle}^b
E_{ab} = -2 \left(\frac{k\cdot u}{a f_K(\chi)}\right)^2 \nabla_{\langle I}
\nabla_{J\rangle} \Psi ,
\label{eq:sachs_jacobi_c}
\end{equation}
where $\nabla_{\langle I} \nabla_{J\rangle} \Psi$ involves the components
of the (spherical) covariant derivatives of the projection of
$\Psi$ at $\lambda$ along the line of sight onto the celestial sphere.
It is convenient to define
$\gamma_{IJ} \equiv \cld_{\la IJ\ra} /
[a f_K(\chi)]$ (or, more generally, $\cld_{\la IJ\ra} /
\sqrt{\det \cld}$, which is independent of $u^a$).
Taking the derivative, and using the
linearized form of Eq.~(\ref{eq:sachs_8f}), we can relate
$\gamma_{IJ}$ to the optical shear:
$\ud \gamma_{IJ} / \ud \lambda = \sigma_{IJ}$.
If we substitute for $\cld_{\la IJ\ra}$ in
Eq.~(\ref{eq:sachs_jacobi_b}), we find
\begin{equation}
\frac{\ud^2 \gamma_{IJ}}{\ud \eta^2} - 2 \frac{\ud \ln f_K(\chi)}{\ud \chi}
\frac{\ud \gamma_{IJ}}{\ud \eta} = - \frac{2}{f_K^2(\chi)} \nabla_{\langle I}
\nabla_{J\rangle} \Psi ,
\label{eq:sachs_jacobi_d}
\end{equation}
where we have switched to conformal time using $a \ud \eta = \ud x^a u_a
= (k\cdot u) \ud \lambda$ and used the zero-order result $\chi = \eta_A - \eta$.
It follows that near to the focus, $\cld_{\la IJ\ra} \sim
\mathcal{O}(\lambda_A - \lambda)^3$~\cite{Seitz:1994xf} and
$\gamma_{IJ} \sim \mathcal{O}(\lambda_A - \lambda)^2$. With this
boundary condition, the solution of Eq.~(\ref{eq:sachs_jacobi_d}) is
\begin{equation}
\gamma_{IJ}(\vnhat) = -2
\int_0^{\chi_*} \ud \chi\,
\frac{f_K(\chi_* - \chi)}{f_K(\chi_*)f_K(\chi)}
\nabla_{\langle I}\nabla_{J\rangle} \Psi(\chi \vnhat ; \eta_A-\chi) ,
\label{eq:sachs_shear}
\end{equation}
where we have integrated back to $\chi_*$. Comparison with
Eq.~(\ref{eq:geo1}) shows that $\gamma_{IJ} = \nabla_{\langle I}
\alpha_{J\rangle}$.

We now turn to the trace of the Jacobi map, which we denote
by $\cld \equiv \sum_I \cld_{II}$. This evolves as
\begin{equation}
\frac{\ud^2 \cld}{\ud \lambda^2} = -\frac{1}{2} \cld R_{ab} k^a k^b
\label{eq:sachs_kappa1}
\end{equation}
to first order. Note that $\cld$ does not vanish in the background
but instead is given by $\cld = 2 a(\lambda) f_K(\chi)$, and the Ricci term in Eq.~(\ref{eq:sachs_kappa1}) also has a zero-order part.
It is very convenient
to remove that zero-order part of the Ricci tensor due to the
expansion by working in the Newtonian
gauge and performing the same conformal transformation that we made in
Sec.~\ref{sec:metric}: $g_{ab} \mapsto \hat{g}_{ab} = \Omega^2 g_{ab}$, where
the conformal factor $\Omega^2 = [a^{-2}(1-2\Phi_\text{N})]$.
The conformal Ricci tensor has non-zero components
\begin{equation}
\hat{R}_{00} = 2 \uD^i \uD_i \Psi \quad , \quad \hat{R}_{ij} = 2K \gamma_{ij}
- 2 \uD_i \uD_j \Psi,
\label{eq:sachs_kappa2}
\end{equation}
where $D_i$ is the three-dimensional covariant derivative constructed from
$\gamma_{ij}$ and $\uD^i = \gamma^{ij} \uD_j$.

Under a conformal
transformation, $k_a = \nabla_a S$ is invariant so that $\hat{k}^a =
\Omega^{-2} k^a$ and so, with $\hat{k}^a = \ud x^a / d\hat{\lambda}$,
$\ud \hat{\lambda} = \Omega^2 \ud \lambda$. For an observer following a path
$x^a(\tau)$, the conformal 4-velocity is $\hat{u}^a = \Omega^{-1} u^a$
and the conformal increment in proper time is
$\ud \hat{\tau} = \Omega \ud \tau$. The Jacobi map is a ratio of proper
distances to angles so, for the same observer, the trace
$\hat{\cld} = \Omega \cld$
where $\Omega$ is evaluated at the source position.

The conformal trace
evolves according to the `hatted' version of Eq.~(\ref{eq:sachs_kappa1}),
where, for an observer at the origin, the Ricci source term is
\begin{equation}
\hat{R}_{ab} \hat{k}^a \hat{k}^b = 2 (\hat{k}\cdot \hat{u})^2\left(
K + \nabla_\perp^2 \Psi + 2\frac{\ud \ln f_K(\chi)}{\ud\chi}
\frac{\partial \Psi}{\partial \chi}\right),
\end{equation}
and $\nabla_\perp^2 = f_K^{-2}(\chi) \nabla^I \nabla_I$ is the (comoving)
transverse Laplacian. We define the first order quantity $\hatkappa\equiv 1-\hat{\cld}/[2 f_K(\chi)]$ where $\chi$ is the perturbed radial distance,
so that Eq.~\eqref{eq:sachs_kappa1} becomes
\begin{equation}
\frac{\ud^2}{\ud \hatlambda^2}\left[2f_K(\chi)\hatkappa\right] - 2\frac{\ud^2f_K(\chi)}{\ud \hatlambda^2} = \hat{\cld}(\hat{k}\cdot \hat{u})^2\left(
K + \nabla_\perp^2 \Psi + 2\frac{\ud \ln f_K(\chi)}{\ud\chi}
\frac{\partial \Psi}{\partial \chi}\right).
\label{eq:sachskapp}
\end{equation}
We now change variables from $\hatlambda$ to $\chi$. To first order we have $\ud\chi/\ud\hatlambda=-(\hat{k}\cdot\hat{u})$ from the null constraint on
$\hat{k}^a$, and the $\chi$ component of the geodesic equation gives
\begin{equation}
\frac{\ud^2f_K(\chi)}{\ud \hatlambda^2} = -(\hat{k}\cdot\hat{u})^2 f_K(\chi)\left( K  + 2\frac{\ud \ln f_K(\chi)}{\ud\chi}\frac{\partial\Psi}{\partial\chi}\right).
\end{equation}
Using these results and substituting $\hat{\cld} =2f_K(\chi)(1-\hatkappa)$,
Eq.~\eqref{eq:sachskapp} simplifies to
\begin{equation}
\frac{\ud^2}{\ud \chi^2}\left[f_K(\chi)\hatkappa\right] =  f_K(\chi) \left(\nabla_\perp^2 \Psi  - K \hatkappa\right),
\end{equation}
which can be rearranged as
\begin{equation}
\frac{\ud}{\ud\chi}\left(f_K^2(\chi)\frac{\ud \hatkappa}{\ud \chi}\right) =   \nabla^I \nabla_I \Psi.
\end{equation}
Near the focus at $A$, $\hat{\cld} = 2\chi + \clo(\chi^2)$ so that
$\kappa=\clo(\chi)$.
Integrating twice with a switch of order, and using the boundary
conditions at $A$, we get
\begin{equation}
\hatkappa(\vnhat) =
\int_0^{\chi_*} \ud \chi\,
\frac{f_K(\chi_* - \chi)}{f_K(\chi_*)f_K(\chi)}
\nabla^I\nabla_{I} \Psi(\chi \vnhat ; \eta_A-\chi) .
\label{eq:sachs_kappa}
\end{equation}
Since $\hat{\cld} = \Omega \cld$ in the non-conformal frame we can define the convergence in terms of the trace of the Jacobi map at first order as
\begin{equation}
\hatkappa(\vnhat) \equiv 1-\frac{\cld(\vnhat) }{2f_K(\chi_*)S(\chi_*\vnhat;\eta_*)} ,
\label{eq:sachs_kappa_def}
\end{equation}
where $S(\chi_*\vnhat;\eta_*)\equiv a(\eta_*)[1+\Phi_\text{N}(\chi_*\vnhat; \eta_*)]$ measures the perturbed scale factor at the source. Note that the $\chi_*$ here in $f_K(\chi_*)$ must be the perturbed value, including the radial displacement due to the time delay. The convergence is measuring only the effect on the angular area due to transverse gradients rather than including the effect of the perturbed volume element at the source or the change in distance relative to the background due to the time delay.


This completes our derivation of the shear and convergence from the Jacobi map in linearly-perturbed
Robertson-Walker cosmologies. It is straightforward to show that the result
is consistent with that implied by the lensing deflection angle and radial
delays derived in the metric approach in Section~\ref{sec:metric}.
Note that derivations in e.g. Refs.~\cite{Blandford:1991,VanWaerbeke:2003uq} appear to have neglected that time delay and $1+\Phi_N$ terms required for a consistent definition of the magnification matrix in terms of the Jacobi map to first order.

\subsubsection{Lensing deflection angle}

The Jacobi map is directly observable and, as we have seen, for weak lensing
the STF part can be related to the derivatives of the lensing deflection
angle, where the latter is calculated by ray-tracing in the conformal
Newtonian gauge. This raises the question, can we determine $\valpha$ directly
from the observable $\gamma_{IJ}$ for weak lensing, i.e.\ can we solve
\begin{equation}
\gamma_{ab} = \nabla_{\langle a} \alpha_{b \rangle}
\label{eq:sachs_10}
\end{equation}
on the sphere? Any vector field can be written as a gradient and a
divergence-free
vector: $\alpha_a = \nabla_a \psi + \epsilon_a{}^b \nabla_b \chi$, where
$\psi$ and $\chi$ are scalar functions on the sphere and $\epsilon_{ab}$
is the alternating tensor on the sphere.
Substituting in
Eq.~(\ref{eq:sachs_10}), we have
\begin{equation}
\gamma_{ab} = \nabla_{\langle a} \nabla_{b\rangle} \psi
- \epsilon^c{}_{\langle a} \nabla_{b \rangle} \nabla_c \chi.
\label{eq:sachs_11}
\end{equation}
Taking the divergence and then taking the divergence or curl again,
we find
\begin{equation}
\nabla^a \nabla^b \gamma_{ab} = \frac{1}{2} \nabla^2 (\nabla^2 + 2)\psi ,
\qquad
\epsilon^{ac} \nabla_c \nabla^b \gamma_{ab} = \frac{1}{2} \nabla^2
(\nabla^2 + 2)\chi .
\label{eq:sachs_12}
\end{equation}
On the full sphere, the (regular) kernel of $\nabla^2(\nabla^2+2)$ is
four-fold degenerate and consists of the $l=0$ and $l=1$ spherical
harmonics. It follows that the deflection field is determined up to
a dipole vector by the observed $\gamma_{ab}$.
For the undetermined dipole vector,
the curl-like part corresponds to a global (rigid) rotation and hence is
unobservable. The $l=1$ gradient part does not contribute to the lensing
shear, but does contribute to the convergence given by $\kappa = -\grad^a\alpha_a/2 = - \grad^2\psi/2$. From Eqs.~\eqref{eq:sachs_kappa} and~\eqref{eq:sachs_shear} we can see that this is consistent with our definition of $\kappa$ in terms of the Jacobi map given in Eq.~\eqref{eq:sachs_kappa_def} for $l>1$. Since $\grad^2$ is only degenerate for the monopole, the observed convergence can be inverted to recover the dipole of the lensing potential that we cannot get from the shear. Note that the measured convergence and hence recovered dipole of the lensing potential and deflection angle do depend on the velocity of the observer, as discussed further below.


\subsection{Gauge dependence and observable dipoles}
\label{sec:gauges}

It is important to understand how the observables depend on the velocity of the observer, $\vv$. First, consider the unlensed CMB. The main $\clo(\vv/c)$ effect is the generation of a significant dipole from the monopole. At next order, $\clo(\vv/c)\clo(\Delta\Theta/\Theta)$, there are additional effects due to the angular aberration of the anisotropies due to the change of frame: there will only be one choice of $\vv$ for which there is no aberration. A boost from this frame will produce a dipole-like rotation of directions on the sky, giving an effective magnification and demagnification in opposite directions, leading to the spacing of the acoustic peaks to be different on opposite hemispheres. In addition $\clo((\vv/c)^2)$ relativistic aberration of the monopole will give rise to a frequency dependent kinematic quadrupole (and higher moments), though this will not concern us here (see Refs.~\cite{Challinor:2002zh,Kamionkowski:2002nd} for more details). The frame in which the dipole vanishes will not in general be the same as the frame in which there is no aberration due to the presence of a primordial dipole contribution, and hence the difference is a gauge-invariant observable.

In linear theory, the observed dipole can be expressed in terms of a gauge-invariant source at last scattering, a gauge-invariant ISW contribution, and a local term that depends on the shear. Hence in the frame in which the shear vanishes, the observed dipole will be just the cosmologically interesting gauge-invariant contributions. The shear vanishes in the Newtonian gauge, and hence the Newtonian gauge is the the one in which observers see no local kinematic dipole.

Now consider the effect of lensing. A dipole lensing convergence has the same effect at lowest order as angular aberration due to local motion, magnifying and demagnifying in opposite hemispheres.  Our calculations of the observed angles of geodesics must include any first-order angular aberration effect.
The explicit results we calculated in the conformal Newtonian gauge show that the Jacobi map contains only terms evaluated at the source or integrated along the line of sight.  This suggests that that there can be no local kinematic angular aberration for a Newtonian gauge observer, consistent with the result above that the kinematic dipole vanishes in this frame. Newtonian gauge observers see no local kinematic dipole and no angular kinematic aberration.

For theoretical work it is very convenient to work in the frame in which there is no local angular aberration. This is the gauge in which the shear vanishes: the Newtonian gauge. This is why we have consistently used the Newtonian gauge, and will continue to do so. In this gauge $\Phi_N$ determines the perturbation to the 3-curvature and the integrated local 3-expansion; $\Psi_N$ determines the acceleration; their difference determines the first-order gauge-invariant Weyl tensor, and their sum depends on the anisotropic stress (see e.g. the appendix of Ref.~\cite{Gordon:2002gv}).

Our actual motion with respect to the CMB will be somewhat non-linear, and the observed dipole and angles need to be corrected to correspond to the Newtonian gauge predictions. The non-linear effects can be taken out easily by boosting to some frame in which the total observed dipole is zero or small~\cite{Menzies:2004vr}. In principle we could then calculate the Newtonian gauge dipole by determining the local shear of the new local spatial hypersurface and calculating the transformation required so that it vanishes. In practice we probably cannot do better than boosting to the frame in which the total dipole is zero, which will differ from the Newtonian gauge by $\sim\clo(10^{-5})$ assuming the primordial Newtonian gauge dipole has the expected amplitude.
We should then be able to measure the Newtonian gauge convergence dipole to an error of $\sim\clo(10^{-5})$, which should be useful since it is expected to be $10^{-4}$--$10^{-3}$: the convergence dipole should be measurable to an error much less than cosmic variance.

In summary, the lensing effect of the CMB is real and observable. Only the dipole convergence is sensitive to the gauge, and the Newtonian gauge dipole convergence is measurable to good accuracy in practice. Changing frame merely correspond to adding some angular aberration to the Newtonian gauge result.

\subsubsection{Other gauge effects}

It is worth commenting on the other potential gauge issues that
arise in a calculation of the weak lensing effect on the CMB. We are
interested in that part of the higher-order CMB anisotropy that
arises (approximately) from the projection of fluctuations
over a spacelike last-scattering surface. Consider making a consistent
second-order calculation of this projection. The possible second-order
effects are then:
(i) the projection of second-order corrections to the source terms (e.g.\
photon energy and momentum density) evaluated
over the zero-order last-scattering surface with the zero-order light cone;
(ii) the projection of the first-order sources over the first-order
perturbed last-scattering surface with the zero-order light cone; and (iii)
the difference in the projection of the first-order sources over the
zero-order last-scattering surface when the first-order perturbed light cone
is used instead of the zero-order one.
The sum of all three
effects is gauge-invariant but the individual effects are not separately
since they depend on how we identify points in the perturbed universe
with those in the homogeneous background (on which we think of the
perturbations as sitting). This identification defines a choice of gauge.
As an extreme example, consider fixing the gauge so that
for our given observation point, the zero-order light cone coincides with
the real perturbed one\footnote{This can be achieved by adopting the
observational coordinates of Ref.~\cite{Stoeger:94}
and references therein}. Then effect (iii) vanishes, which is what we would
naively think of as the lensing effect on the CMB. The light cone
still has an observable perturbed geometry, but in this gauge it arises
only from the local metric fluctuations rather than as an integrated effect.
Note that the gauge transformation to this observational gauge is non-local
so these two viewpoints are consistent. In the observational gauge,
CMB lensing appears in effect (i) above since, under the gauge transformation,
second-order source terms --- that depend non-locally on the fluctuations
along the line of sight --- appear from displacing the first-order source
fields by first-order amounts in the background.

In the generalization of the conformal Newtonian gauge to second-order
(or any other locally-defined gauge),
the second-order fluctuations are sourced by
non-linear physics in the \emph{prior} evolution and any additional
primordial contribution. Of the three secondary effects identified above,
(iii) is expected to dominate though this has not yet been proved rigorously
with a complete second-order calculation. It is this effect that we have
computed here and what we call the lensing effect on the CMB.

\subsection{Lensing of polarization}
\label{subsec:pol_lens}

While CMB photons propagate freely from the last scattering surface, their
number density in phase space (i.e.\ the one-particle distribution function)
is conserved. In terms of the specific intensity $I_\nu$ seen by a field of
observers with velocity $u^a$, the conserved phase space density implies
that $I_\nu / \nu^3$ remains constant~\cite{MTWbook}, where $\nu$ is the photon
frequency. The polarized specific intensity evolves in exactly the same
way provided that it is measured on a basis $E_I^a$ that evolves along the
geodesic as $\clh_{ab} \uD E_I^b / \uD \lambda = 0$~\cite{Challinor:2000as}.
If we trace
a null geodesic back to last scattering in the perturbed universe, there are
several geometric effects that alter the observed polarization from that
in the absence of lensing. First, as with the temperature (or total intensity),
we are probing fluctuations on the last scattering surface at a different
position. The transverse displacement is what is usually referred to as
the gravitational lensing effect and dominates over the effect of the radial
displacement~\cite{Hu:2001yq}. Second, the propagation direction of the perturbed
geodesic when it reaches last scattering differs from that of a photon
propagating along an unperturbed, but displaced, line of sight $\vnhat +
\valpha$. The difference arises because the propagation direction accumulates
only the lensing deflection angles but the lensing displacement $\valpha$
weights the deflections by the ratio of angular diameter distances from
lens to source and observer to source. The difference is of the same order
as the deflection angle ($\sim \text{few arcmin}$) but is much smaller than the
angular scale of the polarization at last scattering which is quadrupolar
and so should not be important (though detailed calculations of this
effect have not been reported in the literature). Note that the same process
affects the Doppler contribution to the temperature anisotropies. The final
geometric process to consider is that in propagating the polarization basis
along the perturbed line of sight $\vnhat$, the resulting vectors at last
scattering will be rotated relative to an unlensed basis transported along
$\vnhat + \valpha$. Provided we obtain the basis at $\vnhat + \valpha$ by
parallel transporting that at $\vnhat$ along the spherical geodesic connecting
the two points, the rotation angle will be $\sim 1\,\text{arcmin}$ and the
effect should also be negligible. In summary, to a good approximation,
the lensed polarization along $\vnhat$ is the same as the unlensed polarization
that would be observed along $\vnhat + \valpha$ on bases that are related
by parallel transport between the two lines of sight. We describe
lensing of CMB polarization in more detail in Section~\ref{sec:pol}.

\subsection{Higher-order lensing}
\label{sec:higherorder}

The lowest weak-lensing approximation gives the deflection angle to first order in the Weyl potential $\Psi$. The unlensed CMB is assumed to be first order, so the lensed CMB $\tilde{\Theta}(\vnhat) = \Theta(\vnhat+\valpha)$ is in a certain sense second order: it is linear perturbations at last scattering lensed by linear perturbations along the line of sight. However $\Theta(\vnhat+\valpha) = \Theta(\vnhat) + \alpha^a \grad_a\Theta(\vnhat) + \dots$, with equality only to lowest order in $\Psi$ (second order in total). It therefore differs from a second-order perturbative expansion of $\Theta$ by higher order terms. As we shall discuss in Section~\ref{sec:temp}, these higher order terms are in fact not negligible. To get an accurate result from a perturbative expansion in $\valpha$ one would need to go to at least third order and probably higher. Of course a fully consistent second-order analysis would also pick up additional physical effects (for example second-order kinetic-SZ and the small effect of
second-order corrections to the sources at last scattering).
However if these corrections are small (or can be added separately), the present analysis is likely to be much more accurate (and certainly very much easier) than evolving the Boltzmann equation to second or higher order.

As we discuss in Section~\ref{sec:nonlin}, late-time non-linear evolution of $\Psi$ can be included within the basic linear-$\Psi$ framework by simply replacing $\Psi$ with its non-linear form. However a full analysis to higher order in $\Psi$ must go beyond the Born approximation, accounting for deviations of the actual photon path from the undeflected path and multiple scattering.

In the first-order approximation, the deflection angles have zero curl and can be described as the gradient of a potential, $\valpha = \grad\psi$. However at next order in $\Psi$ this is no longer true, and in general $\valpha$ has a curl-component, an effect called \emph{field rotation}. This is represented by the rotation $\omega$ in the magnification matrix, Eq.~\eqref{mag_def}. This lens-lens effect appears in the second-order Born approximation from scattering from two sources (hence $\clo(\Psi^2)$). To see this consider applying the magnification matrix of Eq.~\eqref{mag_def} consecutively for two different sources with $\omega=0$ for each source individually. The resultant magnification matrix is given by the product of the two matrices, which is now no longer symmetric in general, and hence the overall effect includes some rotation. In addition to introducing an antisymmetric term in the magnification matrix, $\omega$, the curl of the deflection angle also means that the shear now has a curl component (Eq.~\eqref{eq:sachs_11}). This is the $B$-mode shear often used as a diagnostic in galaxy weak lensing surveys because it is expected to appear only at second order.

The power spectrum and effect of the field rotation have been calculated in Refs.~\cite{Jain:1999ir,Hirata:2003ka}. Because it only appears at second order its power spectrum is fourth order in the potential. It is around $10^{-3}$--$10^{-2}$ times smaller than the convergence power spectrum, with most effect on small scales. It is therefore likely to be much less important than uncertainty in the non-linear potential evolution that gives a much larger effect on these scales. In principle the field rotation does provide a limit to how well the $B$-mode lensing signal can be subtracted (see Section~\ref{sec:unlens}), but is not likely to be important at noise levels $\agt 0.25 \muK$-arcmin~\cite{Hirata:2003ka}. Field rotation can also arise from lensing by gravitational waves~\cite{Dodelson:2003bv,Cooray:2005hm}, though probably at an unobservably small level.

More general studies of second-order corrections to the lensing have been performed in Refs.~\cite{Dodelson:2005zj,Cooray:2002mj,Shapiro:2006em}. For example there is a Born-correction accounting for the transverse displacement of the ray from the undeflected path\footnote{The Born correction usually refers to evaluating the first-order integral expression for the magnification matrix along
the perturbed line of sight. If instead the deflection angle were evaluated
along the perturbed line of sight, and then the angular derivatives were
taken to form the magnification matrix, this would automatically include the
lens-lens coupling.}: this can be included at second order in $\Psi$ by performing a transverse Taylor expansion of the potential about the undeflected path. This and other corrections are generally very small on the scales relevant for the CMB.

\subsection{Lensing in other gravity theories}

Any metric gravitational theory than can be described by the metric of Eq.~\eqref{eq:metric} would have the same predictions for the lensing effect in terms of the Weyl potential that we obtained in General Relativity (GR). How the potential relates to the matter fields can of course change if Einstein's field equation is changed (for an example in the lensing context see Ref.~\cite{Acquaviva:2004fv}).
Other more general theories may give different results. The relativistic version of MOND due to Bekenstein~\cite{Bekenstein:2004ne} is a bi-metric theory, however in this case the photons are still null geodesics, and many of the predictions for lensing are in fact the same as GR~\cite{Zhao:2005za} (though of course the relation to the matter is different).


%% file: potential.tex
In this section we define the lensing potential, an effective integrated potential useful for calculating the effect of weak lensing on the CMB anisotropies. We then calculate the power spectrum of the lensing potential, and discuss non-linear corrections and small scale approximations. In later sections we shall use the lensing potential extensively for computing lensing deflection angles and their covariance.

We recall from Eq.~\eqref{delta_theta} that the deflection angle of a source at conformal distance $\chi_*$ is given in terms of the potential $\Psi$ by
\begin{equation}
\valpha = -2 \int_0^{\chi_*} \d\chi \frac{f_K(\chi_*-\chi)}{f_K(\chi_*)}\vgrad_\perp\Psi(\chi\vnhat; \eta_0 -\chi).
\end{equation}
The
quantity $\eta_0 -\chi$ is the conformal time at which the photon was
at position $\chi \vnhat$.
It is convenient to write $\vgrad_\perp\Psi = (\grad_{\vnhat} \Psi) / f_K(\chi)$ so that
\begin{equation}
\valpha= -2 \int_0^{\chi_*} \d\chi \frac{f_K(\chi_*-\chi)}{f_K(\chi_*)f_K(\chi)}\grad_{\vnhat}\Psi(\chi \vnhat; \eta_0 -\chi),
\end{equation}
where $\grad_\vnhat$ represents the angular derivative, equivalent to the covariant derivative on the sphere defined by $\vnhat$. This result was derived rigorously at lowest order in Eq.~\eqref{eq:geo1}. We then define the \emph{lensing potential},
\begin{equation}
\psi(\vnhat) \equiv -2 \int_0^{\chi_*} \ud \chi\,
\frac{f_K(\chi_*-\chi)}{f_K(\chi_*)f_K(\chi)} \Psi(\chi \vnhat; \eta_0 -\chi),
\label{psin}
\end{equation}
so that the deflection angle is given by $\grad_\vnhat \psi$. From now on we write this simply as $\grad\psi$. Note that the lensing potential appears to be formally divergent because of the $1/\chi$ term near $\chi=0$. However this divergence only affects the monopole potential, which does not contribute to the deflection angle. We may therefore set the monopole term to zero, and the remaining multipoles will be finite, at which point the lensing potential field is well defined.

For the CMB we can approximate recombination as instantaneous so that
the CMB is described by a single source plane at $\chi = \chi_*$. We neglect the very small effect of late-time sources, including reionization, so a single 2D map of the lensing potential on the sphere contains all the required information. For scales on which the potential $\Psi$ is Gaussian, the lensing potential will be Gaussian. On smaller scales non-linear evolution can introduce non-Gaussianity even for Gaussian primordial fields, however on acoustic scales this is a small correction so we defer discussion of non-linear evolution to a later section.

For a flat universe $f_K(\chi) = \chi$, and for simplicity we shall assume flatness from now on. The geometry only enters into the calculation of the lensing potential (and unlensed CMB) power spectra, and the generalization to a non-flat universe is straightforward. Once the lensing potential has been computed deflection angles etc. may be derived without further reference to the FRW geometry.

The lensed CMB temperature in a direction $\vnhat$, $\tT(\vnhat)$, is given by the unlensed temperature in the deflected direction, $\tT(\vnhat) = T(\vnhat') = T(\vnhat + \valpha)$ where $\valpha$ is a deflection angle. At lowest order it is a pure gradient, $\valpha = \grad\psi$.

\subsection{Power spectrum of the lensing potential}

For a Gaussian lensing potential its power spectrum and the cross-correlation with the CMB contain all the information we need to describe fully the statistics of the lensed CMB. We shall therefore now derive a result for the lensing potential angular power spectrum in terms of the power spectrum of the potential.

We expand 3-D fields into harmonic space using the Fourier convention
\begin{equation}
\Psi(\vx;\eta) = \int \frac{\ud^3 \vk}{(2\pi)^{3/2}}\,
\Psi(\vk;\eta) e^{i \vk \cdot \vx},
\label{phi:FFT}
\end{equation}
and define the power spectrum for the assumed statistically homogeneous potential by
\begin{equation}
\langle \Psi(\vk;\eta) \Psi^*(\vk';\eta') \rangle =
\frac{2\pi^2}{k^3} \clp_\Psi(k;\eta,\eta') \delta(\vk-\vk'),
\label{phi:Pk}
\end{equation}
where $\eta$ denotes conformal time. Using Eq.~\eqref{psin} the angular correlation function for the lensing potential is then
\begin{multline}
\la \psi(\vnhat) \psi(\vnhat') \ra =\\ 4  \int_0^{\chi_*} \ud \chi\,
\int_0^{\chi_*} \ud \chi'\,  \left(\frac{\chi_*-\chi}{\chi_*\chi}\right)
\left(\frac{\chi_*-\chi'}{\chi_*\chi'}\right) \int \frac{\ud^3 \vk}{(2\pi)^3} \frac{2\pi^2}{k^3} \clp_\Psi(k;\eta,\eta') e^{i\vk\cdot \vx}e^{-i\vk\cdot \vx'}.
\end{multline}
We now use the result
\begin{equation}
e^{i\vk\cdot \vx} = 4\pi \sum_{lm} i^l j_l(k\chi) Y^*_{lm}(\vnhat) Y_{lm}(\hat{\vk})
\end{equation}
where $j_l(r)$ is a spherical Bessel function given in terms of standard Bessel functions by $j_l(r)=(\pi/2r)^{1/2} J_{l+1/2}(r)$. The angular integral over $\hat{\vk}$ can then be done using the orthogonality of the spherical harmonics and we have
\begin{multline}
\la \psi(\vnhat) \psi(\vnhat') \ra =
16\pi\sum_{l l'm m'}\int_0^{\chi_*} \ud \chi\,
\int_0^{\chi_*} \ud \chi'\,  \left(\frac{\chi_*-\chi}{\chi_*\chi}\right)
\left(\frac{\chi_*-\chi'}{\chi_*\chi'}\right) \\
\times\int \frac{\ud k}{k} j_l(k\chi)j_{l'}(k\chi')\clp_\Psi(k;\eta,\eta') Y_{lm}(\vnhat) Y_{l'm'}^*(\vnhat')\delta_{ll'}\delta_{mm'}.
\label{psiCorr}
\end{multline}
The lensing potential can be expanded in spherical harmonics as
\begin{equation}
\psi(\vnhat) = \sum_{lm} \psi_{lm} Y_{lm}(\vnhat),
\end{equation}
and for a statistically isotropic field the angular power spectrum $C_l^\psi$ is defined by
\begin{equation}
\la \psi_{lm} \psi_{l'm'}^* \ra = \delta_{ll'}\delta_{mm'} C_l^\psi.
\label{Cphidef}
\end{equation}
Taking the spherical harmonic components of Eq.~\eqref{psiCorr} we therefore get
\begin{equation}
C_l^\psi = 16\pi \int \frac{\ud k}{k}\, \int_0^{\chi_*} \ud \chi\,
\int_0^{\chi_*} \ud \chi'\, \clp_\Psi(k;\eta_0-\chi,\eta_0-\chi')
j_l(k\chi) j_l(k\chi') \left(\frac{\chi_*-\chi}{\chi_*\chi}\right)
\left(\frac{\chi_*-\chi'}{\chi_*\chi'}\right).
\label{cpsi}
\end{equation}

\begin{figure}
\begin{center}
\psfig{figure=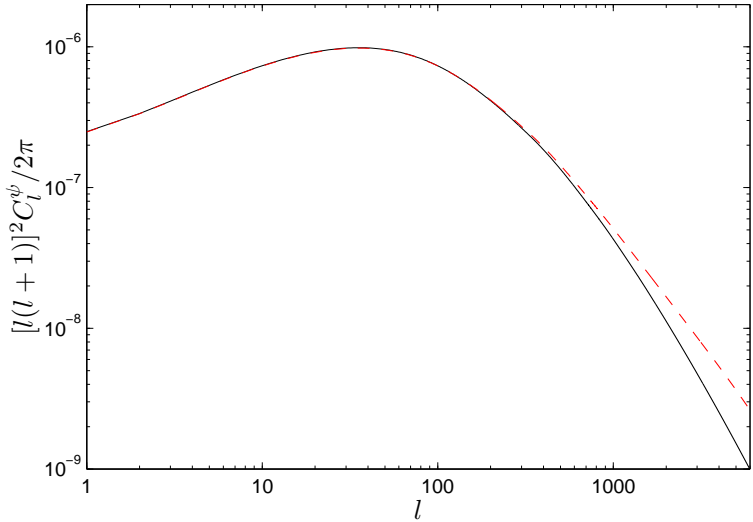,width=10cm}
\caption{The power spectrum of the deflection angle (given in terms of the lensing potential $\psi$ by $\grad\psi$) for a concordance $\Lambda$CDM model.
The linear theory spectrum (solid) is compared with the same model including non-linear corrections (dashed) from \HALOFIT~\cite{Smith:2002dz}.
\label{CPhi}}
\end{center}
\end{figure}

In linear theory we can define a transfer function $T_\Psi(k;\eta)$ so that $\Psi(\vk;\eta) = T_\Psi(k;\eta) \primvar(\vk)$ where $\primvar(\vk)$ is the
primordial comoving curvature perturbation (or other variable for isocurvature modes). We then have
\begin{eqnarray}
C_l^\psi  &=& 16\pi \int \frac{\ud k}{k}\, \clp_{\primvar}(k) \left[\int_0^{\chi_*} \ud \chi\,
T_\Psi(k;\eta_0-\chi) j_l(k\chi) \left(\frac{\chi_*-\chi}{\chi_*\chi}\right)\right]^2,
\label{cpsi_transfer}
\end{eqnarray}
where the primordial power spectrum is $\clp_{\primvar}(k)$. Given some primordial power spectrum this can be computed easily numerically using Boltzmann codes such as \CAMB\footnote{\url{http://camb.info}}~\cite{Lewis:1999bs}. Since it is deflection angles that are physically relevant, it is usual to plot the power spectrum of the deflection angle $\grad\psi$ given by $l(l+1)C_l^\psi$, as shown for a typical model in Fig.~\ref{CPhi}. Note that for $l\ge 1$ the Bessel functions go to zero at the origin, $j_l(k\chi) \rightarrow 0$ as $\chi\rightarrow 0$, so the $l\ge 1$ power spectrum is finite and well defined.

\begin{figure}
\begin{center}
\psfig{figure=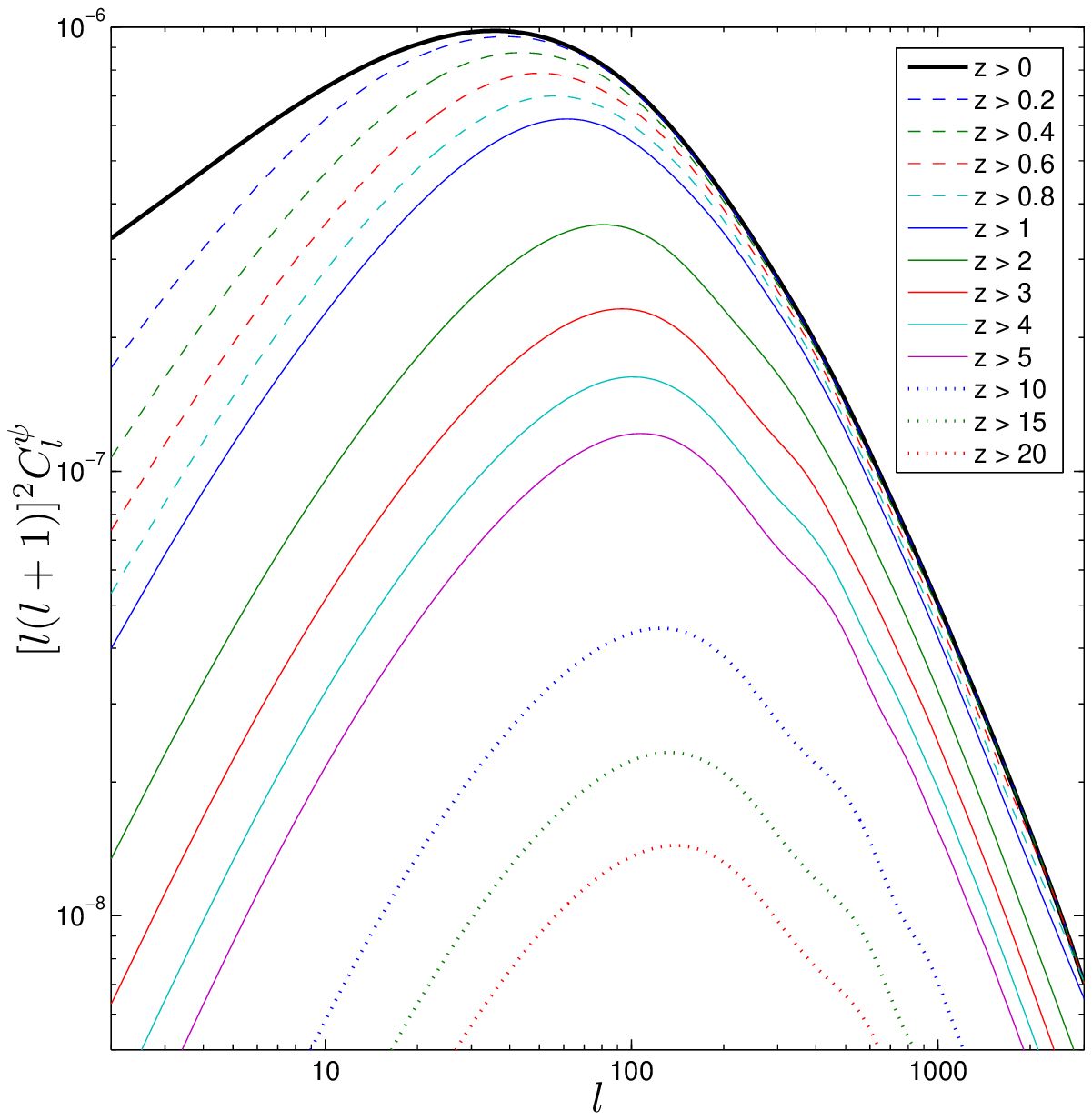,width=7cm}
\psfig{figure=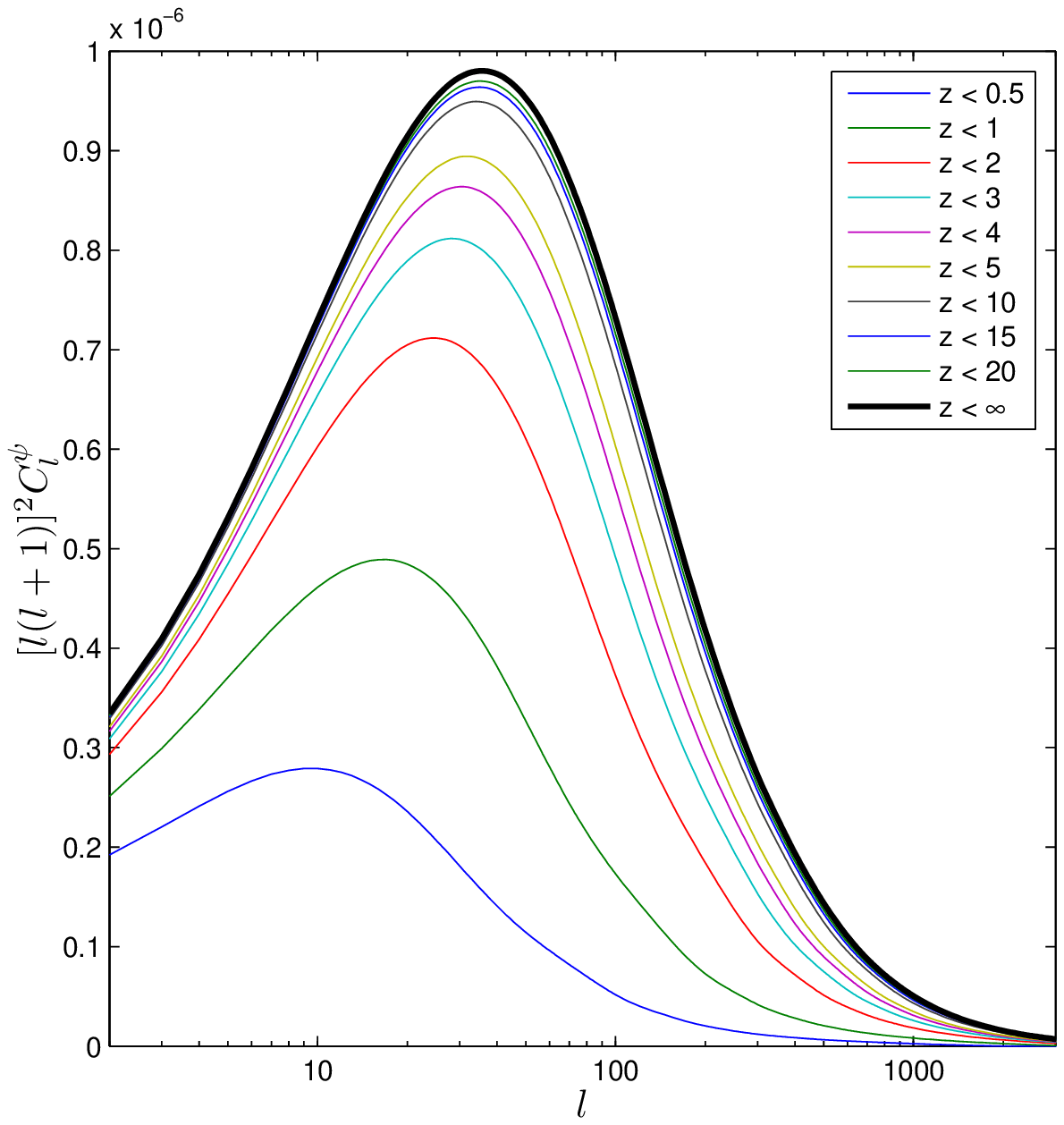,width=7cm}
\caption{Cumulative contribution of different redshifts to the power spectrum of the lensing potential for a concordance $\Lambda$CDM model. Note we have used a log scale for $C_l^\psi$  in the left-hand plot, but linear in the right-hand plot.
\label{phi_z}}
\end{center}
\end{figure}
\begin{figure}
\begin{center}
\psfig{figure=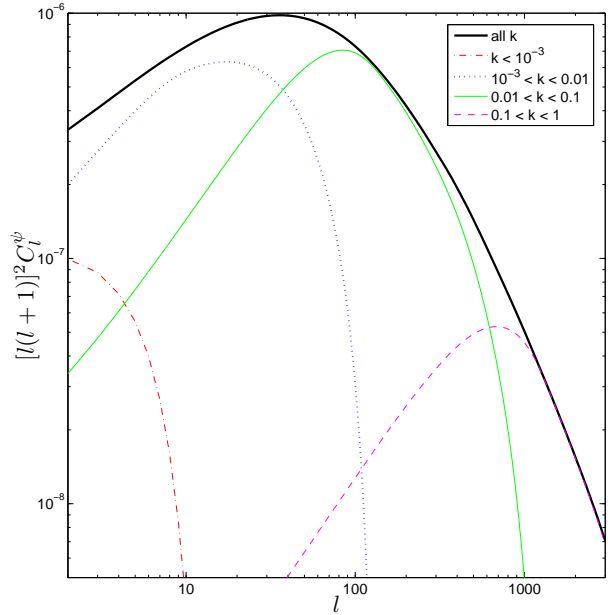,width=8cm}
\caption{Contributions of different wavenumbers $k$ (in $\Mpc^{-1}$) to the power spectrum of the lensing potential for a concordance $\Lambda$CDM model.
\label{phi_k}}
\end{center}
\end{figure}

The last scattering surface is a long way away, so the lensing potential has contributions out to quite high redshift as show in Fig.~\ref{phi_z}. Nearby low redshift potentials only contribute to the large-scale lensing, so the spectrum is only quite weakly sensitive to late time non-linear evolution. The contributions from different wavenumber ranges are shown in Fig.~\ref{phi_k}.

\begin{figure}
\begin{center}
\psfig{figure=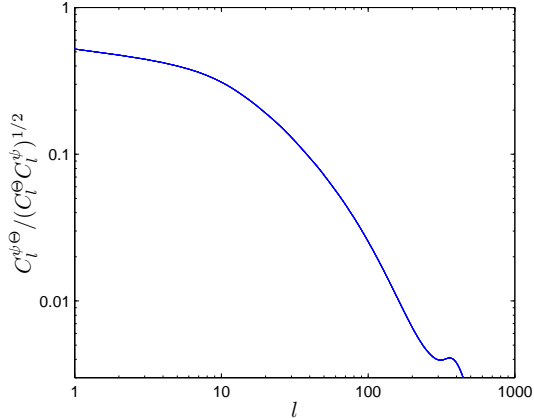,width=7cm}
\caption{Correlation of the lensing potential with the CMB temperature.}
\label{phiTcorr}
\end{center}
\end{figure}

The cross-correlation with the temperature can be worked out using an analogous derivation~\cite{Hu:2000ee}, and a typical result is shown in Fig.~\ref{phiTcorr}. On very large scales the correlation is significant because of the ISW contribution to the large-scale temperature anisotropy caused by time varying potentials along the line of sight. However on smaller scales where the ISW contribution is much smaller, and the lensing potential becomes almost totally uncorrelated to the temperature. The deflection power spectrum peaks at $l\sim 60$, at which point the correlation is $\alt 10\%$. For most applications, for example computing the lensed CMB power spectra, this small correlation can be safely neglected~\cite{Linder:1996rb}. However the significant large-scale correlation does mean that we already know quite a lot about the large-scale lensing potential from observations of the CMB temperature.

Note the lensing potential has physically relevant modes down to $l=1$ (see Section~\ref{sec:gauges}).

\subsection{Limber approximation}

At high $l$ the power spectrum $\clp_\Psi(k)$ varies slowly compared to the
spherical Bessel functions in Eq.~(\ref{cpsi}), which pick out the scale
$k \sim l/\chi$. Using
\begin{equation}
\int k^2 \ud k \, j_l(k\chi) j_l(k\chi') = \frac{\pi}{2\chi^2}
\delta(\chi-\chi') ,
\end{equation}
we can Limber-approximate $C_l^\psi$ as
\begin{equation}
C_l^\psi \approx \frac{8\pi^2}{l^3} \int_0^{\chi_*} \chi \ud \chi\,
\clp_\Psi(l/\chi;\eta_0-\chi) \left(\frac{\chi_*-\chi}{\chi_*\chi}\right)^2.
\end{equation}
This approximation is rather good, and is used by the numerical code~\CMBFAST\footnote{\url{http://cmbfast.org}}. On small scales it is numerically much more efficient than Eq.~\eqref{cpsi_transfer} due to the reduced dimensionality of the required integral.

The power spectrum of the potential can be related to the power spectrum of the density perturbations using the Poisson equation, Eq.~\eqref{Poisson}. For a flat universe in matter or dark energy domination
\begin{equation}
\clp_\Psi(k;\eta) = \frac{9\Omega_m^2(\eta)H^4(\eta)}{4} \frac{\clp_{\bar\delta}(k;\eta)}{k^4} =\frac{9\Omega_m^2(\eta)H^4(\eta)}{8\pi^2} \frac{P(k;\eta)}{k},
\end{equation}
where $\clp_{\bar\delta}(k;\eta)$ is the power in the fractional comoving density perturbation, $P(k;\eta)$ is the conventionally defined matter power spectrum, $H(\eta)$ is the Hubble parameter, and $\Omega_m(\eta)$ is the fraction of the energy density in matter.

\subsection{Non-linear evolution}
\label{sec:nonlin}

The most important assumption we have made so far is that the lensing potential is linear and Gaussian. On small scales this will not be quite correct due to non-linear evolution. The main effect is to change the $\Psi$ power spectrum on small scales, and can be estimated from numerical simulations. For simple models, fits to numerical simulation like the \HALOFIT\ code of Ref.~\cite{Smith:2002dz} can be used
to compute an
approximate non-linear, equal-time power spectrum given an accurate numerical linear power spectrum at some redshift. \HALOFIT\ is expected to be accurate at the few percent level for standard $\Lambda$CDM models with power-law primordial power spectra. For more general models, for example with an evolving dark energy component, further simulations are required (see e.g.~\cite{McDonald:2005gz}). Given a method for computing the non-linear power spectrum, a good approximation is simply to scale the potential transfer functions $T_\Psi(k,\eta)$ of Eq.~\eqref{cpsi_transfer} so that the power spectrum of the potential $\Psi$ has the correct non-linear form at that redshift:
\begin{equation}
T_\Psi(k,\eta) \rightarrow T_\Psi(k,\eta) \sqrt{\frac{\clp^\text{non-linear}_\Psi(k;\eta)}{\clp_\Psi(k;\eta)}}.
\label{Tnonlin}
\end{equation}
Since non-linear effects on $C_l^\psi$ are only important where the Limber
approximation holds, Eq.~\eqref{Tnonlin} should be very accurate.

The effect of the non-linear evolution on the power spectrum of the lensing potential is shown in Fig.~\ref{CPhi}. Although there is very little effect on scales where the power peaks ($l\sim 60$), non-linear evolution significantly increases the power on small scales.
For future reference, this changes the lensed temperature power spectrum $\tilde{C}_l^{\Theta}$ by $\alt 0.2\%$ for $l\sim 2000$, but there are percent level changes on smaller scales. The effect on the $B$-mode polarization power spectrum is more dramatic, giving a $>6\%$ increase in power on all scales. On scales beyond the peak in the unlensed $B$-mode power
($l\agt 1000$) the extra non-linear power becomes more important, producing an order unity change in the $B$-mode spectrum on small scales. On these scales the assumption of Gaussianity is probably not very good, and the accuracy will also be limited by the precision of the non-linear power spectrum.

%% file: TempCl.tex
Observations of the CMB temperature power spectrum are becoming increasingly accurate. The anisotropies at last scattering are expected to be closely Gaussian, and the power spectrum, $C_l$, can be worked out accurately in linear theory. However we observe the lensed temperature field, which will have a modified power spectrum (as well as having a non-Gaussian structure). This effect must be taken into account to obtain accurate cosmological parameter constraints from CMB observations.

We first describe a simple derivation of the lensed $C_l$ using a series expansion in the deflection angle, which turns out to be a good way to understand the most important effects~\cite{Hu:2000ee}. However it is not very accurate because on small scales the lensing deflection will be of the same order as the wavelength being deflected, and a perturbative expansion in the deflection angle is then not a good approximation. We therefore also discuss a non-perturbative calculation of the lensing effect by considering the correlation function~\cite{Seljak:1996ve,Challinor:2005jy}. Another approach to the lensed power spectrum is described in Ref.~\cite{Metcalf:1997ih}.

A calculation of the lensed power spectrum to within cosmic variance must take account of sky curvature, however this effect is small so we postpone discussion of curvature corrections to Section~\ref{sec:curv}. Non-linear evolution must also be included to get the small-scale spectrum accurately, and can be included in the Gaussian approximation by using a lensing potential computed using the non-linear power spectrum as described in Section~\ref{sec:nonlin}.

 \subsection{Simple lowest-order calculation}
\label{harmonic_flat}
As a first calculation of the effect on the CMB, we compute the lensing power spectrum at lowest order in $C_l^\psi$. To do this we work in harmonic space, using the symmetric flat-sky 2D Fourier transform convention for the temperature field:
\begin{eqnarray}
\Theta(\vx) = \int \dFT{\vl} \Theta(\vl) e^{i\vl\cdot \vx},\qquad
\Theta(\vl) = \int \dFT{\vx} \Theta(\vx) e^{-i\vl\cdot \vx}.
\label{Fourier}
\end{eqnarray}
We shall assume that the temperature fluctuations are statistically isotropic, which means that any statistical average cannot depend on where on the sky (or in which orientation) it is evaluated. The correlation function $\xi$ of the temperature at two points can therefore only depend on the separation between the points
\begin{equation}
\la \Theta(\vx) \Theta(\vx') \ra = \xi(|\vx - \vx'|).
\end{equation}
With this assumption the covariance of the unlensed Fourier components is
\begin{align}
\la \Theta(\vl) \Theta^*(\vl') \ra & = \int \dFT{\vx} \int \dFT{\vx'} e^{-i\vl\cdot\vx} e^{i\vl'\cdot\vx'} \xi(|\vx-\vx'|) \nonumber\\
& = \int \dFT{\vx} \int \dFT{\vr} e^{i(\vl'-\vl)\cdot\vx}e^{i\vl'\cdot \vr} \xi(r) \nonumber\\
& = \delta(\vl'-\vl)\int \ud^2\vr  e^{i\vl\cdot\vr} \xi(r).
\label{xi_relation}
\end{align}
In the second line we changed variables to $\vr = \vx -\vx'$, and have defined $r\equiv |\vr|$. We define the power spectrum $C_l^\Theta$ so that
\begin{align}
C_l^\Theta &= \int \ud^2\vr  e^{i\vl\cdot\vr} \xi(r)=  \int  r\ud r \int \ud\phi_\vr\, e^{i l r \cos(\phi_\vl -\phi_\vr)} \xi(r) = 2\pi \int r\ud r\, J_0(l r) \xi(r).
\label{flatCl-corr}
\end{align}
Here we have used the Bessel function $J_n(r)$ that arises from integrating the result
\begin{equation}
e^{i r\cos\phi} = \sum_{n=-\infty}^\infty  i^n J_n(r) e^{i n \phi} = J_0(r) + 2 \sum_{n=1}^\infty i^n J_n(r)\cos(n\phi).
\label{BessJ}
\end{equation}
The power spectrum for our statistically isotropic temperature field is therefore diagonal in $\vl$, and is given by
\begin{equation}
\langle \Theta(\vl) \Theta^*(\vl') \rangle = C_l^\Theta \delta(\vl-\vl').
\end{equation}
We can perform a similar Fourier transform for the lensing potential $\psi$, and define the corresponding power spectrum $C_l^\psi$ for the flat sky (equivalent to Eq.~\eqref{Cphidef} on small scales).

\begin{figure}
\begin{center}
\psfig{figure=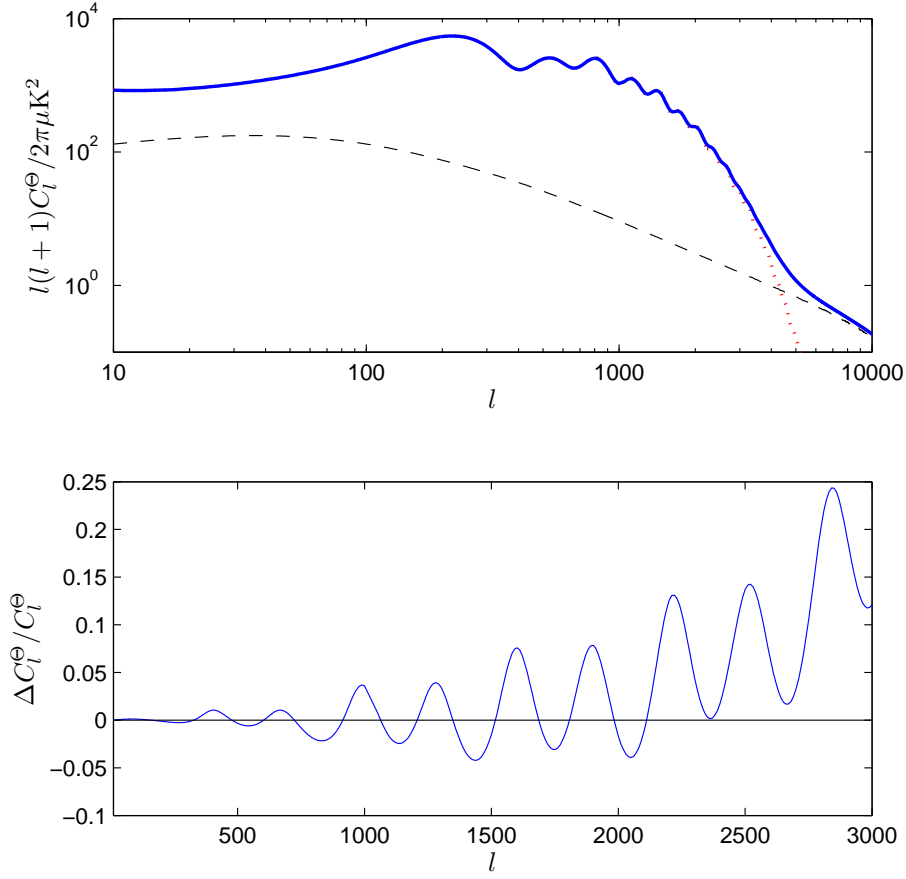,width=12.0cm}
\caption{Top: the lensed temperature power spectrum (solid) and the unlensed spectrum (dotted), compared to the large $l$ asymptotic result of Eq.~\eqref{TT_highl_approx} (dashed). Bottom: the fractional change in the power spectrum due to lensing. Both plots are for a typical concordance $\Lambda$CDM model.
\label{lensedT}}
\end{center}
\end{figure}

Lensing re-maps the temperature according to
\begin{eqnarray}
\tTheta(\vx) &=& \Theta(\vx') = \Theta(\vx + \vgrad\psi) \nonumber \\
&\approx& \Theta(\vx) + \grad^a \psi(\vx) \grad_a \Theta(\vx) + \half \grad^a \psi(\vx) \grad^b \psi(\vx) \grad_a\grad_b \Theta(\vx) + \dots
\end{eqnarray}
As discussed, the series expansion is not a good approximation on all scales. However it can be used to get qualitatively correct results for the lensed $C_l$, and is useful for giving a simple derivation to aid understanding.
Introducing the Fourier transform of the lensing potential, $\psi(\vl)$, we have
\begin{equation}
\vgrad\psi(\vx) = i \int \dFT{\vl} \vl \psi(\vl) e^{i\vl\cdot \vx}, \qquad
\vgrad\Theta(\vx) = i \int \dFT{\vl} \vl \Theta(\vl) e^{i\vl\cdot \vx}.
\end{equation}
Taking the Fourier transform of $\tTheta(\vx)$ and substituting we get the Fourier components to second order in $\psi$
\begin{align}
\tTheta(\vl) \approx &\,\Theta(\vl) - \int \dFT{\vl'}  \vl'\cdot (\vl-\vl') \psi(\vl-\vl')\Theta(\vl') \nonumber\\
          & - \frac{1}{2} \int \dFT{\vl_1} \int \dFT{\vl_2}
 \vl_1\!\cdot\![\vl_1 +\vl_2-\vl]\, \vl_1\!\cdot\! \vl_2 \Theta(\vl_1)\psi(\vl_2)\psi^*(\vl_1+\vl_2-\vl).
\label{T_series_l}
\end{align}
We now work out the power spectrum of the lensed field $\tTheta(\vx)$ to lowest order in $C_l^\psi$.
We can neglect the $\la \Theta(\vl) \psi(\vl)\ra$ correlation
since the small scale temperature at last scattering is almost uncorrelated with the late-time potentials responsible for the lensing, and the correlation only has a tiny effect on the result.
The covariance for the lensed temperature remains diagonal due to statistical isotropy, so that
\begin{equation}
\la \tTheta(\vl)\tTheta^*(\vl')\ra = \delta(\vl-\vl') \tC^\Theta_l.
\end{equation}
At lowest order in $C_l^\psi$ there are contributions from the square of the first order term in $\tTheta$, and also a cross-term from the zeroth and second order terms. Using $\psi(\vl) = \psi^*(-\vl)$, we get
\begin{align}
\tC^\Theta_l
\approx   C_l^\Theta +
 \int \frac{\ud^2 \vl'}{(2\pi)^2} \left[ \vl'\cdot(\vl-\vl')\right]^2 C_{|\vl-\vl'|}^\psi C_{l'}^\Theta
 - C^\Theta_l \int \frac{\ud^2 \vl'}{(2\pi)^2} (\vl\cdot\vl')^2 C^\psi_{l'}.
\end{align}
This simplifies to the final result~\cite{Hu:2000ee}
\begin{equation}
\tC^\Theta_l
\approx  (1-l^2 R^\psi) C_l^\Theta +
 \int \frac{\ud^2 \vl'}{(2\pi)^2} \left[ \vl'\cdot(\vl-\vl')\right]^2 C_{|\vl-\vl'|}^\psi C_{l'}^\Theta.
\label{TT_lensed_series}
\end{equation}
Here we have defined half the total deflection angle power
\begin{equation}
R^\psi\equiv \frac{1}{2} \la |\grad\psi|^2\ra = \frac{1}{4\pi}\int \frac{\ud l}{l} \,l^4 C^\psi_l, \label{R_def}
\end{equation}
where for typical models $R^\psi\sim 3\times 10^{-7}$, corresponding to an rms deflection of $\sim 2.7\, \arcmin$. The integral term has the form of a convolution of the unlensed temperature spectrum with the lensing potential power spectrum. This convolution serves to smooth out the main peaks in the unlensed spectrum, which is the main qualitative effect on the temperature spectrum on large scales. The effect is shown in Fig.~\ref{lensedT}, and is several percent at $l \agt 1000$. On small scales where there is little power in the unlensed CMB the convolution transfers power from large scales to small scales to increase the small-scale power.

\subsubsection{Scale-invariant spectrum}

The deflection angle power spectrum $l^4 C_l^\psi$ peaks on large scales $l\sim 50$ as shown in Fig.~\ref{CPhi}.  For $l\gg 50$ the bulk of the convolution term in Eq.~\eqref{TT_lensed_series} therefore comes from scales with $l\sim l'$. Consider what happens if the temperature power spectrum is scale invariant on relevant scales, $l^2 C_l^\Theta = \text{const}$. The lensed power spectrum is then given by
\begin{eqnarray}
\tC^\Theta_l
&\approx&  (1-l^2 R^\psi) C^\Theta_l +
l^2 C^\Theta_l \int \frac{\ud^2 \vl'}{(2\pi)^2}\frac{ \left[ \vl'\cdot(\vl-\vl')\right]^2}{l'{}^2 } C_{|\vl-\vl'|}^\psi \nonumber\\
&=&  (1-l^2 R^\psi) C^\Theta_l +
 l^2 C^\Theta_l \int  \frac{\ud^2 \vl_1}{(2\pi)^2} \frac{[(\vl_1-\vl)\cdot \vl_1]^2}{(\vl-\vl_1)^2} C^\psi_{l_1} \nonumber\\
&=&  (1-l^2 R^\psi) C^\Theta_l +
 l^2 C^\Theta_l \int \frac{\ud l_1}{l_1} \,\frac{l_1^4 C^\psi_{l_1}}{4\pi}  + l^2 C^\Theta_l \int_l^\infty \frac{\ud l_1}{l_1} \,\frac{l_1^4 C^\psi_{l_1}}{4\pi}\left(1-(l/l_1)^2\right) \nonumber\\
 &=&  C^\Theta_l \left[ 1 + l^2  \int_l^\infty \frac{\ud l_1}{l_1} \,\frac{l_1^4 C^\psi_{l_1}}{4\pi}\left(1-(l/l_1)^2\right)\right].
\label{TT_lensed_scaleinv}
\end{eqnarray}
The remaining integral is generally small, and the lensed spectrum only deviates from scale invariant at the $\clo(10^{-3})$ level. If there were no lensing power at $l>l_0$, scale invariance would be preserved on scales $l>l_0$: a large-scale lensing mode magnifies and demagnifies small-scale structures, which has no effect if the structures are scale invariant. Lensing of the CMB is important because the acoustic oscillations and small scale damping give a well defined non-scale-invariant structure to the power spectrum.

\subsubsection{White spectrum}

Though of limited relevance for CMB lensing, it is also interesting to consider the lensing effect on a white spectrum with $C_l^\Theta = \text{const}$. In this case we have~\cite{Mandel:2005xh}
\begin{eqnarray}
\tC^\Theta_l
&\approx&  (1-l^2 R^\psi) C_l^\Theta +
C_{l}^\Theta \int \frac{\ud^2 \vl_1}{(2\pi)^2} \left[ \vl_1\cdot(\vl+\vl_1)\right]^2 C_{l_1}^\psi\nonumber \\
&=&  C_l^\Theta + C_{l}^\Theta \int \frac{\ud l_1}{l_1} \frac{ l_1^6 C_l^\psi}{2\pi}\nonumber \\
&=& C_l^\Theta\left(1 + 4\la \kappa^2\ra\right),
\label{TT_lensed_white}
\end{eqnarray}
where we used $\kappa = - \grad^2\psi/2$ to identify the variance of the convergence. Thus a white spectrum lenses to a white spectrum scaled by a factor depending on the total power in the lensing convergence.

\subsubsection{Small-scale limit}

The fact that $l^2 R^\psi \sim \clo(1)$ at $l\sim 2500$ in Eq.~\eqref{TT_lensed_series} serves as a reminder that the lowest-order series-expansion result is not likely to be accurate on all scales. This is because on scales with $l\sim 2000$ the deflection angles are becoming comparable to the wavelength of the unlensed fields, so the Taylor expansion is no longer a good approximation.
For accurate results at all $l$ we need to perform a calculation that does not rely on a series expansion. In principle this could be done by extending the above calculation to all orders~\cite{Cooray:2003ar}, however in practice it is most easily done by considering the correlation function as we discuss in the next section. See also Ref.~\cite{Mandel:2005xh} for an interesting alternative approach.

Although the series expansion is not accurate for $l\sim 2000$, on
 very small scales, $l \gg 3000$, the series expansion result is actually accurate again. This is because the perturbations in the unlensed spectrum are nearly wiped out by diffusion damping, so the unlensed spectrum has very little power and can be described accurately on small scales by a single gradient term. Although the deflection angles are larger than the unlensed wavelengths on these scales, a large bulk deflection is unobservable so the effect on the observed small scales is the same \emph{as if} the lensed field could be described by $\tT \approx \vgrad\psi \cdot \vgrad T$. In particular if we set $C_l^\Theta = 0$ in the first term of Eq.~\eqref{TT_lensed_series} and use $l' \ll l$ we obtain~\cite{Zaldarriaga:2000ud,Hu:2000ee}
\begin{eqnarray}
\tC_l^\Theta &\approx&  \int \frac{\ud^2 \vl'}{(2\pi)^2} \left[ \vl'\cdot(\vl-\vl')\right]^2 C_{|\vl'-\vl|}^\psi C_{l'}^\Theta\nonumber \\
&\approx&  C_{l}^\psi \int \frac{\ud^2 \vl'}{(2\pi)^2} \left[ \vl'\cdot \vl \right]^2 C_{l'}^\Theta\nonumber \\
      &\approx&  l^2 C_l^\psi \int \frac{\ud l_2}{l_2} \frac{l_2^4 \, C_{l_2}^\Theta}{4\pi}\nonumber\\
      &\approx& l^2 C_l^\psi R^\Theta.
\label{TT_highl_approx}
\end{eqnarray}
Here we defined
\begin{equation}
R^\Theta \equiv \frac{1}{2} \la |\grad T|^2\ra = \frac{1}{4\pi}\int \frac{\ud l}{l} \,l^4 C^\Theta_l \sim 10^9\muK^2
\label{RT_def}
\end{equation}
similarly to Eq.~\eqref{R_def} to measure the power in the unlensed CMB temperature gradient. The lensed CMB power is therefore proportional to the unlensed gradient power $R^\Theta$, and to the power in the deflection angle $l^2 C_l^\psi$.

This can be understood as follows. Small scale clumps (or voids) in the lensing potential disturb the smooth background CMB gradient by the varying deflection angle over the clump, giving a clump-scale feature in the observed temperature. The amplitude of this wiggle is proportional to both the strength of the background gradient, and the size of the deflection angles. The wiggle is on the same scale as the clump, and the lensed power spectrum therefore follows the scale dependence of the deflection angle. See  Section~\ref{subsec:clusters} for further discussion of the lensing effect from small scale fields, and in particular for the signal from non-linear structures such as clusters.

 \subsection{Lensed correlation function}
\label{corr_flat}

\begin{figure}
\begin{center}
\psfig{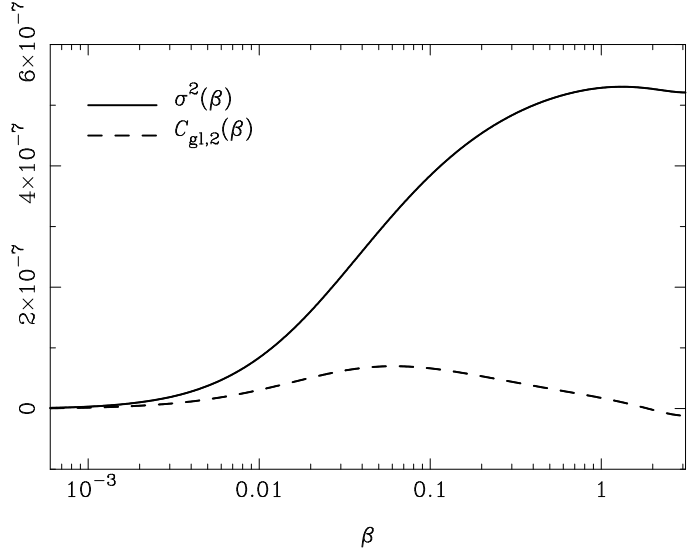}
\caption{
The functions $\sigma^2(\beta)\equiv \Cgl(0)-\Cgl(\beta)$ [solid] and $\Cgltwo(\beta)$ [dashed]
as a function of angular separation $\beta$ (in radians) for a typical
concordance model. For accuracy the results are
calculated using the full-sky generalization, and use the linear power spectrum for $C_l^\psi$.
\label{sigmaplot}}
\end{center}
\end{figure}

For a more accurate calculation of the lensing effect on all scales we now calculate the lensed correlation function in the flat-sky
limit, broadly following the method of Ref.~\cite{Seljak:1996ve}. The correlation function is invariant under displacements: it only depends on the separation of the two points. Unlike the series expansion approach it is therefore insensitive to bulk unobservable shifts of the unlensed CMB. The result only depends on the correlation structure of the \emph{difference} of deflection angles, which is what is directly observable.

Lensing re-maps the temperature according to
\begin{equation}
\tilde{\Theta}(\vx) = \Theta(\vx + \valpha),
\end{equation}
where now to get an accurate result on all scales we do not expand in $\valpha=\vgrad\psi$.
The lensed correlation function $\xil(r)$ is given by
\begin{eqnarray}
\xil(r) &\equiv&
\langle \tilde{\Theta}(\vx) \tilde{\Theta}(\vx') \rangle\nonumber \\
&=& \langle \Theta(\vx+\valpha) \Theta(\vx'+\valpha') \rangle \nonumber\\
&=&
\int \dFT{\vl} \int \dFT{\vl'} \la e^{i\vl\cdot (\vx + \valpha)} e^{-i\vl'\cdot(\vx'+\valpha')} \ra_\valpha\, \la \Theta(\vl)\Theta(\vl')^*\ra_\Theta
 \nonumber\\
&=& \int \frac{\ud^2 \vl}{(2\pi)^2}\, C_l^\Theta e^{i \vl \cdot \vr}
\langle e^{i\vl \cdot (\valpha - \valpha')} \rangle_\valpha.
\label{flat_lensed_corr}
\end{eqnarray}
As before we have defined $\vr \equiv \vx-\vx'$ and neglected the small correlation between the CMB temperature and the lensing potential. The remaining average is over the deflection angle. For a Gaussian variate $y$ with zero mean and variance $\sigma_y^2\equiv \la y^2\ra$ we have
\begin{align}
\la e^{iy} \ra &= \frac{1}{\sqrt{2\pi}\sigma_y} \int^\infty_{-\infty} \ud y\, e^{iy}e^{-y^2/2\sigma_y^2} =
\frac{1}{\sqrt{2\pi}\sigma_y} \int^\infty_{-\infty} \ud y \,e^{-(y-i\sigma_y^2)^2/2\sigma_y^2}
e^{-\sigma_y^2/2}\nonumber \\
&= e^{-\la y^2\ra/2}.
\label{gauss_avg}
\end{align}
Since we are assuming $\valpha$ is a Gaussian field, $\vl \cdot (\valpha - \valpha')$ is a Gaussian variate, and the expectation value in Eq.~\eqref{flat_lensed_corr} therefore reduces to
\begin{eqnarray}
\left\la e^{i\vl \cdot (\valpha - \valpha')} \right\ra &=&
\exp\left(- \frac{1}{2} \left\langle [\vl \cdot (\valpha-\valpha')]^2 \right\rangle\right).
\label{expavg}
\end{eqnarray}
To compute the average in the exponential we need to work out the correlation tensor
\begin{equation}
\langle \alpha_i \alpha_j' \rangle = \langle \grad_i\psi(\vx) \grad_j\psi(\vx') \rangle =
\int \frac{\ud^2 \vl}{(2\pi)^2} l_i l_j C_l^\psi e^{i\vl\cdot \vr}.
\end{equation}
By symmetry, the correlator can only depend on $\delta_{ij}$ and the trace-free
tensor $\hat{r}_{\langle i} \hat{r}_{j \rangle}\equiv \hat{r}_i \hat{r}_j - \half\delta_{ij}$ where $\hat{\vr} = \vr/r$.
We evaluate these parts by taking the trace to give
\begin{align}
\Cgl(r) &\equiv \la \valpha \cdot \valpha'\ra  \nonumber\\
 &= \int \dFT{\vl} l^2 C_l^\psi  e^{i\vl\cdot \vr}  \nonumber\\
 &=\frac{1}{2\pi} \int \ud l\, l^3 C_l^\psi J_0(lr),
\end{align}
and contracting with $\hat{r}_{\la i} \hat{r}_{j\ra}$ to give
\begin{align}
\Cgltwo(r) &\equiv -2\hat{r}_{\la i} \hat{r}_{j\ra} \langle \alpha^i \alpha'{}^j \rangle  \nonumber\\
   &= -2\int \frac{\ud^2\vl}{(2\pi)^2} \hat{r}_{\la i} \hat{r}_{j\ra} l^i l^j C_l^\psi e^{i\vl\cdot \vr}  \nonumber\\
   &=-2\int \frac{\ud^2\vl}{(2\pi)^2} \frac{l^2 \cos 2\phi}{2} C_l^\psi e^{i\vl\cdot \vr} \nonumber\\
&= \frac{1}{2\pi} \int \ud l\, l^3 C_l^\psi J_2(lr).
\label{Cgltwo_flatdef}
\end{align}
Here we defined $\phi$ as the angle between $\vl$ and $\vr$, and used Eq.~\eqref{BessJ} to express the integrals in terms of Bessel functions. Using these results we have
\begin{equation}
\langle \alpha_i \alpha_j' \rangle = \frac{1}{2}
\Cgl(r) \delta_{ij} - \Cgltwo(r) \hat{r}_{\langle i}
\hat{r}_{j \rangle}.
\label{alpha_cov}
\end{equation}
We can now compute the required expectation value in Eq.~\eqref{expavg}:
\begin{eqnarray}
\left\la [\vl \cdot (\valpha-\valpha')]^2 \right\ra &=& l^i l^j\la (\alpha_i-\alpha'_i)(\alpha_j-\alpha'_j)\ra \nonumber\\
&=& l^2 [\Cgl(0) - \Cgl(r)] + 2l^i l^j \hat{r}_{\langle i} \hat{r}_{j \rangle}\Cgltwo(r)\nonumber\\
&=&l^2 [\sigma^2(r) + \cos 2\phi\, \Cgltwo(r)].
\label{flat_expect}
\end{eqnarray}
We have defined the quantity $\sigma^2(r) \equiv \Cgl(0)-\Cgl(r)=\half\la (\valpha-\valpha')^2\ra$, which is half the variance of the
relative deflection of the two points.

Putting this all together we get the final result for the lensed correlation function
\begin{eqnarray}
\xil(r) &=&
\int \frac{\ud^2 \vl}{(2\pi)^2} C_l^\Theta e^{i \vl \cdot \vr}
 \exp\left(-\frac{1}{2}\la \left[\vl \cdot (\valpha - \valpha')\right]^2\ra\right)\nonumber
\\
&=&\int \frac{\ud^2 \vl}{(2\pi)^2} C_l^\Theta e^{i l r \cos\phi}
\exp\left(-\frac{1}{2}l^2 [\sigma^2(r) + \cos 2\phi\, \Cgltwo(r)]\right) \nonumber\\
&=&\int \frac{\ud^2 \vl}{(2\pi)^2} C_l^\Theta e^{i l r \cos\phi}
e^{-l^2\sigma^2(r)/2}\left( 1 - \frac{l^2}{2} \cos 2\phi\, \Cgltwo(r) + \dots \right)\nonumber \\
&=& \int  \frac{\ud l}{l} \, \frac{l^2C_l^\Theta}{2\pi} \,e^{-l^2 \sigma^2(r) /2}
\left[ J_0(lr) + \frac{1}{2}l^2 \Cgltwo(r) J_2(lr) + \dots \right].
\label{flatt}
\end{eqnarray}
Here for simplicity we expanded in $\Cgltwo$ before integrating over $\phi$.  Since $\Cgltwo$ is significantly less than $\sigma^2$ (as
shown in Fig.~\ref{sigmaplot}), the perturbative expansion in
$\Cgltwo$ converges much faster than one in $\sigma^2$ and this is a good approximation for the scales of most interest. Note that it is important to keep $\exp(-l^2 \sigma^2/2)$ as the main non-perturbative term in $C_l^\psi$.

\begin{figure}
\begin{center}
\psfig{figure=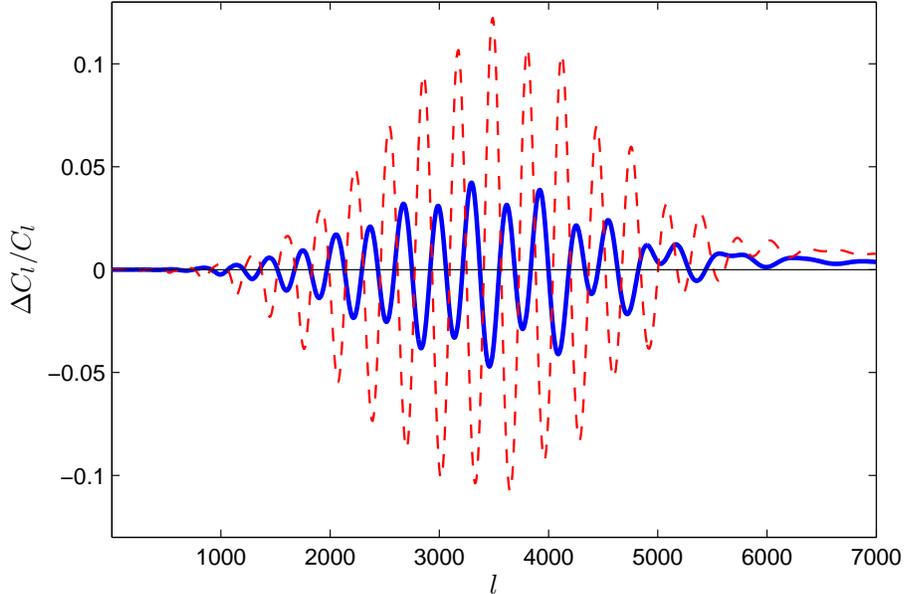,width=12.0cm}
\caption{Fractional difference between the lensed $C_l$ results using the series-expansion method to lowest order in $C_l^\psi$ compared to the more accurate correlation function result (temperature (solid) and $E$-polarization (dashed; see Section~\ref{sec:pol})).
\label{HuT}}
\end{center}
\end{figure}

The lensed correlation function is related to the lensed $C_l$ by Eq.~\eqref{flatCl-corr}. The calculation of the
correlation function and the inversion to the power spectrum is numerically very efficient, requiring only one dimensional sums and integrals. It is used in the numerical codes \CMBFAST~\cite{Seljak:1996is}, \CMBEASY~\cite{Doran:2003sy} and \CAMB~\cite{Lewis:1999bs}. Results to the next order in
$\Cgltwo$ are given in~\cite{Challinor:2005jy}, along with the generalization to a non-flat sky.  This is sufficient for a calculation to within cosmic variance on scales where
the non-linear power spectrum can be computed accurately and non-Gaussian corrections can be neglected. The full result is implemented in the public \CAMB\ code.

The difference between the result using the accurate correlation-function method and the lowest-order series-expansion result is shown in Fig.~\ref{HuT}. The series expansion result is accurate on large scales where the deflection angle is much smaller than the wavelength of the modes, and also on very small scales where the unlensed CMB has little power and can be accurately approximated by a single gradient term. In the intermediate regime the series-expansion result is not accurate and the non-perturbative correlation function method must be used. Note that the errors are percent level by $l\sim 2000$, amounting to a significant fraction of the lensing effect on these scales.

\subsubsection{Exact result}

Although the series expansion we used in Eq.~\eqref{flatt} is fully adequate for an accurate flat-sky calculation in most cases relevant for the CMB, for completeness we note that the integral in Eq.~\eqref{flatt} can actually also be evaluated exactly as an infinite sum. Substituting $r \rightarrow i r$ in Eq.~\eqref{BessJ} and using $J_n(i r) = i^n I_n(r)$ where $I_n$ is a modified Bessel function we have
\begin{equation}
e^{-r \cos\phi} = \sum_{n=-\infty}^\infty (-1)^n I_n(r) e^{i n \phi} = I_0(r) + 2 \sum_{n=1}^\infty (-1)^n  I_n(r) \cos(n\phi).
\label{I_n_expand}
\end{equation}
Using this and Eq.~\eqref{BessJ} to expand the exponentials in Eq.~\eqref{flatt}, and integrating over $\phi$, leads to the exact flat-sky weak lensing result for uncorrelated Gaussian fields with the temperature on a single source plane
\begin{equation}
\xil(r) =  \int \frac{\ud l}{l} \, \frac{l^2C_l^\Theta}{2\pi} e^{-l^2\sigma^2(r)/2} \sum_{n=-\infty}^\infty  I_n\left[l^2 \Cgltwo(r)/2\right] J_{2 n}(l r).
\label{flatt_exact}
\end{equation}
Note that for $|n|>0$ there are two equal contributions from $n=|n|$ and $n=-|n|$. Also note that although $I_n(r) \rightarrow e^r/\sqrt{2\pi r}$ for large $r$, because $\sigma^2(r) \ge \Cgltwo(r)$ the combination $e^{-l^2\sigma^2(r)/2} I_n(l^2 \Cgltwo(r)/2)$ goes to zero as $l\rightarrow \infty$.
For a fixed maximum $l$, only a low number of terms in $|n|$ near zero are required to achieve good accuracy. This general form may be useful for calculating the lensing effect on very small scales $l \gg 10^3$ when $C_l^\Theta$ has small scale power and a low order expansion in $\Cgltwo$ is insufficient. One non-CMB example where this may be useful is calculating the lensing effect on the 21-cm power spectrum~\cite{Mandel:2005xh}.

\subsubsection{Conservation of total variance}

The variance at a point $\xil(0) = \la |\tTheta(\vx)|^2\ra$ can be obtained by noting that $\Cgltwo(0)=\sigma^2(0)=0$ and $I_n(0) = \delta_{n0}$. It follows from Eq.~\eqref{flatt_exact} that the total power is conserved by lensing
\begin{equation}
\xil(0) = \xi(0) = \int \frac{\ud l}{l}\, \frac{l^2 C_l^\Theta}{2\pi}   = \int \frac{\ud l}{l}\, \frac{l^2  \tC_l^\Theta}{2\pi} = \la |\tTheta(\vx)|^2\ra = \la |\Theta(\vx)|^2\ra.
\end{equation}
This reflects the fact that weak lensing only alters photon directions, and hence the spatial correlation structure, but does not change the variance in any given direction. We are observing the same photons, just in a slightly different direction.

\subsubsection{Relation between lensed and unlensed correlation functions}

We can also calculate the lensed correlation function in terms of the unlensed one using Eqs.~\eqref{flatt} and~\eqref{flatCl-corr}:
\begin{eqnarray}
\xil(r) &=& \int \frac{\ud^2\vl}{(2\pi)^2} \int \ud^2 \vr'\, e^{-i\vl\cdot\vr'} \xi(r') e^{i\vl\cdot \vr} \exp\left(-\frac{1}{2}l^2 [\sigma^2(r) + \cos 2(\phi_\vl-\phi_\vr)\, \Cgltwo(r)]\right)\nonumber\\
&=& \int \ud^2 \vr'\, \xi(r') \int \frac{\ud^2\vl}{(2\pi)^2} e^{i\vl\cdot(\vr-\vr')}e^{-l^2\sigma^2(r)/2}e^{-(l_{r1}^2-l_{r2}^2)\Cgltwo(r)/2}\nonumber\\
&=& \int \frac{\ud^2 \vr'\, \xi(r')}{2\pi\sqrt{\sigma^4(r)-\Cgltwo^2(r)}} \exp\left(-\frac{(\vr-\vr')^2}{2[\sigma^2(r)+\Cgltwo(r)]}-\frac{r'{}^2\sin^2(\phi_\vr-\phi_{\vr'})\Cgltwo(r)}{[\sigma^4(r)-\Cgltwo^2(r)]}\right)
.
\end{eqnarray}
In the middle line, we have resolved $\vl$ into its components $l_{r1}$
and $l_{r2}$ parallel and perpendicular to $\vr$.
In an isotropic approximation where $\Cgltwo(r)=0$, this is an isotropic Gaussian convolution of the two-dimensional correlation function, with a smoothing width $\sigma^2(r)$ that depends on $r$. This smoothing of features in the correlation function corresponds in harmonic space to the smoothing of the $C_l^\Theta$ peaks. The angular integral can also be done to express the answer in terms of modified Bessel functions, but the result it not especially enlightening or useful.

\subsection{Comparison with other CMB temperature secondaries}

\begin{figure}
\begin{center}
\psfig{figure=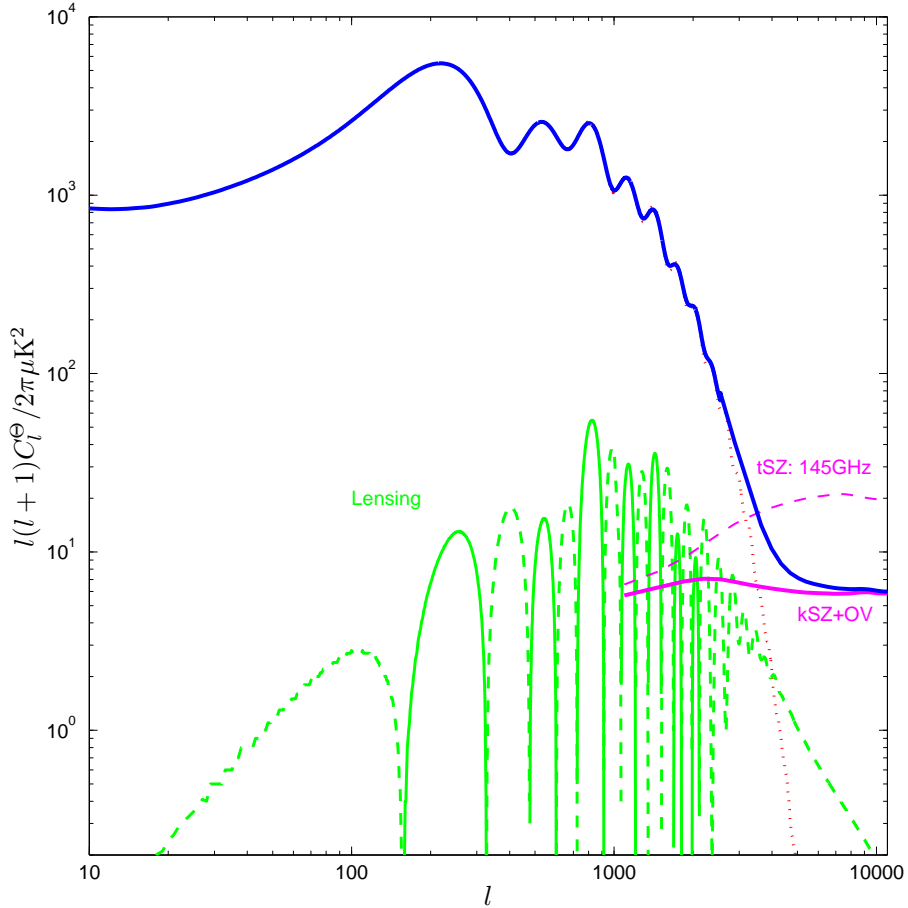,width=12.0cm}
\caption{Contributions to the total blackbody-spectrum temperature anisotropy $C_l^\Theta$ (thick solid) from CMB lensing and kinetic SZ (kSZ) plus Ostriker-Vishniac (OV) for a typical $\Lambda$CDM model with inhomogeneous reionization. The kSZ+OV signal is taken from a model in Ref.~\cite{Zahn:2005fn}, and can vary depending on parameters. The spectrally-distinguishable thermal SZ signal is shown dashed, though this also depends strongly on parameters. On very small scales kSZ+OV dominates both the unlensed (dotted) and lensed CMB power spectrum.
\label{nonlinCompare}}
\end{center}
\end{figure}

 In addition to weak lensing, there are a variety of other non-linear effects on the CMB.
 The thermal  Sunyaev-Zel'dovich~\cite{SZ,Cooray:2004cd} (SZ) effect due to scattering from hot electrons can be significant on small scales. This signal has a frequency spectrum different to that of the linear CMB and hence can in principle easily be distinguished. However there are other non-linear effects which, like CMB lensing, have the same frequency spectrum as the primordial anisotropies. These are a potential source of complication when analysing the lensed CMB as they cannot be easily distinguished, and may indeed be correlated with the lensing signal.

The kinetic SZ effect is the main such effect for the temperature anisotropies~\cite{Amblard:2004ih,Zahn:2005fn,Santos:2003jb}, arising from scattering from electrons in a bulk flow of a non-linear perturbation with respect to the CMB. It is closely related to the second-order Ostriker-Vishniac effect~\cite{Hu:1995sv}, they are both essentially Doppler terms determined by the electron velocity projected along the line of sight direction. These effects combine  with additional signals due to patchy reionization~\cite{Knox:1998fp} to give a significant non-linear anisotropy from around the epoch of reionization and afterwards.  Current uncertainties in the reionization history and morphology make detailed predictions for the spectrum difficult, and at the moment this uncertainty gives rise to a percent-level uncertainty in the expected total $C_l^\Theta$ at $l=2000$~\cite{Zahn:2005fn}. Fig.~\ref{nonlinCompare} shows a sample comparison.
Precision cosmology from the temperature damping tail will require accurate modelling of both lensing and the kinetic SZ effect, and they may indeed be significantly correlated~\cite{Amblard:2004ih}.  For the polarization spectra the kinetic SZ effect is much smaller~\cite{Hu:1999vq,Zahn:2005fn}, and lensing is likely to be the dominant effect.

Other non-linear signals include the Rees-Sciama effect, a late time ISW-like anisotropy generated by photons propagating though potential wells that are evolving because of non-linear growth or their bulk motion~\cite{ReesSciama}. This is generally subdominant on all scales~\cite{Seljak:1995eu}, and hence is not a major concern. There is also a second-order time-delay effect due to the varying potentials along the line of sight. However the effect turns out to be much weaker than the lensing contributions, and can probably be neglected~\cite{Seljak:1994wa,Hu:2001yq}.

%% file: pol.tex
The presence of a non-zero photon quadrupole at last scattering means that there will be an interesting polarization signal. As for the temperature, the observed polarization field will have been lensed by potential gradients along the line of sight, and we need to model this effect. The polarization orientation along a lensed light beam is simply transported in the manner described in
Section~\ref{subsec:pol_lens}. At lowest order this means that the orientation of the polarization in the observed direction $\vnhat$ and the lensed direction $\vnhat'$ will be the same with respect to the geodesic connecting the two points. The effect of lensing may therefore be modelled in a manner similar to the temperature lensing, with the added complication that the polarization is a tensor field that is parallel-transported along deflection vectors.

We start with a brief review of the properties of the polarization field, and explain the decomposition into physically distinct $E$ and $B$ modes. We then calculate the effect of lensing in the flat sky approximation, both using the approximate series expansion and correlation function methods. We conclude this section with a discussion of the important impact of CMB lensing on the ability to observe primordial gravitational waves.

\subsection{Stokes' Parameters}

A CMB photon detector measures an electric field $\vE$ perpendicular to the direction of observation $\vnhat$. From the observation we define the intensity matrix
\begin{eqnarray}
\clp_{ab} = C \la E_a E_b^* \ra
\end{eqnarray}
for some constant $C$ that we will choose so that $\clp_{ab}$ is measured in temperature units. The angle brackets here denote a time average. The rank-two correlation tensor can be decomposed into irreducible parts
\begin{equation}
\clp_{ab} = P_{ab}   + \half\delta_{ab} I  + V_{[ ab ]},
\end{equation}
where $P_{ab} = \clp_{\la ab\ra}$ is the symmetric trace-free part, the trace $I$ is the total power intensity, and $V_{[ab]} =\half V\epsilon_{ab}$ is the antisymmetric part. The antisymmetric part will only be non-zero if there is a phase lag between $E_1$ and $E_2$, and therefore describes circular polarization. Since the Thomson scattering responsible for the CMB signal does not induce circular polarization we shall neglect the $V$ Stokes' parameter here. The intensity we have already discussed in the section on the lensing of the temperature field. Here we discuss lensing of the polarization described by $P_{ab}$.

It is often useful to describe the polarization with respect to a fixed basis in the plane orthogonal to the direction of observation. For some choice of orthonormal directions $\ve_1$ and $\ve_2$ the components of the trace-free symmetric polarization tensor can be written
\begin{equation}
P_{ij} =\frac{1}{2} \begm Q & U \\ U & -Q \enm.
\end{equation}
This defines the two Stokes' parameters $Q$ and $U$, which depend implicitly on the basis in which they are evaluated (physically, on the orientation of the polarizer when the observation is made). The $Q$ parameter can be thought of as the difference in power measured when the polarizer is aligned with $\ve_1$ compared to when it is aligned with $\ve_2$. The $U$ Stokes' parameter measures the difference in power between the $45^\circ$ directions $\ve_1 \pm \ve_2$. So $Q$ and $U$ change sign under a rotation by $90^\circ$, and into each other under a rotation by $45^\circ$.

The eigenvalues of $2P_{ij}$ are $\pm (Q^2 + U^2)^{1/2}$. The eigenvectors make an angle $\half\tan^{-1}(U/Q)$ to $\ve_1$, so it is sometimes convenient to think of the polarization field as a field of headless vectors of length $(Q^2 + U^2)^{1/2}$ at an angle $\half\tan^{-1}(U/Q)$ to the $\ve_1$ vectors. The headless vectors represent the basis-independent physical orientation in which a polarizer would give maximum signal.

It is also useful to consider the components of $P_{ab}$ with respect to the complex vectors $\ve_\pm  = \ve_1 \pm i \ve_2$
\begin{eqnarray}
P &\equiv& \ve_+^a \ve_+^b P_{ab} = Q + i U \nonumber\\
P^* &=& \ve_-^a \ve_-^b P_{ab} = Q - i U.
\end{eqnarray}
So instead of using $P_{ab}$, we can if we wish describe the polarization in terms of a complex field given by the components with respect to a particular basis $\ve_\pm$.

On the flat sky we use the basis $\ve_\pm = \ve_x \pm i \ve_y$.  Consider what happens if we rotate our basis by an angle $\gamma$ clockwise (e.g. rotating $\ve_x$ in the direction away from $\ve_y$). We can write this rotation in a convenient complex form as
\begin{eqnarray}
\ve_\pm \equiv \ve_x \pm i \ve_y &\rightarrow& \ve_x' \pm i \ve_y ' \nonumber\\
      &=& (\cos\gamma\,\ve_x -\sin\gamma\,\ve_y) \pm i(\sin\gamma\,\ve_x + \cos\gamma\,\ve_y) \nonumber\\
      &=& e^{\pm i\gamma}(\ve_x \pm i \ve_y) = e^{\pm i\gamma} \ve_\pm.
\end{eqnarray}
It follows that in the rotated basis
\begin{eqnarray}
P' = \ve_+^a{}' \ve_+^b{}' P_{ab} = e^{2i\gamma} P.
\end{eqnarray}
Under the transformation $\ve_+ \rightarrow e^{i\gamma} \ve_+$ a quantity ${}_s\eta$ is called spin-$s$ if
\begin{equation}
{}_s\eta \rightarrow {}_s\eta e^{i s\gamma}.
\end{equation}
So the $P^{(x,y)}$ we defined above has spin $2$, where the $(x,y)$ superscript denotes that $Q$ and $U$ are measured with respect to the $\ve_x$ and $\ve_y$ basis.

On the sphere, the natural basis consists of the $\ve_\theta$ and $\ve_\phi$ vectors of the spherical-polar coordinate system. Looking from inside the sphere, $\ve_\theta$ points along great circles from the north to south pole, and $\ve_\phi$ points to the left orthogonal to $\ve_\theta$. There is therefore potential for sign confusion because $(\ve_\theta, \ve_\phi)$ form a right-handed set (viewed from inside), whereas $(\ve_x, \ve_y)$ define a left-handed set, so we must be careful with how we define things. The rotation
\begin{equation}
\ve_\theta + i\ve_\phi \rightarrow e^{i\gamma} (\ve_\theta + i\ve_\phi)
\end{equation}
corresponds to an anticlockwise rotation about the outward normal $\vr$. If we measure $Q$ and $U$ with respect to the $(\ve_\theta,\ve_\phi)$ basis $P^{(\theta,\phi)}$ will be spin 2 as before. However if we define $Q$ and $U$ with respect to the left-handed set analogous to $(\ve_x, \ve_y)$ using $(\ve_\theta, -\ve_\phi)$, the complex $P^{(\theta,-\phi)}$ then has spin -2. Both conventions have been used in the literature. On the sphere we shall define $Q$ and $U$ with respect to $\ve_\theta$ and $\ve_\phi$, so $P$ is spin 2. We note, however,
that this differs from the astronomical (IAU) standard which defines
$Q$ and $U$ on $-\ve_\theta$ and $\ve_\phi$.

\subsection{$E$ and $B$ polarization}

We can write a polarization tensor field $P_{ab}$ in terms of two scalar fields $P_E$ and $P_B$ as
\begin{equation}
P_{ab} = \grad_{\la a}\grad_{b\ra} P_E + \epsilon^c{}_{(a}\grad_{c)}\grad_b P_B,
\end{equation}
where $P_E$ describes the `$E$-mode' gradient-like polarization, and $P_B$ described the `$B$-mode' curl-like polarization. The angle brackets around the indices denote the symmetric trace-free part, and $\grad_a$ is a covariant derivative in the surface on which we measure $P_{ab}$ (i.e. orthogonal to the observation direction $\vnhat$).
If the fields are defined over a surface without boundary (as the polarization is on the full sphere), this decomposition is unique because the derivatives are invertible. What happens on a surface with boundaries (for example observations over only part of the sky) is more complicated, and discussed in detail in Refs.~\cite{Lewis:2001hp,Bunn:2002df,Lewis:2003an}.

Locally pure $E$ and $B$ quantities can be extracted by taking two derivatives to form the pure $E$ quantity $\grad^{\la a}\grad^{b\ra} P_{ab}$ and pure $B$ quantity $\epsilon_c{}^{(a}\grad^{c)}\grad^b P_{ab}$. However since integration of these terms to recover the underlying $P_E$ and $P_B$ is inherently non-local, it is usually more convenient to work directly in harmonic space.
The scalar fields $P_E$ and $P_B$ can be expanded in terms of scalar harmonics $Q_k$, giving
\begin{equation}
P_{ab} =\frac{1}{\sqrt{2}} \sum_k \left[ E_k Q^G_k{}_{ab} + B_k Q^C_k{}_{ab} \right],
\end{equation}
where we have defined gradient and curl tensor harmonics~\cite{Kamionkowski:1996ks}
\begin{equation}
Q^G_k{}_{ab} = N_k \grad_{\la a}\grad_{b\ra} Q_k \qquad  Q^C_k{}_{ab} = N_k \epsilon^c{}_{(a}\grad_{b)}\grad_c Q_k
\end{equation}
and $N_k$ is some normalization. The $E_k$ and $B_k$ are the harmonic components of the gradient-type and curl-type terms, and the factor of $1/\sqrt{2}$ is inserted for later consistency.\footnote{Note that some authors define harmonic coefficients so they differ by a sign from those here~\cite{Zaldarriaga:1996xe}, changing the sign of the temperature $E$-polarization cross-correlation power spectrum. The sign here is different to the output of CAMB and CMBFAST.}

We can also expand $P$, the complex spin 2 component, in harmonics as follows.
The alternating tensor $\epsilon_{ab}$ can be written as in terms of $\ve_\pm$ as
\begin{equation}
\epsilon^{ab} = -i\, \ve^{[a}_-\ve^{b]}_+.
\end{equation}
Using the fact that $\ve_+^2=0$ and $\ve_+\cdot \ve_- = 2$, we then have
\begin{eqnarray}
P &=& \ve_+^a \ve_+^b P_{ab}  \nonumber\\
  &=& \ve_+^a \ve_+^b \frac{1}{\sqrt{2}} \sum_k N_k \left[ E_k \grad_a\grad_b Q_k + B_k \epsilon^c{}_{(a}\grad_{b)}\grad_c Q_k \right]  \nonumber \\
 &=& \frac{1}{\sqrt{2}} \sum_k N_k \left[ E_k \ve_+^a \ve_+^b\grad_a\grad_b Q_k + \frac{i}{2}\ve_-{}_a\ve_+^c\ve_+^a \ve_+^b \grad_b\grad_c Q_k \right]   \nonumber\\
 &=& \sum_k \frac{N_k}{\sqrt{2}} (E_k + i B_k) \ve_+^a \ve_+^b \grad_a\grad_b Q_k  \nonumber\\
 &=& \sum_k (E_k + i B_k) {}_2 Q_k.
\end{eqnarray}
Here we have defined the spin-2 harmonics, ${}_2 Q_k$, given by
\begin{equation}
{}_2 Q_k \equiv \frac{1}{\sqrt{2}}\ve_+^a \ve_+^b Q_k^G{}_{ab} \equiv \frac{N_k}{\sqrt{2}} \ve_+^a \ve_+^b \grad_a\grad_b Q_k.
\end{equation}
Why don't we just expand the complex field spin-2 field $P$ directly in terms of scalar harmonics?
We certainly could do this, but the harmonic coefficients would then be spin 2, and therefore still depend on the choice of local basis orientations. Instead, we want to expand in spin-2 harmonics, so that the harmonic coefficients $E_k$ and $B_k$ are related to physically relevant basis-independent scalar quantities.

The complex conjugate $P^* = Q -iU$ has spin -2, and can be expanded analogously in terms of spin $-2$ harmonics ${}_{-2} Q_k$ giving
\begin{equation}
P^* = \sum_k (E_k - i B_k) {}_{-2} Q_k
\end{equation}
where
\begin{equation}
{}_{-2} Q_k \equiv \frac{N_k}{\sqrt{2}} \ve_-^a \ve_-^b \grad_a\grad_b Q_k.
\end{equation}

\subsubsection{Flat-sky harmonics}

On the flat sky we use harmonics $Q_k = e^{i\vl\cdot \vx}$, so the tensor harmonics are
\begin{eqnarray}
Q^G_{ab}(\vl) &=& - \sqrt{2}\,\, \vlhat_{\la a} \vlhat_{b\ra}  e^{i\vl\cdot \vx} \\
Q^C_{ab}(\vl) &=& - \sqrt{2}\,\, \epsilon^c{}_{(a} \vlhat_{b)}  \vlhat_c  e^{i\vl\cdot \vx}
\end{eqnarray}
where we used the normalization factor $N_l = \sqrt{2}/l^2$ required so that the tensor harmonics are normalized.  The spin-2 harmonics are therefore given by
\begin{eqnarray}
{}_2 Q(\vl) &=& l^{-2}\,\ve_+^a \ve_+^b \grad_a\grad_b e^{i\vl\cdot\vx} \nonumber\\
        &=&  l^{-2}\, (\partial_x + i\partial_y)^2 e^{i\vl\cdot\vx}\nonumber\\
        &=&  - (\cos\phi_\vl + i \sin\phi_\vl)^2 e^{i\vl\cdot\vx} \nonumber\\
        &=& -e^{2i\phi_\vl}e^{i\vl\cdot\vx}.
\end{eqnarray}
We can therefore expand the polarization field as
\begin{eqnarray}
 P &=& Q+iU = -\int \dFT{\vl} \left[(E(\vl) + i B(\vl)\right] e^{2i\phi_\vl}e^{i\vl\cdot\vx} \\
 P^* &=& Q-iU = -\int \dFT{\vl} \left[(E(\vl) - i B(\vl)\right] e^{-2i\phi_\vl}e^{i\vl\cdot\vx}.
\end{eqnarray}
The $e^{\pm 2i\phi_\vl}$ factors serve to rotate between a natural basis defined in terms of $\vl$, and the fixed basis defined in terms of $\ve_x$. The inverse relations are
\begin{eqnarray}
E(\vl) + i B(\vl) &=& -\int \dFT{\vx} P e^{-2i\phi_\vl}e^{-i\vl\cdot\vx} \nonumber\\
E(\vl) - i B(\vl) &=& -\int \dFT{\vx} P^* e^{2i\phi_\vl}e^{-i\vl\cdot\vx} .
\label{pol_transform}
\end{eqnarray}

\subsubsection{Full-sky harmonics}
\label{spinYlm}

On the full sky we can expand scalar fields in terms of scalar harmonics $Q_k = Y_{lm}$, and therefore define spin 2 harmonics~\cite{Zaldarriaga:1996xe}
\begin{equation}
{}_2 Y_{lm} \equiv \frac{N_l}{\sqrt{2}} \ve_+^a \ve_+^b \grad_a\grad_b Y_{lm}
\end{equation}
and similarly for the spin -2 case. We cannot give a neat closed form expression for the ${}_2 Y_{lm}$, but they can be calculated recursively as can the spherical harmonics themselves~\cite{Lewis:2001hp}. The normalization may be shown to be
\begin{equation}
N_l = \sqrt{\frac{2(l-2)!}{(l+2)!}}.
\end{equation}

\subsection{Lensed polarization power spectra}

\subsubsection{Lowest-order calculation}

We start by doing a simple first-order calculation, analogous to the one we did for the temperature field in Section~\ref{harmonic_flat}. The lensed polarization tensor $\tP_{ab}$ can be expanded in the deflection angle as
\begin{eqnarray}
\tP_{ab}(\vx) &=& P_{ab}(\vx+\vgrad\psi) \nonumber \\
   &\approx& P_{ab}(\vx) + \grad^c \psi \grad_c P_{ab}(\vx) + \frac{1}{2}  \grad^c \psi \grad^d \psi \grad_c\grad_d P_{ab}(\vx) + \dots.
\end{eqnarray}
As in the temperature case, we cannot expect a second-order truncation to be very accurate on all scales, however it is useful for a simple analytical analysis that correctly describes the main aspects of polarization lensing.
Because the $\ve_\pm$ vectors are independent of position on the flat sky, they commute with derivatives, so the same expression holds if $P_{ab}$ is replaced with its spin-2 components $P = \ve_+^a\ve_+^b P_{ab}$.

Scalar mode perturbations produce no $B$-mode polarization at linear order, in which case $B(\vl)=0$. The lensing effect of tensor modes is expected to be small, so we are mostly interested in the $B$-modes generated from lensing a purely $E$-mode unlensed field. To make our life a little simpler, we shall therefore set $B(\vl)=0$. Performing a harmonic expansion we then have the lensed harmonic coefficients
\begin{align}
\tE(\vl) &\pm i \tB(\vl) \approx \, E(\vl) - \int \dFT{\vl'}  \vl'\cdot (\vl-\vl')e^{\pm2i(\phi_{\vl'}-\phi_\vl)} \psi(\vl-\vl')E(\vl') \nonumber\\
          & - \frac{1}{2} \int \dFT{\vl_1} \int \dFT{\vl_2} e^{\pm2i(\phi_{\vl'}-\phi_\vl)}
              \vl_1\!\cdot\![\vl_1 +\vl_2-\vl]\, \vl_1\!\cdot\! \vl_2 E(\vl_1)\psi(\vl_2)\psi^*(\vl_1+\vl_2-\vl).
%
\end{align}
For a statistically isotropic field the unlensed power spectra are given by
\begin{equation}
\la E(\vl) E^*(\vl') \ra = \delta(\vl-\vl') C_l^E \qquad \la B(\vl) B^*(\vl') \ra = \delta(\vl-\vl') C_l^B
\qquad \la E(\vl) \Theta^*(\vl') \ra = \delta(\vl-\vl') C_l^X,
\end{equation}
where we have assumed that statistical parity invariance holds, so that $B$
in uncorrelated with $\Theta$ or $E$.
Here we are taking $C_l^B=0$, but will calculate the corresponding non-zero lensed $B$-mode power spectrum. Keeping terms to order $C_l^\psi$ we therefore get
\begin{eqnarray}
\tC_l^E + \tC_l^B &=& C_l^E + \int \frac{\ud^2 \vl'}{(2\pi)^2}\left[\vl'\cdot (\vl-\vl')\right]^2 C_{|\vl-\vl'|}^\psi C^E_{|\vl'|}
- C_l^E \int \frac{\ud^2 \vl'}{(2\pi)^2} (\vl'\cdot\vl)^2 C^\psi_{l'}
\\
\tC_l^E - \tC_l^B &=& C_l^E + \int \frac{\ud^2 \vl'}{(2\pi)^2}\left[\vl'\cdot (\vl-\vl')\right]^2 e^{4i(\phi_{\vl'}-\phi_\vl)}C_{|\vl-\vl'|}^\psi C^E_{|\vl'|} - C_l^E \int \frac{\ud^2 \vl'}{(2\pi)^2} (\vl'\cdot\vl)^2 C^\psi_{l'}\\
\tC_l^X &=&  C_l^X + \int \frac{\ud^2 \vl'}{(2\pi)^2}\left[\vl'\cdot (\vl-\vl')\right]^2 e^{2i(\phi_{\vl'}-\phi_\vl)}C_{|\vl-\vl'|}^\psi C^X_{|\vl'|} - C_l^X \int \frac{\ud^2 \vl'}{(2\pi)^2} (\vl'\cdot\vl)^2 C^\psi_{l'}.
\end{eqnarray}
The integrals are actually real because the non-complex terms are symmetric as $\vl'$ is rotated in either direction from $\vl$, so we can replace the exponential with a $\cos$-term. This gives the final result for the lensed spectra to lowest order in $C_l^\psi$~\cite{Hu:2000ee}
\begin{eqnarray}
\tC_l^E  &=& (1-l^2R^\psi)C_l^E + \int \frac{\ud^2 \vl'}{(2\pi)^2}\left[\vl'\cdot (\vl-\vl')\right]^2 C_{|\vl-\vl'|}^\psi C^E_{|\vl'|} \cos^2 2(\phi_{\vl'}-\phi_\vl)  \\
\tC_l^B  &=&  \int \frac{\ud^2 \vl'}{(2\pi)^2}\left[\vl'\cdot (\vl-\vl')\right]^2 C_{|\vl-\vl'|}^\psi C^E_{|\vl'|}
\sin^2 2(\phi_{\vl'}-\phi_\vl) \\
\tC_l^X &=&  (1-l^2R^\psi)C_l^X +  \int \frac{\ud^2 \vl'}{(2\pi)^2}\left[\vl'\cdot (\vl-\vl')\right]^2 C_{|\vl-\vl'|}^\psi C^X_{|\vl'|}\cos 2(\phi_{\vl'}-\phi_\vl).
\end{eqnarray}
Here we have used the definition of $R^\psi$ from Eq.~\eqref{R_def} as in the temperature case. The more general result allowing for unlensed $B$-modes can be obtained as a straightforward generalization, and is given in Ref.~\cite{Hu:2000ee}. We will also include the $B$-terms in the next section when we do a more accurate calculation in terms of the correlation function.

\begin{figure}
\begin{center}
\psfig{figure=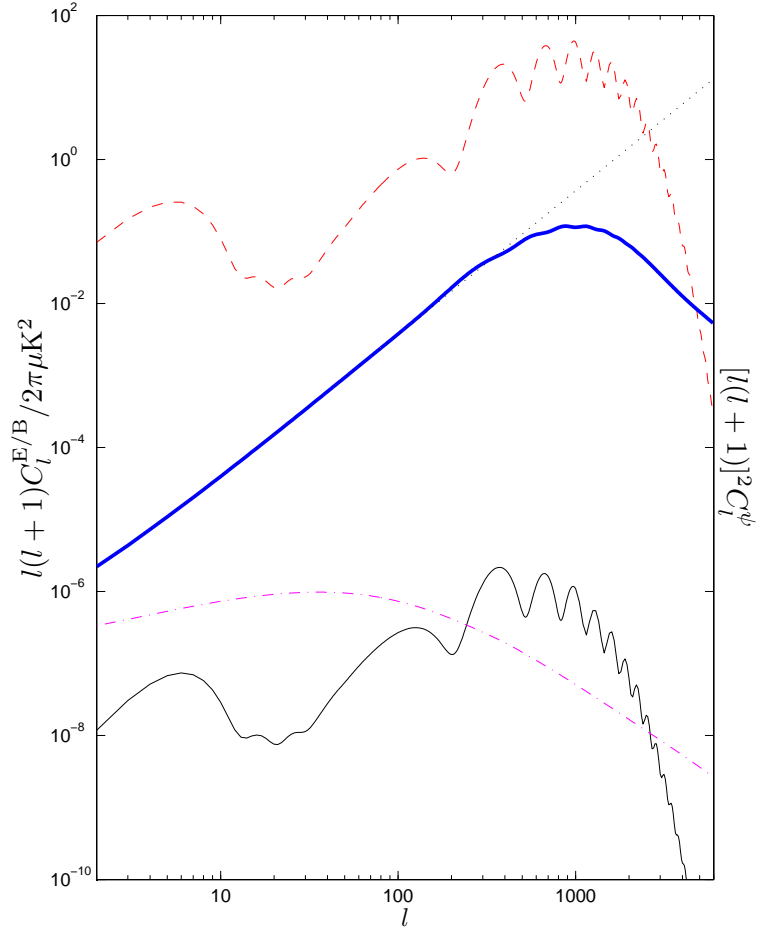,width=10cm}
\caption{The unlensed $E$-power spectrum $l(l+1)C_l^E/2\pi$ (top, dashed), lensing potential power spectrum $[l(l+1)]^2C_l^\psi$ (dot-dashed), and the contribution to the large-scale lensed $B$-mode power spectrum from each $\log l$ given by half their product (thin solid). The dotted line is the large-scale white spectrum $l(l+1)C_l^B/2\pi$ given by Eq.~\eqref{Bwhite}, which can be compared to the full numerical result (thick solid). The lensed $B$-mode power spectrum has a close-to-white spectrum at $l\ll 1000$.
\label{EtoB}}
\end{center}
\end{figure}

\begin{figure}
\begin{center}
\psfig{figure=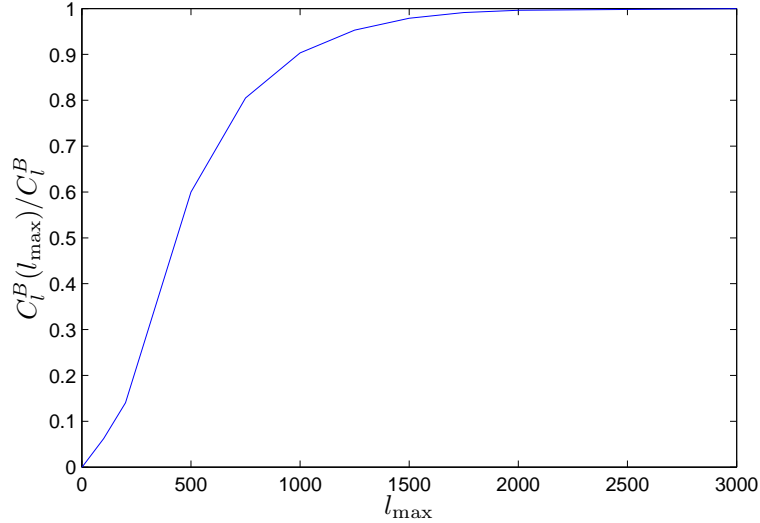,width=10cm}
\caption{The contributions to the low-$l$ ($l\sim 10$) $C_l^B$ power spectrum from lensing potential with $l \le l_\text{max}$.
 A significant fraction of the power comes from the tail of $C_l^\psi$ at high $l$.
\label{BBlmax}}
\end{center}
\end{figure}

Since the tensor modes would only have a significant signal on quite large scales it is interesting to consider how large the $B$-mode lensing signal should be there. If we take $|\vl|\ll |\vl'|$ we have
\begin{eqnarray}
\tC_l^B  &\sim&  \int \frac{\ud^2 \vl'}{(2\pi)^2}\, l'{}^4  C_{l'}^\psi \,C^E_{l'} \sin^2 2(\phi_{\vl'}-\phi_\vl) \nonumber\\
      &=& \frac{1}{4\pi}\int \frac{\ud l'}{l'} \,\,l'{}^4  C_{l'}^\psi\,\, l'{}^2C^E_{l'},
\label{Bwhite}
\end{eqnarray}
independent of $l$, corresponding to a white-noise spectrum for $\tC_l^B$ on large scales. This is actually a rather good approximation because there is little $E$-mode power on large scales, so the spectrum is indeed close to white on scales $l \ll 1000$, as shown in Fig.~\ref{EtoB}. For standard constant spectral index $\Lambda$CDM models allowed by the current data $C_l^B \sim 2\times 10^{-6}\muK^2$.

The scales responsible for the large-scale $B$-mode signal cover a wide range in $l$, as shown in Figs~\ref{EtoB} and~\ref{BBlmax}. To see why this should be so, consider a $E$-pattern at last scattering with some fixed spatial frequency $k_E$. If this is then lensed by an aligned lensing potential mode with $k^\perp_\psi \sim k_E$ (perpendicular to the line of sight), the mode coupling
produces power on scales down to $|k^\perp_\psi - k_E|$. This is the power that will be observed in the $B$-mode on large scales (and also in the $E$-mode if it weren't dominated by the much larger unlensed signal from reionization and last scattering). There are contributions to the large scale $B$ modes from all scales on which there is non-zero $E$ and lensing potential power. Since the relevant $E$ and $\psi$ modes are dominant on small scales (the last-scattering $E$ is small on large scales), they have a short correlation length. It follows that each contribution to the lensed $B$ from different coincidentally-aligned patches of  $E$ and $\psi$ will be nearly spatially uncorrelated. Uncorrelated random fields in real space correspond to a white spectrum in harmonic space, hence the prediction for a very nearly white $B$-mode spectrum on large scales where there is little $E$-mode power ($l\alt 1000$).

As we have shown a purely scalar $E$-mode polarization signal at last scattering will give rise to non-zero $B$-modes after lensing by the potential along the line of sight. This has important consequences for the detectability of primordial tensor modes via their $B$-mode signal, as discussed in the next section.

\begin{figure}
\begin{center}
\psfig{figure=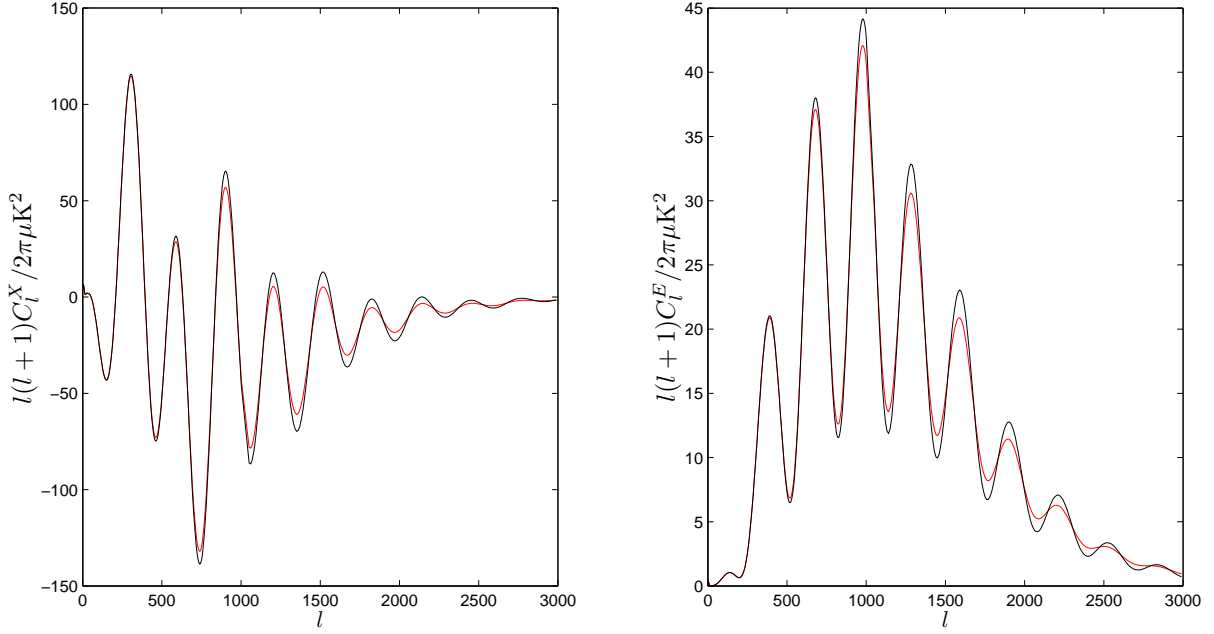,width=16cm}
\caption{The unlensed $T$-$E$ correlation and $E$-mode power spectra compared to the smoother lensed results.
\label{lensedE}}
\end{center}
\end{figure}

\begin{figure}
\begin{center}
\psfig{figure=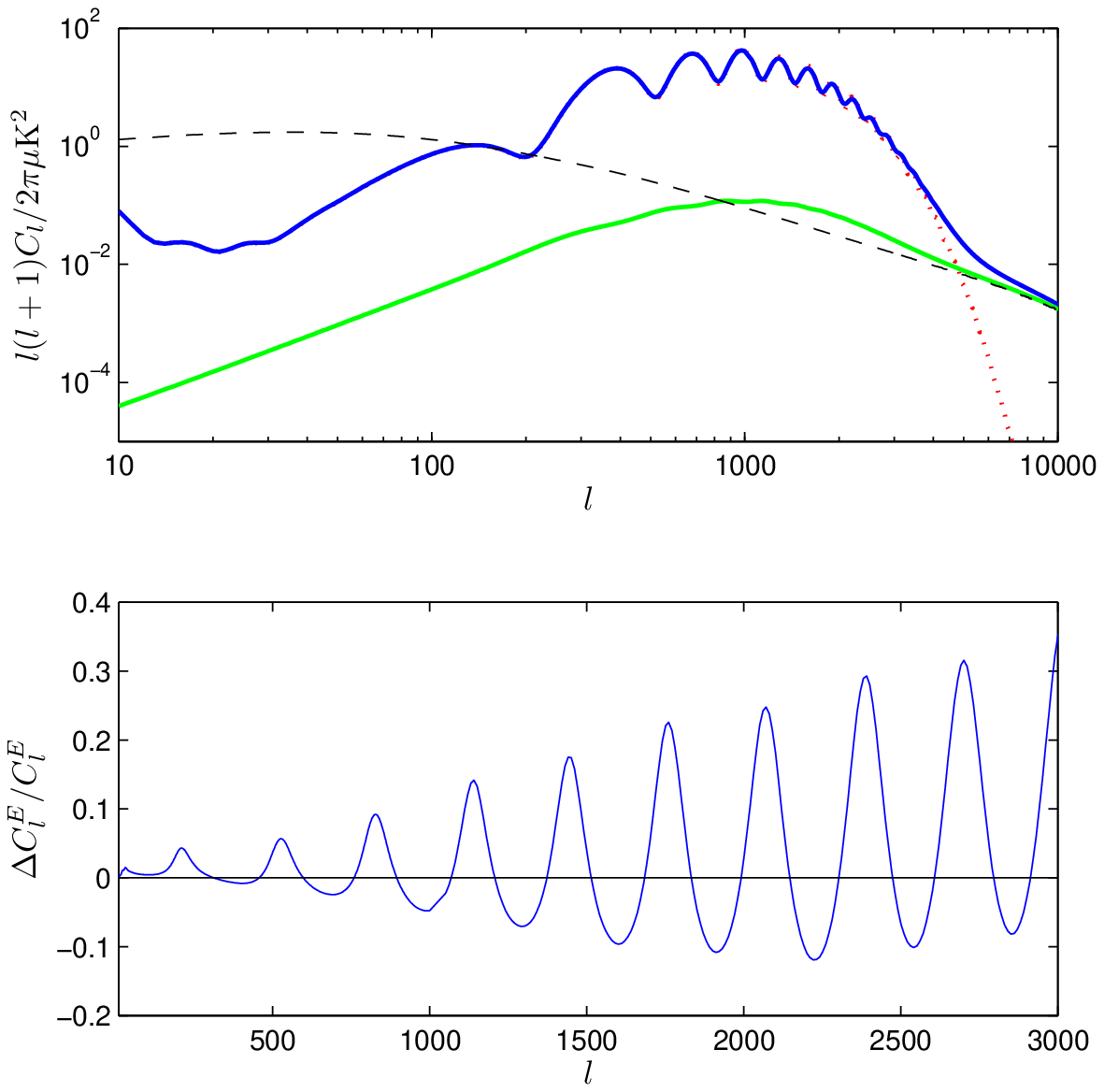,width=12cm}
\caption{Top: the lensed $E$ (top solid) and lensed $B$ (bottom solid) power spectra, compared to the unlensed $E$ spectrum (dotted) and the asymptotic result of Eq.~\eqref{EE_highl_approx} (dashed). Bottom: the fractional change in the $E$ power spectrum due to lensing. All results are for a fiducial standard $\Lambda$CDM cosmology.
\label{lensedEB}}
\end{center}
\end{figure}

The effect of lensing on the $E$-mode power spectrum is more like the effect on the temperature: it is essentially a convolution with the lensing potential, which smooths out the acoustic peaks. Because the unlensed polarization peaks are rather sharper than in the temperature case, this actually means that the effect of lensing is quantitatively more important on the $C_l^E$ spectrum. The fractional change in the spectrum is shown in Fig.~\ref{lensedEB}, and is clearly seen by eye in Fig.~\ref{lensedE}. Lensing must therefore be modelled in order to get accurate results when doing parameter estimation from polarized power spectra. On very small scales the unlensed polarization is damped and there is very little unlensed power. We can therefore apply a similar argument to Eq.~\eqref{TT_highl_approx} that we used for the temperature spectrum on small scales: the unlensed polarization field can be accurately described by a single gradient term. Using $l' \ll l$, we get the $l\gg 3000$ approximation
\begin{eqnarray}
\tC_l^E &\approx& \, C_{l}^\psi \int \frac{\ud^2 \vl'}{(2\pi)^2}\left[\vl'\cdot \vl\right]^2  C^E_{|\vl'|} \cos^2 2(\phi_{\vl'}-\phi_\vl)\nonumber\\
   &=& \frac{1}{2} l^2 C_l^\psi R^E
\label{EE_highl_approx}
\end{eqnarray}
for the $E$-polarization and
\begin{eqnarray}
\tC_l^B &\approx& \, C_{l}^\psi \int \frac{\ud^2 \vl'}{(2\pi)^2}\left[\vl'\cdot \vl\right]^2  C^E_{|\vl'|} \sin^2 2(\phi_{\vl'}-\phi_\vl)\nonumber\\
   &=& \frac{1}{2} l^2 C_l^\psi R^E \nonumber\\
   &=& \tC_l^E
\label{BB_highl_approx}
\end{eqnarray}
for the $B$-polarization. So on small scales $\tC_l^E\sim \tC_l^B$, and the power is proportional to the deflection angle power at that scale, as shown in Fig.~\ref{lensedEB}. Here
\begin{equation}
R^E \equiv \frac{1}{4\pi}\int \frac{\ud l}{l} \,l^4 C^E_l = \la |\grad Q|^2\ra = \la |\grad U|^2\ra \sim 2\times 10^7\muK^2
\label{RE}
\end{equation}
measures the total power in the gradient of the unlensed polarization field and is defined analogously to Eq.~\eqref{R_def} when there are no unlensed $B$-modes. The corresponding rms unlensed gradient is $ \la |\grad Q|^2\ra^{1/2} = \la |\grad U|^2\ra^{1/2} \sim 1.3\,\muK\, \arcmin^{-1}$. The fact that $E$ and $B$ have equal power on small scales is a special case of more general results for lensing of polarization gradients by small lenses (see Section~\ref{subsec:cluster_pol}).

\subsubsection{Lensed polarization correlation functions}

As in the case of the temperature spectrum, the series expansion in the deflection angle that we have used in the previous section is not expected to be very accurate on small scales, so for a more accurate calculation of the lensed power spectra we need a non-perturbative calculation, most easily performed via the correlation function. The calculation is rather similar to the one for the temperature we did in Section~\ref{corr_flat}, so we shall not labour the similar parts of the derivations here. The calculation was first done in Ref.~\cite{Zaldarriaga:1998ar}, though here we include new fully non-perturbative results as well as lowest terms in the series expansion.

We shall work from the spin-2 polarization field $P$. The scalar correlation function between polarization at $\vx$ and $\vx'$ should be independent of the basis used to define $P$ at the two points. To do this, we want to describe the polarization in the physically relevant basis defined by $\vr \equiv \vx - \vx'$. If $\vr$ makes an angle $\phi_r$ to the $\ve_x$ axis, this amounts to rotating the basis by an angle $\phi_r$ anticlockwise at each point, giving $P_r(\vx) = e^{-2i\phi_r} P(\vx)$. In this physical basis we can then define the basis-independent correlation functions
\begin{eqnarray}
 \xi_+(r) &\equiv& \la P^*_r(\vx) P_r(\vx') \ra = \la P^*(\vx) P(\vx') \ra \\
 \xi_-(r) &\equiv&  \la P_r(\vx)\, P_r(\vx') \ra =  \la e^{-4i\phi_r} P(\vx) P(\vx') \ra \\
\xi_X(r) &\equiv&  \la P_r(\vx)\, \Theta(\vx') \ra =  \la e^{-2i\phi_r} P(\vx) \Theta(\vx') \ra.
\end{eqnarray}
In tensor notation the first is proportional to
$\la P_{ab}(\vx) P^{ab}(\vx') \ra$ and the second to the spin-4 component of
$\langle P_{ab}(\vx) P_{cd}(\vx')$, however on the sphere one has to be careful to define what one means by contracting tensors at different points\footnote{The required operation is to parallel-transport one of the tensors along the geodesic connecting the two points.}, so we shall stick with the complex representation that generalizes more obviously to a spherical calculation.
We note that $\xi_+(r)$ is real because of statistical rotational invariance.
In terms of the Stokes' parameter defined in the physical basis by $P_r = Q_r + i U_r$ the correlation functions are
\begin{eqnarray}
 \xi_+(r) &=& \la Q_r(\vx)Q_r(\vx') + U_r(\vx)U_r(\vx') \ra \\
 \xi_-(r) &=& \la Q_r(\vx)Q_r(\vx') - U_r(\vx)U_r(\vx') \ra + i\la Q_r(\vx)U_r(\vx')+ U_r(\vx)Q_r(\vx')\ra \\
\xi_X(r) &=& \la Q_r(\vx)\Theta(\vx')\ra + i \la U_r(\vx)\Theta(\vx')\ra  .
\end{eqnarray}
If we assume that statistical parity invariance also holds, the correlators
must be invariant under reflection of the fields. Reflecting in $\vr$,
the temperature and $Q_r$ Stokes' parameter remain the same, but
$U_r$ changes sign, so that $\la Q_r(\vx)U_r(\vx')+ U_r(\vx)Q_r(\vx')\ra$ and $\la U_r(\vx) \Theta(\vx')\ra$ would change sign. For a statistically parity-invariant ensemble we therefore expect the imaginary parts to vanish, and we shall neglect them from now on.

The calculation for the lensing correlation function $\xil_+(r)$ is exactly analogous to that for the temperature in Section~\ref{corr_flat} with $C_l^\Theta$ replaced by $C_l^E + C_l^B$,
\begin{eqnarray}
\xil_+(r) &=& \la P^*(\vx + \valpha) P(\vx' + \valpha') \ra \nonumber\\
&=&  \frac{1}{2\pi} \int l \ud l\,
(C_l^E + C_l^B)e^{-l^2\sigma^2/2}\sum_{n=-\infty}^\infty  I_n(l^2\Cgltwo/2) J_{2n}(l r) \nonumber \\
&=& \frac{1}{2\pi} \int l \ud l\, (C_l^E +C_l^B) \,e^{-l^2 \sigma^2(r) /2}
\left[ J_0(lr) + \frac{1}{2}l^2 \Cgltwo(r) J_2(lr) +\dots\right].
\label{lensed_xiplus}
\end{eqnarray}
The calculation for $\xil_-(r)$ is however a little different
because of the rotation factor $e^{-4i\phi_r}$; we have
\begin{eqnarray}
\xil_-(r) &=& \la e^{-4i\phi_r} P(\vx + \valpha) P(\vx' + \valpha') \ra \nonumber\\
&=& \int \frac{\ud^2 \vl}{(2\pi)^2}\, (C_l^E- C_l^B) e^{i \vl \cdot \vr} e^{4i(\phi_l-\phi_r)}
\langle e^{i\vl \cdot (\valpha - \valpha')} \rangle_\valpha \nonumber\\
 &=& \int \frac{\ud^2 \vl}{(2\pi)^2} (C_l^E- C_l^B) e^{i \vl \cdot \vr} \cos 4\phi\,
\exp\left(-\frac{1}{2}l^2 [\sigma^2(r) + \cos 2\phi\, \Cgltwo(r)]\right) \nonumber\\
 &=& \int \frac{\ud^2 \vl}{(2\pi)^2} (C_l^E- C_l^B) e^{i l r \cos\phi} e^{-l^2 \sigma^2(r) /2} \left[2\frac{\ud^2}{\ud (l^2 \Cgltwo/2)^2}-1\right]\exp\left(-\frac{1}{2}l^2 \cos 2\phi\, \Cgltwo(r)\right)
 \nonumber\\
 &=& \frac{1}{2\pi} \int l \ud l\,(C_l^E- C_l^B) e^{-l^2 \sigma^2(r) /2}
 \sum_{n=-\infty}^\infty  \left[2I_n''(l^2\Cgltwo/2)-I_n(l^2\Cgltwo/2)\right] J_{2n}(l r) \nonumber\\
 &=&
 \frac{1}{2\pi} \int l \ud l\, (C_l^E - C_l^B)
 e^{-l^2 \sigma^2(r) /2}
\left( J_4(lr) + \frac{1}{4}l^2 \Cgltwo(r)[J_2(lr)+J_6(lr)]+\dots\right),
\label{lensed_ximinus}
\end{eqnarray}
where as before we defined $\phi = \phi_l-\phi_r$ and dashes denote derivatives with respect to the argument. The cross-correlation derivation is similar, giving
\begin{eqnarray}
\xil_X(r) &=& \la e^{-2i\phi_r} \Theta(\vx + \valpha) P(\vx' + \valpha') \ra \nonumber\\
 &=& -\int \frac{\ud^2 \vl}{(2\pi)^2} C_l^X e^{i \vl \cdot \vr} \cos 2\phi\,
\exp\left(-\frac{1}{2}l^2 [\sigma^2(r) + \cos 2\phi\, \Cgltwo(r)]\right) \nonumber\\
 &=& \frac{1}{2\pi} \int l \ud l\,C_l^X e^{-l^2 \sigma^2(r) /2}
 \sum_{n=-\infty}^\infty  I_n'(l^2\Cgltwo/2) J_{2n}(l r) \nonumber\\
 &=&
\frac{1}{2\pi} \int l \ud l\, C_l^X
 e^{-l^2 \sigma^2(r) /2}
\left( J_2(lr) + \frac{1}{4}l^2 \Cgltwo(r)[J_0(lr)+J_4(lr)]+\dots\right).
\label{lensed_X}
\end{eqnarray}

The series expansion to next order in $\Cgltwo$ and the spherical generalizations needed for a calculation accurate to within cosmic variance are given in Ref.~\cite{Challinor:2005jy} (see also Section~\ref{sec:curv}).

The relation between the correlation functions and the power spectra follow from a similar argument to Eq.~\eqref{xi_relation}, or using the orthogonality of the Bessel functions combined with the unlensed limit of Eqs.~\eqref{lensed_xiplus},~\eqref{lensed_ximinus} and~\eqref{lensed_X}:
\begin{eqnarray}
C_l^E + C_l^B &=&  2\pi \int r\ud r\, J_0(l r) \xi_+(r)\\
C_l^E - C_l^B &=&  2\pi \int r\ud r\, J_4(l r) \xi_-(r)\\
C_l^X  &=&  2\pi \int r\ud r\, J_2(l r) \xi_X(r).
\end{eqnarray}
The correlation function method is significantly more accurate than the lowest-order series-expansion result at $ 1000 \alt l \alt 6000$ as shown in Fig.~\ref{HuT}, and must be used for an accurate calculation of the lensed power spectra.

\subsection{$B$-mode confusion: implications for constraining inflation}
\label{sec:Bconfusion}

\begin{figure}
\begin{center}
\psfig{figure=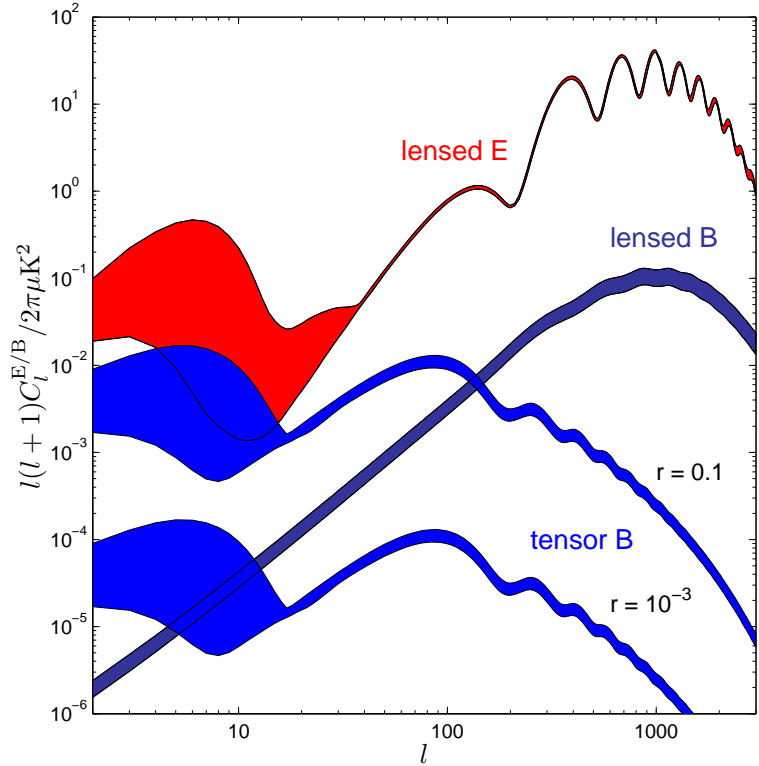,width=10cm}
\caption{The 95\%-confidence regions for the CMB polarization power spectra given a 2005 compilation of CMB and large scale structure data from Ref.~\cite{MacTavish:2005yk}. The parameters varied are those for a constant spectral index $\Lambda$CDM model with insignificant neutrino mass and sharp reionization. The two tensor results are the $B$-mode power spectra expected from for a scale-invariant tensor-mode power spectrum with $A_T= 4r \times 10^{-9}$ and two possible values of $r$. The effect of tensors on the cosmological parameter constraints was neglected, and a prior $\zre>6$ was assumed.
\label{ClRange}}
\end{center}
\end{figure}

Much future work on the CMB will be directed towards the detection of any primordial gravitational waves (tensor modes). This is because tensor modes produce a distinct $B$-mode polarization signal that can serve as a `smoking gun' for primordial gravitational waves if it can be distinguished from the lensing signal~\cite{Kamionkowski:1996ks,Seljak:1997gy}. Any detection of tensor modes would be of great interest as a way of discriminating different classes of models of inflation and other models of the early universe~\cite{LL,Mukhanov92,Steinhardt01,Linde:2005he}.

The primordial tensor modes are described by the transverse traceless part of the metric tensor, $h_{ij}$, which is expected to be Gaussian in most models. The power spectrum $P_h(k)$ of the super-horizon modes after inflation (or more generally, near the start of the radiation dominated era) is defined by
\begin{equation}
\la h^{ij} h_{ij} \ra = \int \frac{dk}{k} P_h(k).
\end{equation}
It is usually parameterized as
\begin{equation}
P_n(k) = A_T \left(\frac{k}{k_0}\right)^{n_T}
\end{equation}
for some choice of pivot scale $k_0$, where most inflationary models predict that $n_T$ is close to being constant. In the context of slow-roll inflation, values of $r$ consistent with current upper limits imply the tensor spectral index $n_T \approx 0$. For simplicity we shall assume $n_T=0$, which is likely to be a good approximation over the relevant range of scales. The pivot-point power $A_T$ is used to measure the amplitude of the primordial tensor modes. For standard single-field slow-roll models of inflation, the tensor amplitude is related approximately to the amplitude of the inflaton potential when the mode left the horizon by~\cite{Starobinskii:79,LL}
\begin{equation}
V_* \approx \frac{3A_T}{128} m_{\text{Pl}}^4 \approx A_T [4.8\times 10^{18} \GeV]^4,
\end{equation}
where $m_{\text{Pl}}$ is the Planck mass. For standard $\Lambda$CDM models a tensor amplitude of $A_T \sim 4\times 10^{-9}$ would contribute about the same power as the scalar modes to the large scale CMB anisotropy. If we define a `scalar/tensor' ratio
\begin{equation}
r \equiv A_t /(4\times 10^{-9}) \sim V_*/[3.8\times 10^{16} \GeV]^4,
\end{equation}
it will be approximately the ratio of the tensor to scalar contributions to the large-scale CMB temperature power. There are many different definitions of $r$ in the literature\footnote{Common definitions for $r$ are the ratio of the CMB temperature quadrupole powers, or a ratio of some scalar primordial power to the tensor power at a given scale. The definitions differ by order unity factors that depend on the cosmological parameters or choice of scalar variable.}, so for clarity we have defined it here in terms of the unambiguous tensor mode power $A_T$. The smallest tensor amplitude clearly detectable within the cosmic variance of the CMB temperature signal is around $A_T\sim 4\times 10^{-10}$ ($r\sim 0.1$), so the CMB polarization $B$-mode `smoking gun'~\cite{Seljak:1997gy} is needed to do much better than this.

The problem is that we have no idea what the tensor amplitude is, and it could be small. As we showed in Eq.~\eqref{Bwhite} the lensing of purely scalar $E$-mode polarization will produce a nearly white lensed $B$-mode spectrum $C_l^B \sim 2\times 10^{-6}\muK^2$ (corresponding to a standard deviation averaged in a
pixel of $1\,\text{arcmin}$ of $\surd C_l^B \sim 5 \muK\text{-}\arcmin$). If the $B$-modes from the tensor modes become comparably small, the lensing signal becomes an important source of confusion. For the Planck satellite the experimental noise gives $N_l^B \sim 4\times 10^{-4}\muK^2$ ($\surd N_l^B \sim 70 \muK\text{-}\arcmin$), which is large enough that the lensing `noise' is only a very small correction. However for future more sensitive observations, or those coving a smaller area of sky,  the lensing signal may become a limiting factor. The range of $C_l^B$ allowed by current data is shown in Fig.~\ref{ClRange}, along with various possible tensor $B$-mode spectra with different amplitudes. The large scale white noise power expected from lensing is currently constrained to be $C_l^B = (2.0\pm 0.2)\times 10^{-6}\muK^2$ in standard $\Lambda$CDM models.

Treating the lensing signal as a Gaussian noise, for a nearly full-sky observation with no experimental noise the lowest tensor amplitude that can be distinguished is about $A_T\sim (1\text{--}4)\times 10^{-12}$ (depending on the reionization redshift), corresponding to $r\sim (0.2\text{--}1)\times 10^{-3}$ and an energy scale of inflation $\sim (5\text{--}7)\times 10^{15}\GeV$~\cite{Hu:2001fb,Lewis:2001hp}. As observations approach the lensing limit, they need to survey more and more of the sky in order to beat down the lensing variance: the presence of the lensing signal therefore pushes observational strategies towards favouring large sky coverage~\cite{Hu:2001fb}. The issue is complicated by the mixing of $E$ and $B$ modes when only part of the sky is observed~\cite{Lewis:2001hp,Bunn:2002df,Lewis:2003an,Amarie:2005in}, which also favours larger sky coverage. Since the tensor-mode power is mostly on large scales ($l\alt 200$), this lensing noise limit can be reached with low-resolution experiments. However, as we shall discuss in Section~\ref{sec:unlens}, high-resolution observations can in principle be used subtract most of the lensing signal, and therefore obtain significantly better limits on the primordial gravitational waves.


%% file: nongauss.tex

Inflationary and other models predict a very nearly Gaussian spectrum of primordial perturbations. Since evolution to last scattering is essentially linear, the CMB anisotropies should also be very nearly Gaussian. Furthermore the lensing potential is most sensitive to large-scale modes of the density field which are also Gaussian. On small scales where non-linear evolution is important there will be many perturbations along the line of sight, reducing the effect of any non-Gaussianity in the lensing potential by the central-limit theorem. However the \emph{lensed} CMB---a non-linear function of the two nearly Gaussian fields---will not be Gaussian. In this section we discuss some of the properties of this non-Gaussian signature, assuming for simplicity the unlensed CMB and lensing potential are exactly Gaussian. Deviations from Gaussianity due to non-linear evolution can be assessed using numerical simulations~\cite{Jain:1999ir,Amblard:2004ih}. We shall not discuss other potentially important non-Gaussian signatures, such as that due to the SZ component~\cite{Cooray:2001wa}.

\subsection{Three-point function (bispectrum)}

For an unlensed sky, the temperature three-point function vanishes because there are an odd number of products and the distribution of $\Theta$ is an even function, so $\la \Theta(\vl_1)\Theta(\vl_2)\Theta(\vl_3)\ra = 0$. To consider the leading-order non-Gaussian signal for the lensed field we use the lowest-order series-expansion result from Eq.~\eqref{T_series_l}:
\begin{equation}
\tTheta(\vl) \approx \,\Theta(\vl) - \int \dFT{\vl'}  \vl'\cdot (\vl-\vl') \psi(\vl-\vl')\Theta(\vl').
\end{equation}
The three point function then has non-zero even terms, and we have to lowest order in $C_l^{\Theta\psi}$~\cite{Zaldarriaga:2000ud}
\begin{eqnarray}
\la \tTheta(\vl_1)\tTheta(\vl_2)\tTheta(\vl_3)\ra &\approx&  - \int \dFT{\vl'}  \vl'\cdot (\vl_3-\vl')  \la\Theta(\vl_1)\psi(\vl_3-\vl')\ra \la\Theta(\vl_2) \Theta(\vl')\ra + \text{perms} \nonumber\\
&\approx& -\frac{1}{2\pi} \delta(\vl_1+\vl_2+\vl_3)\left( C^{\Theta\psi}_{l_1} C^{\Theta}_{l_2} \vl_1\cdot\vl_2
+ \text{perms} \right),
\end{eqnarray}
where $\text{`perms'}$ indicates the five other permutations of $\vl_1\dots\vl_3$. The three point function therefore directly probes the large-scale temperature-lensing potential correlation. The cross-correlation on large scales is due to the integrated Sachs-Wolfe contribution to the primary anisotropy. This comes from low redshift, and hence is correlated with the matter fluctuations contributing to the lensing potential.  The bispectrum can therefore be used as a probe of the perturbation growth and expansion history of the universe at low redshift, and hence can be used to help constrain the dark energy and curvature~\cite{Seljak:1998nu,Goldberg:1999xm,Hu:2001fb,Giovi:2003ri,Giovi:2004te,Gold:2004ee}.

On small scales where all the $\vl$ are large the cross-correlation is negligible (see Fig.~\ref{phiTcorr}), and the three point function vanishes even for the lensed sky. Note that because the lensed field is linear in $\Theta$, the three point function is exactly zero if there is no cross-correlation---this result does not depend on the series expansion. On very small scales our assumption of Gaussianity of the lensing potential will not be quite correct and there may be additional contributions to the bispectrum because of non-linear evolution (and hence skewness) of $\psi$~\cite{Pen:2003vw,Cooray:2000uu}: the distribution must be skewed in non-linear theory because the matter densities are constrained to be larger than zero. Late-time non-linear effects other than lensing  may also introduce a non-zero bispectrum, in particular through correlations of CMB secondaries with the lensing potential, especially SZ~\cite{Goldberg:1999xm,Komatsu:2001rj,Kesden:2002jw}.

\subsection{Four-point function (trispectrum)}

The four-point function is the first higher-order correlation function that has an interesting signal on small scales where the lensing potential can be taken to be statistically independent of the temperature~\cite{Bernardeau:1996aa,Hu:2001fa}. For the Gaussian unlensed fields we have
\begin{equation}
\la \Theta(\vl_1)\Theta(\vl_2)\Theta(\vl_3)\Theta(\vl_4) \ra = C^\Theta_{l_1}C^\Theta_{l_3}\delta(\vl_1+\vl_2)\delta(\vl_3+\vl_4) + \text{perms}.
\end{equation}
The non-Gaussian signature is therefore in the connected part of the correlation function that vanishes for Gaussian fields~\cite{Zaldarriaga:2000ud}:
\begin{multline}
\la \tTheta(\vl_1)\tTheta(\vl_2)\tTheta(\vl_3)\tTheta(\vl_4) \ra_c \equiv  \la \tTheta(\vl_1)\tTheta(\vl_2)\tTheta(\vl_3)\tTheta(\vl_4) \ra -  \left[\tC^\Theta_{l_1}\tC^\Theta_{l_3}\delta(\vl_1+\vl_2)\delta(\vl_3+\vl_4) + \text{perms}\right] \\
 \approx \frac{1}{2(2\pi)^2}\delta(\vl_1+\vl_2+\vl_3+\vl_4)\left[ C^\psi_{|\vl_1+\vl_3|} C^\Theta_{l_3} C^\Theta_{l_4}(\vl_1+\vl_3)\cdot \vl_3 \, (\vl_2+\vl_4)\cdot \vl_4 + \text{perms} \right],
\label{eq:non_gauss_trispectrum}
\end{multline}
where `perms' denotes all permutations of $\vl_1 \ldots \vl_4$. The four-point function therefore probes the lensing potential power spectrum.  It should be clearly distinguishable from cosmic variance for observations with high sensitivity at $l\agt 1000$ (see Ref.~\cite{Hu:2001fa}).
We shall return to the issue of extracting information about the lensing potential from non-Gaussian signatures in Section~\ref{sec:recon}.

The $n$-point functions can also be evaluated in real space, for example the connected four point correlation is $\la \tTheta(\vx_1)\tTheta(\vx_2)\tTheta(\vx_3)\tTheta(\vx_4) \ra_c$. These forms are often more convenient for partial-sky observations as they are manifestly local. However the number of different pixel correlations is very large, and computing all possible quadrilaterals would be practically impossible. For this reason practical estimators tend to concentrate on particular subsets of geometries, for example when two points are the same~\cite{Bernardeau:1996aa}. For lensing purposes the main interest is in reconstructing the lensing potential, and methods adapted for this purpose are described in Section~\ref{sec:recon}.

Secondary anisotropies such as SZ will have some correlation with the small-scale lensing potential, introducing additional non-Gaussian terms and modifying the expectation for the trispectrum~\cite{Kesden:2002jw}.

The fact that lensing induces a non-zero four-point function also changes the covariance of lensed power spectrum estimates relative to the uncorrelated Gaussian expectation $\la |\Delta \hat{C}_l|^2\ra = 2C_l^2/(2l+1)$. This is essentially because the large scale lensing modes correlate the $C_l$, reducing the effective number of independent $\tTheta_{lm}$ observations. The effect is however rather small ($\sim 2\%$) at Planck sensitivity, though potentially becoming more important on smaller scales~\cite{Hu:2001fa}. The effect is, however,
significant for $B$-mode polarization as we describe in
Section~\ref{subsec:non_gauss_pol}.

\subsection{Connected $n$-point correlation functions (cumulants)}
For a Gaussian field the $n$-point correlation functions are equal to their disconnected parts for $n>2$. The connected part of the functions for $n > 2$ therefore define an infinite series of higher-order non-Gaussian signals, also known as the cumulants~\cite{Winitzki:1998jq}. Since the lensed field is linear in the unlensed $\Theta$, all the odd-cumulants are zero both with and without lensing (neglecting small large scale cross-correlations). The even cumulants for $n\ge 6$ depend at $\clo(C_l^\psi)$ on $n > 2$ cumulants of the lensing potential, which are zero if the lensing potential is Gaussian~\cite{Kesden:2002jw}. However there will be non-zero higher-order terms.

\subsection{Polarization}
\label{subsec:non_gauss_pol}

The polarization fields are essentially uncorrelated with the lensing
potential so the $n$-point functions with $n$ odd are zero\footnote{There will be a tiny correlation from polarization generated by CMB quadrupole scattering at clusters.}. The trispectrum
for $E$-mode polarization is similar to that for the temperature anisotropies:
\begin{eqnarray}
\langle \tilde{E}(\vl_1)\tilde{E}(\vl_2) \tilde{E}(\vl_3) \tilde{E}(\vl_4)
\rangle_c &\approx& \frac{1}{2(2\pi)^2}\delta(\vl_1 + \vl_2 + \vl_3 + \vl_4)
\nonumber \\
&&\mbox{} \times
\left[C^\psi_{|\vl_1+\vl_3|}C^E_{l_3}C^E_{l_4}W_E(\vl_1,-\vl_3)
W_E(\vl_2,-\vl_4) + \text{perms}\right],
\end{eqnarray}
where $W_E(\vl,\vl')\equiv \vl'\cdot(\vl-\vl')\cos 2(\phi_{\vl}-\phi_{\vl'})$
and we have ignored any unlensed $B$-modes. For the lens-induced $B$-modes,
at leading order these are first order in $\psi$ so the trispectrum
is $\clo(C_l^\psi)^2$; we find
\begin{eqnarray}
\langle \tilde{B}(\vl_1) \tilde{B}(\vl_2) \tilde{B}(\vl_3) \tilde{B}(\vl_4)
\rangle_c & \approx & \frac{1}{4} \delta( \vl_1 + \vl_2 + \vl_3 + \vl_4)
\nonumber \\
&&\mbox{} \times
\int\, \frac{\ud^2 \vL}{(2\pi)^4} \left[ C_L^E C^E_{|\vl_1+\vl_3-\vL|}
C^\psi_{|\vl_1-\vL|}C^\psi_{|\vl_2+\vL|} W_B(\vl_1,\vL)W_B(\vl_2,-\vL)
\right. \nonumber \\
&&\mbox{} \phantom{xxxx} \times \left.
W_B(\vl_3,\vl_1+\vl_3-\vL)W_B(\vl_4,\vL-\vl_1-\vl_3) + \text{perms} \right],
\label{eq:non_gauss_btri}
\end{eqnarray}
where $W_B(\vl,\vl')\equiv \vl'\cdot(\vl-\vl')\sin 2(\phi_{\vl}-\phi_{\vl'})$.
The lensed $E$ and $B$ modes still have opposite parity and so statistical
parity invariance (with isotropy)
demands that they are (exactly) uncorrelated. However, they are no
longer independent and all of the mixed cumulants (i.e.\ those with
one, two or three lensed $B$-modes) are non-zero.
We shall only give the result here for the
cumulant involving two lensed $E$-modes and two lensed $B$-modes since
this is responsible for correlating the $E$ and $B$-mode power spectra.
At leading order we have
\begin{eqnarray}
\langle \tilde{E}(\vl_1) \tilde{E}(\vl_2) \tilde{B}(\vl_3) \tilde{B}(\vl_4)
\rangle_c & \approx & \frac{1}{(2\pi)^2} \delta(\vl_1+\vl_2+\vl_3+\vl_4)
\nonumber \\
&&\mbox{} \hspace{-0.2\textwidth} \times
C_{l_1}^E C_{l_2}^E  \left[C_{|\vl_1+\vl_3|}^\psi W_B(\vl_3,-\vl_1)
W_B(\vl_4,-\vl_2) + C_{|\vl_2+\vl_3|}^\psi W_B(\vl_3,-\vl_2)W_B(\vl_4,-\vl_1)
\right].
\end{eqnarray}
Trispectra can also be formed between the temperature and polarization fields
but we shall not consider these here. For a discussion of the spherical
generalization of the polarization trispectra, see Ref.~\cite{Okamoto:2002ik}.

As with the temperature, the non-zero trispectra alter the covariance of
polarization power spectra estimates. The most important effect is in
the covariance of the $B$-mode power since all of the signal comes from
lensing~\cite{Smith:2004up}. The flat-sky covariance of the lensed
$B$-mode spectrum is approximately
\begin{equation}
\text{cov}(\tilde{C}^B_l , \tilde{C}^B_{l'}) =
\frac{2(\tilde{C}^B_l)^2}{2l} \delta_{ll'}
+ \frac{1}{4\pi} \int \ud \phi_\vl \ud \phi_{\vl'} T_c(\vl,-\vl,\vl',-\vl'),
\end{equation}
where the connected trispectrum $T_c(\vl_1,\vl_2,\vl_3,\vl_4)$ is given by the
factor on the right of Eq.~(\ref{eq:non_gauss_btri}) that multiplies
the delta function. The presence of this term increases the diagonal
of the covariance and correlates the power spectra over a broad range of
scales~\cite{Smith:2004up,Smith:2005ue}. It is essential to include properly
this non-Gaussian covariance in parameter estimation from future lensed
$B$-mode spectrum estimates, both to avoid under-estimation
of the errors and biases in the estimated
parameters~\cite{Smith:2005ue}. The number of independent $B$-modes produced by
lensing is lower than if the field were Gaussian and this reduction
demands that (more) careful attention be paid to the
approximate form of the likelihood that is assumed for parameter
estimation to avoid further bias.

\subsection{Other non-Gaussian signatures}

Isotropic Gaussian random fields have many special properties, deviations from which constitute many different possible non-Gaussian signatures~\cite{BBKS,Bond87}. The lensing signal is expected to show up in many of these. For example the hot and cold spot ellipticities have a very specific probability distribution for Gaussian fields~\cite{Bond87}, so additional ellipticities caused by lensing shear should show up clearly as an excess of more elliptical shapes~\cite{Bernardeau:1998mw}. Likewise the hot and cold spot correlation function should be changed in a characteristic way~\cite{Takada:1999mv,Takada:2001af}.

\subsection{Distinguishing primordial non-Gaussianity}

Lensing changes the spatial correlation structure of the CMB field, but does not change the distribution at a point. It follows that any primordial non-Gaussian signature in the one-point function will be preserved by lensing: points are moved around, but the temperature distribution of the points is unchanged for an observation with infinite resolution. However beam effects do complicate the analysis, so lensing does change the beam-convolved one-point function~\cite{Winitzki:1998jq,Kesden:2002jw}.

The three-point function (bispectrum) from lensing ideally only has a cross-correlation signal on large scales, but in practice has an additional signal from SZ-lensing correlation on small scales. These are important sources of confusion for a primordial bispectrum, but may be distinguished by their distinctive scale dependence~\cite{Komatsu:2001rj}.
The lensing also has a significant four-point function (trispectrum)~\cite{Lesgourgues:2004ew}, but again could be distinguished from primordial signals by the approximately known amplitude and different scale dependence~\cite{Brunier:2006fu}. The non-zero lensing trispectrum also means that the variance of the bispectrum is increased from the Gaussian result: lensing makes it harder to distinguish a small scale primordial bispectrum from cosmic variance~\cite{Babich:2004yc}.

Confusion with kinetic SZ and point source signals may be more important than lensing, though different sources should have different scale dependence~\cite{Komatsu:2001rj}. The CMB polarization may be much cleaner (SZ signals are higher order), and if sufficiently sensitive CMB polarization data can be used some primordial non-Gaussian signals may ultimately be lensing limited. However since non-Gaussian scalar modes only produce linear $E$-modes, if precision $B$-polarization data is available it may be possible to use the information in the $B$ modes to reconstruct the lensing field (see Sec.~\ref{sec:recon}), effectively removing some of the non-Gaussian signal due to an unknown lensing potential.

%% file: recon.tex

A fixed large-scale lensing potential will lens smaller background unlensed CMB anisotropies in a characteristic way. By measuring a large number of these anisotropies it should therefore be possible to extract information about the lensing deflection field. 
The unlensed CMB field is, of course, unobservable, but its statistics are very well understood. Statistical measures of the lensed CMB fields should therefore be able to constrain the lensing potential~\cite{Zaldarriaga:1998te,Hu:2001fa,Hu:2001tn}. On small scales, where there is little unlensed CMB power, the lensed CMB  measures the small-scale lensing potential more or less directly.

The lensing potential can also be partly reconstructed using large-scale structure tracers such as galaxy lensing or 21cm emission~\cite{Cooray:2002ng,Sigurdson:2005cp,Zahn:2005ap}.
The 21cm power spectrum is relatively featureless, and hence contains relatively little information from a given redshift about the lensing potential. However the possibility of using different redshift slices in principle allows 21cm reconstruction to be useful~\cite{Zahn:2005ap}.

Here we concentrate on reconstruction from the CMB temperature and polarization, and  assume the primordial fields are Gaussian (if there is a primordial non-Gaussian signature this can complicate the analysis~\cite{Lesgourgues:2004ew}). In this section we discuss various ways of estimating the lensing potential. In Section~\ref{sec:unlens} we discuss the application to delensing and how accurately it can be done in principle and practice. Applications to cluster CMB lensing are discussed further in Section~\ref{subsec:clusters}.

\subsection{Temperature quadratic estimator}

For a flat-sky statistically-isotropic ensemble, harmonics with different $\vl$ are uncorrelated, $\la \Theta(\vl) \Theta^*(\vl')\ra = \delta(\vl-\vl') C_l^\Theta$. However the actual sky we observe is of course not isotropic, and in particular for a given fixed lensing potential, the distribution of the observed temperature will not be isotropic. This suggests that we may be able to use the quadratic off-diagonal terms of the $\psi$-fixed correlation $\la \tTheta(\vl) \tTheta(\vl')\ra_\Theta$ to constrain the lensing potential in our sky realization~\cite{Hu:2001tn,Hu:2001kj,Hu:2001fa,Okamoto03,Cooray:2002py}. This essentially amounts to using correlations between lots of temperature anisotropies around each lensing clump to constrain the potential of the clump.

We use the series expansion in the deflection angle (Eq.~\eqref{T_series_l}) to lowest order in the lensing potential. Averaging over realizations of the unlensed temperature field $\Theta$ gives
\begin{eqnarray}
\la \tTheta(\vl) \tTheta^*(\vl-\vL) \ra_\Theta &=& \delta(\vL) \,C_l^\Theta - \int \dFT{\vl'}\bigl[ \vl'\cdot(\vl-\vl')\psi(\vl-\vl')\la \Theta(\vl')\Theta^*(\vl-\vL) \ra \nonumber\\
 &&\phantom{\delta(\vL) C_l^\Theta - \int \dFT{l'}}
 +\vl'\cdot(\vl-\vL-\vl') \psi^*(\vl-\vL-\vl')
\la \Theta(\vl) \Theta^*(\vl')\ra\bigr] + \mathcal{O}(\psi^2)\nonumber\\
&=&\delta(\vL)\, C_l^\Theta + \frac{1}{2\pi}\left[ (\vL-\vl)\cdot \vL \,C^\Theta_{|\vl-\vL|} + \vl\cdot\vL\, C^\Theta_l\right] \psi(\vL)  + \mathcal{O}(\psi^2).
\end{eqnarray}
The $\vL\ne 0$ correlation therefore probes the lensing potential. The $\psi(\vL=0)$ mode of the lensing potential is not observable (zero gradient), so below we implicitly consider only $|\vL|>0$. To estimate the lensing potential on our sky we perform a weighted average of the off-diagonal terms, defining the quadratic estimator
\begin{equation}
\hat{\psi}(\vL) \equiv N(\vL) \int \dFT{\vl} \tTheta(\vl) \tTheta^*(\vl-\vL) g(\vl,\vL),
\end{equation}
where $g(\vl,\vL)$ is some weighting function. In order for the estimator to be unbiased at lowest order we want $\la \hat{\psi}(\vL) \ra_\Theta = \psi(\vL)$, so the normalization is
\begin{equation}
N(\vL)^{-1} = \int \frac{\ud^2 \vl}{(2\pi)^2} \left[ (\vL-\vl)\cdot \vL \,C^\Theta_{|\vl-\vL|} + \vl\cdot\vL\, C^\Theta_l\right] g(\vl,\vL).
\end{equation}
We are then free to choose $g$ to maximize the signal to noise.
To zeroth order in $C_l^\psi$ the variance is given by $\la|\hat{\psi}(\vL)-\psi(\vL)|^2\ra\sim \la|\hat{\psi}(\vL)|^2\ra$ and the non-Gaussian connected part of the four-point function (Eq.~\eqref{eq:non_gauss_trispectrum}) can be neglected. We then have
\begin{equation}
\la \hat{\psi}^*(\vL)\hat{\psi}(\vL')\ra = \delta(\vL-\vL')2N(\vL)^2\int \frac{\ud^2 \vl}{(2\pi)^2} \Ctot_l \Ctot_{|\vl-\vL|} [g(\vl,\vL)]^2 + \clo(C_l^\psi),
\end{equation}
where $\Ctot_l = \tC^\Theta_l + N_l$ and $N_l$ is the noise contribution.
Minimizing the leading-order variance gives a weight function~\cite{Hu:2001tn}
\begin{equation}
g(\vl,\vL) = \frac{(\vL-\vl)\cdot \vL \,C^\Theta_{|\vl-\vL|} + \vl\cdot\vL\, C^\Theta_l}{2\Ctot_l \Ctot_{|\vl-\vL|}}.
\end{equation}
Here we have chosen the arbitrary normalization of $g$ so that the lowest order `noise' on the reconstructed potential is determined by
\begin{equation}
\delta(\mathbf{0})
\la|\hat{\psi}(\vL)|^2\ra^{-1}  = N(\vL)^{-1} = \int \frac{\ud^2 \vl}{(2\pi)^2} \frac{  \left[ (\vL-\vl)\cdot \vL \,C^\Theta_{|\vl-\vL|} + \vl\cdot\vL\, C^\Theta_l\right]^2 }{ 2\Ctot_l \Ctot_{|\vl-\vL| }}.
\end{equation}
The delta function here should be interpreted in terms
of the sky area in the usual way: $\delta(\mathbf{0})=f_{\text{sky}}/\pi$.

These results allow us to estimate the lensing potential to within some `noise' determined by a combination of cosmic variance from the finite number of temperature modes and observational noise. Note that the estimator is only valid to lowest order in the lensing potential; this means that in reality the estimator is biased due to the significance of higher-order terms. In particular the $\clo(C_l^\psi)$ terms we have neglected will be significant when attempting to estimate the lensing potential power spectrum from the variance.  These terms need to be accounted for via a correction to get unbiased results~\cite{Cooray:2002py}, and a full analysis is given in Ref.~\cite{Kesden:2003cc}.

The estimator can be generalized to the curved sky~\cite{Okamoto03}, but for small sky areas curved-sky corrections are likely to be small compared to the neglect of higher order terms. Note also that this is just a quadratic estimator, it is not the optimal estimator --- see the discussion of Bayesian methods below.

Combining the above results, the harmonic-space result for the estimator becomes
\begin{equation}
\hat{\psi}(\vL) = N(\vL)\, \vL\cdot\int \dFT{\vl} \frac{\vl C^\Theta_l \tTheta(\vl)}{\Ctot_l} \frac{\tTheta(\vL-\vl)}{  \Ctot_{|\vL-\vl|}}.
\end{equation}
The integral is a harmonic-space convolution, which may therefore be written alternatively as a real-space product, giving~\cite{Hu:2001tn}
\begin{eqnarray}
\hat{\psi}(\vL) &=& -i N(\vL)\, \vL\cdot \int \dFT{\vx} e^{-i\vL\cdot\vx} F_1(\vx)\grad F_2(\vx) \nonumber\\
                &=& - N(\vL)\int \dFT{\vx} e^{-i\vL\cdot\vx} \grad \cdot\left[F_1(\vx)\grad F_2(\vx)\right]
\end{eqnarray}
where we defined two filtered fields given in harmonic space by
\begin{equation}
F_1(\vl) \equiv \frac{\tilde{\Theta}(\vl)}{\Ctot_l} \qquad F_2(\vl) \equiv  \frac{\tilde{\Theta}(\vl) C^\Theta_l}{\Ctot_l}.
\end{equation}
This tells us that the estimator of the potential is related to correlations between the gradient of the temperature (encoded in $\grad F_2(\vx)$) and the small-scale filtered anisotropy $F_1(\vx)$. These should be correlated because on small scales the unlensed CMB is approximately a temperature gradient, and the small-scale lensed anisotropies come from disturbing this gradient with the deflection angle from the lensing potential. The particular choice of filters improves the signal to noise over similar estimators used in earlier work~\cite{Zaldarriaga:1998te}.
This result can be generalized to the curved sky case~\cite{Okamoto03} for application to large-area observations. The representation of the estimator in terms of real-space products of filtered fields is numerically efficient because the integrals are now in the form of Fourier transforms which can be performed quickly using FFTs. On the curved sky they can be performed in terms of spherical harmonic transforms, which are still quite efficient because the azimuthal integrals can be performed using FFTs.

In practice the reconstruction may not be so straightforward. In addition to the error from the use of the series expansion, the lensing signal can also be complicated by contributions from the kinetic-SZ effect, and non-linear evolution may be important on small scales. If not taken into account these can significantly bias the result obtained from the simple quadratic estimator derived above~\cite{Kesden:2003cc,Amblard:2004ih}. The kinetic SZ signal comes predominantly from large clusters, so masking out these sources using thermal SZ measurements can help to reduce this source of reconstruction error (at the expense of slightly reducing the lensing signal)~\cite{Amblard:2004ih}.

\subsection{Polarization quadratic estimators}

\begin{figure}
\begin{center}
\psfig{figure=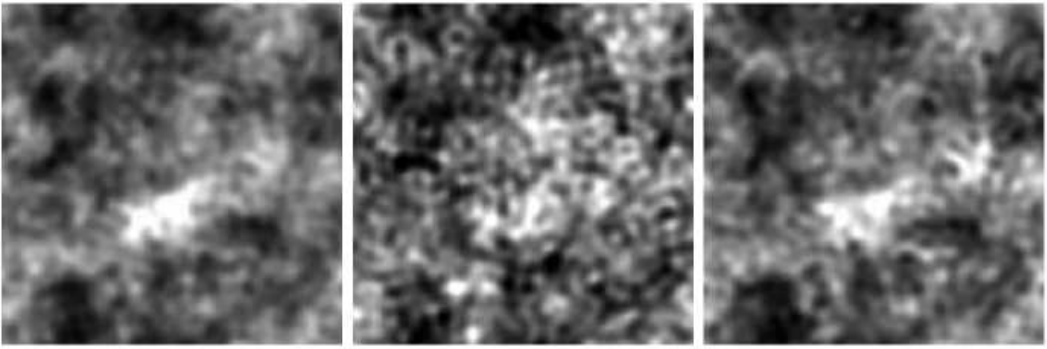,width=14cm}
\caption{A simulated reconstruction of the magnitude of the deflection angle from Ref.~\cite{Hu:2001kj}. The left panel shows the deflection from the realization used in the simulation. The centre panel shows the field reconstructed using the temperature quadratic estimator. The right hand panel shows the better result from using the $EB$ quadratic estimator. The simulation is for an idealized CMB observation with isotropic white-noise variance $N_l^\Theta=N^E_{l}/2=N^B_{l}/2 = 8.5\times 10^{-8} \muK^2$ ($\surd{N_l^\Theta} = 1 \muK$-arcmin) and beam full-width half-maximum of $4\,\arcmin$. The frames are $10^\circ\times 10^\circ$.
\label{HuTTEB}}
\end{center}
\end{figure}

Precision observations of the lensed CMB polarization field have the potential to improve greatly the lensing reconstruction from the temperature alone~\cite{Benabed:2000jt,Guzik:2000ju,Hu:2001kj,Okamoto03,Hirata:2003ka}. The reason why the polarization is potentially much better is that the sky should have zero $TB$ and $EB$ correlations without lensing: they have no cosmic variance from the unlensed fields on small scales. These terms can therefore be used to construct good quadratic estimators for the lensing potential along the lines of the temperature estimator. In addition, lens-induced $B$-modes typically arise from smaller-scale lenses than for temperature, allowing reconstruction of the lensing potential out to smaller scales.

For two CMB fields $X$ and $Y$, the estimators are of the form~\cite{Hu:2001kj}
\begin{equation}
\hat{\psi}_{XY}(\vL) = N_{XY}(\vL) \int \dFT{\vl} \tX(\vl) \tY^*(\vl-\vL) g_{XY}(\vl,\vL).
\end{equation}
The $BB$ estimator gives very little information when there are no unlensed $B$-modes: to get a good estimator we need to correlate the lens-induced $B$ modes with the $E$ modes. For the case $X=E$ and $Y=B$ the minimum-variance estimator has
\begin{equation}
g_{EB}(\vl,\vL) = \sin (2\phi) \frac{\vl\cdot\vL\, C^E_l+(\vL-\vl)\cdot \vL \,C^B_{|\vl-\vL|}}{\Ctot{}^E_l \Ctot{}^B_{|\vl-\vL|}},
\end{equation}
where $\phi$ is the angle between $\vl$ and $\vL-\vl$ and on small scales we expect the unlensed $C^B_l=0$. The corresponding lowest-order noise is given by
\begin{equation}
\label{polnoise}
\delta(\mathbf{0})\la|\hat{\psi}_{EB}(\vL)|^2\ra^{-1}  = N_{EB}(\vL)^{-1} = \int \frac{\ud^2 \vl}{(2\pi)^2} \frac{  \left[ \vl\cdot\vL\, C^E_l + (\vL-\vl)\cdot \vL \,C^B_{|\vl-\vL|}\right]^2 \sin^2 (2
\phi) }{ \Ctot_l{}^E \Ctot{}^B_{|\vl-\vL| }}.
\end{equation}
As for temperature, this weighting is optimal only for the leading-order
part of the variance (which is what is given in the last equation).
Results for the other polarization estimators take a similar form and can be found in Ref.~\cite{Hu:2001kj}, and efficient forms are given in Ref.~\cite{Okamoto03}. In the general case all the possible estimators may be combined to give a total minimum-variance estimator for the lensing potential.

For low-noise observations the $EB$ estimator turns out to give the highest signal-to-noise reconstruction~\cite{Hu:2001kj}. We can see the reason for this in Eq.~\eqref{polnoise} --- for low noise levels $\Ctot{}_l^B$ is small, and given entirely by the lensing contribution to the $B$-modes, so the noise is small. A simulated reconstruction comparing the temperature and $EB$ results for a simulated low-noise observation is shown in Fig.~\ref{HuTTEB}.

One might wonder why it is the \emph{lensed} fields that are determining the noise level. This is in fact only a property of the quadratic estimator: more general methods have cosmic variance determined essentially by the \emph{unlensed} fields.  The unlensed $B$ modes are expected to vanish on small scales, so in this case the noise on the reconstruction should be tiny. We therefore now consider a better, more Bayesian, analysis of the reconstruction problem.

\subsection{Bayesian methods}

\begin{figure}
\begin{center}
\psfig{figure=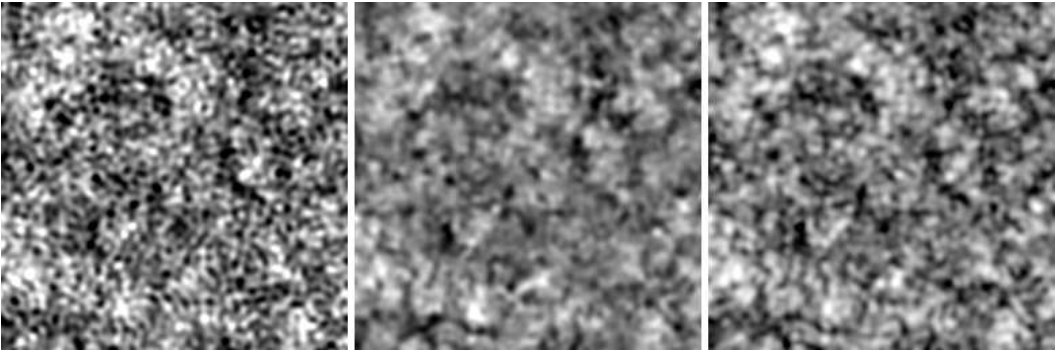,width=14cm}
\caption{A simulated reconstruction of the lensing convergence $\kappa$ from Ref.~\cite{Hirata:2003ka} using CMB polarization. The left panel shows a realization of the convergence used in the simulation. The centre panel shows the Wiener filtered field reconstructed using the quadratic estimator. The right hand panel shows the improved iterative Bayesian estimator of Ref.~\cite{Hirata:2003ka}. The simulation is for an idealized CMB observation with isotropic white-noise variance $N^E_{l}=N^B_{l} = 1.7\times 10^{-7} \muK^2$ ($\surd N^E_l = 1.41\muK\,\arcmin$, about $50$ times lower than Planck) and beam full-width half-maximum of $4\, \mu$K-arcmin. The frames are $8^\circ 32'$ in width, and show only the $l\le1600$ modes for clarity. The Wiener filtered field appears smoother because the filtering suppresses the poorly recovered small-scale modes.
\label{HirataSeljak}}
\end{center}
\end{figure}

Assuming the unlensed fields, lensing potential and noise are all Gaussian, we can write down exactly what the probability of a given observation should be. From this we can use Bayes' theorem to solve for the posterior distribution of the fields given the observed data. This probability function is all we can ever know about the fields, and therefore contains the most information we can hope to extract. Useful estimators such as the maximum-likelihood can in principle be found once the full probability function is known.

We write the unlensed CMB field as a vector in pixel or harmonic space, $\vX$. Here $\vX$ is in general taken to have components for the temperature, $E$ and $B$ polarization fields. The observation consists of the lensed fields plus noise, $\vX^\tot = \tvX + \vn$. Given a theoretical covariance of the unlensed fields $\mS = \la \vX \vX^\dag\ra$ determined by the power spectra, the posterior distribution of the lensing potential and unlensed fields is given by
\begin{equation}
-2 \log P(\vpsi,\vX|\vX^\tot,\mS,\mN) = [\vX^\tot - \mLambda(\psi)\vX]^\dag \mN^{-1}
[\vX^\tot - \mLambda(\psi)\vX] -2\log P(\vpsi,\vX|\mS) + \text{const}.
\label{PpsiX}
\end{equation}
Here $\mN$ is the (assumed Gaussian) noise covariance and we have defined the lensing operator $\mLambda(\psi)$ so that $\vXt = \mLambda(\psi)\vX$. In real space $\mLambda(\psi)$ just represents a point re-mapping matrix determining how points move around under the lensing deflection angle. Note that $\mLambda(\psi)$ is a non-linear function of $\psi$ unless we make the (in general poor) first-order series-expansion approximation.
In the absence of polarization observations the prior is given by
\begin{equation}
-2 \log P(\vpsi,\vTheta|\mS)   =\begm \vTheta^\dag & \vpsi^\dag \enm \begm \mC^{\Theta} &
\mC^{\Theta\psi} \\ \mC^{\Theta\psi} & \mC^{\psi} \enm^{-1} \begm \vTheta \\ \vpsi \enm
\label{bayes_full}
\end{equation}
assuming isotropic Gaussian unlensed fields. The $\Theta$--$\psi$ correlation
is about $50\%$ for the quadrupole, but falls to less
than $1\%$ by $l\sim 150$ (see Fig.~\ref{phiTcorr}). The prior therefore contains significant information about the potential given the large-scale unlensed temperature $\vTheta$ (which is very close to the lensed large-scale $\tilde{\vTheta}$).
From now on we neglect the information about the potential that can be learnt directly from the correlation with the large scale temperature. We shall concentrate instead on what can be learnt from the lensing signal, taking the temperature and lensing fields to be uncorrelated.

For a fixed potential, the posterior distribution of the unlensed field $\vX$ is Gaussian (the log probability is quadratic in $\vX$ as long as the prior is Gaussian).
If we are only interested in reconstructing the lensing potential, we can integrate out the unlensed field $\vX$. A simple way to get at the answer is to consider averaging over $\vX$ and the noise $\vn$, but keeping the potential $\psi$ fixed, to give the variance
\begin{equation}
\mC_g \equiv \la \vX^\tot \vX^\tot{}^\dag \ra_{\vX,\vn} = \mLambda(\psi)\la \vX\vX^\dag\ra\mLambda(\psi)^\dag +\mN = \mLambda(\psi) \mS \mLambda(\psi)^\dag + \mN.
\end{equation}
Marginalized over noise and unlensed $\vX$ realizations we therefore have
\begin{equation}
-2 \log P(\psi|\vX^\tot,\mS,\mN) = \vX^\tot{}^\dag  \mC_g^{-1} \vX^\tot + \ln \det \mC_g  -2\log P(\psi|\mS).
\end{equation}
 This can also be obtained by a rather longer analytical integration of Eq.~\eqref{PpsiX}. In principle this equation tells us all we need to know about the posterior distribution of the lensing potential.
 However the marginalized posterior distribution is now a complicated non-Gaussian function of $\psi$, so even finding the maximum-likelihood point is non-trivial. Refs.~\cite{Hirata:2002jy,Hirata:2003ka} develop an approximate iterative scheme for solving for the maximum-likelihood point and an estimate of the local error. A more fully Bayesian approach might be to try to use sampling techniques to probe the full posterior distribution. However sampling from high-dimensional non-Gaussian distributions is in general difficult, and currently no workable methods have been developed.

Since the Bayesian approach is optimal (uses all the information) we could expect it to be better than the quadratic estimator. Indeed this is the case, with the maximum-likelihood Bayesian estimator being superior in all cases~\cite{Hirata:2002jy,Hirata:2003ka}. For the temperature the improvement is actually quite small. However for the polarization it turns out that a Bayesian estimator can do very much better. This is essentially because the quadratic estimator has cosmic variance from the \emph{lensed} fields, which are non-zero because lensing induces $B$-mode polarization by lensing $E$. The maximum-likelihood estimator knows there should be zero $B$-modes on small scales in the unlensed fields, and the cosmic variance error is then essentially zero, significantly reducing the noise on the estimator. Figure~\ref{HirataSeljak} shows a low-noise simulated reconstruction using the iterative method compared to the quadratic estimator. We discuss this in more detail in the next section where we consider how well the lensing $B$-modes can be removed by delensing using estimators of the lensing potential.

%% file: unlens.tex

Analysing the full statistics of the lensed sky is complicated.
For the near future, parameter estimation using the lensed CMB power spectrum is likely to be accurate enough~\cite{Lewis:2005tp}. However a more optimal analysis should account for the full non-Gaussian distribution of the lensed sky, or, equivalently, the full posterior distribution of the unlensed sky and lensing potential. If the lensing potential could be reconstructed accurately, the sky could be delensed simply by mapping points back according to the deflection angle. This would be useful as the unlensed sky has much simpler statistics than the lensed sky. The smoothing effect of lensing also potentially hides any small scale structure in the CMB power spectrum~\cite{Hu:2003vp}, so delensing would be useful to uncover this information.

With polarization observations the lensing effect is especially important:
 as discussed in Section~\ref{sec:Bconfusion} the $B$-mode polarization generated by lensing may dominate the signal expected from primordial gravitational waves.
For full-sky observations, the lensing $B$-modes act as a source of noise and
limit primordial tensor-mode detections to ratios $r\agt \clo(10^{-3})$~\cite{Hu:2001fb,Lewis:2001hp}. If only part of the sky is observed the lensing variance is larger and the limiting ratio is higher.
It is possible that $B$-modes will be detected at $r>10^{-3}$, however they could easily be smaller.
For future $B$-mode polarization observations delensing will be critical to do better than this; the ultimate observable limit on tensor modes may well depend on how well the lensed $B$-mode signal can be subtracted, assuming
that instrument and astrophysical foregrounds can be adequately cotrolled.
Even if a larger $B$-mode signal is detected, we will probably still want to remove the lensing confusion, for example to measure the spectral index and amplitude to better precision and test for Gaussianity.  Since $B$-mode subtraction relies on using small-scale anisotropies to reconstruct the lensing potential, future observations aiming at low tensor amplitude will need to have high resolution as well as high sensitivity.

Although delensing (if doable) would be useful, if we can model the full statistical distribution of the lensed sky accurately enough it is not actually needed: an optimal analysis could be done by using the full true likelihood function. However in practice this is likely to be very difficult, and delensing methods would in any case be very useful as a way of identifying deviations from the expected model.

With observations of just the one lensed sky is it possible to learn about \emph{both} the unlensed sky \emph{and} the lensing potential? An observation of the CMB temperature in $N$ pixels provides just $N$ numbers, so clearly in general it is not possible to solve exactly for the $2N$ numbers needed to describe both the unlensed temperature and the lensing potential fields. However since we know a lot about the statistical properties of both fields, it should be possible to constrain both fields statistically. In particular, for the temperature we could ideally assume that there is very little unlensed power on small scales, so the unlensed CMB is close to a pure temperature gradient. Assuming we can measure the background gradient at some point, the lensing deflection angles aligned with the gradient can then be reconstructed exactly by noting that
the change in the observed temperature due to lensing is
$\Delta\Theta(\vx) = \vgrad \Theta\cdot \valpha(\vx)$. Orthogonal deflection angles cannot be reconstructed, so we can only measure \emph{half} of the possible information in the deflection field. However the deflection angle field is expected to be the gradient of the lensing potential, $\valpha=\vgrad\psi$, so its curl is zero: in harmonic space the lensing signal is $\Delta \Theta(\vl) = i\vgrad \Theta\cdot \vl\, \psi(\vl)$.
Hence all modes $\psi(\vl)$ of the lensing potential can me measured except for the zero-measure modes with orthogonal $\vl$ so that $\vgrad \Theta \cdot \vl=0$. In the presence of noise modes with $\vl$ nearly parallel to the direction of the gradient will be constrained much better those nearly orthogonal.
On larger scales our prior is that the unlensed CMB is a \emph{Gaussian} field with an (approximately) known power spectrum. Any deviations from consistency with this prior gives extra information that can be used to constrain at least some modes of the lensing potential. For example a large-scale lensing mode would correlate the shapes of background anisotropy blobs over the scale of the mode.

When polarization information is available the situation is much improved. To see this on small scales, consider the damping tail where both the primordial $E$ and $B$-modes are expected to be very small. There are now two unlensed gradient fields, the gradients of the $Q$ and $U$ Stokes' parameters. In some fixed basis the excess in the lensed fields are $\Delta Q(\vx) = \vgrad Q\cdot \valpha(\vx)$ and $\Delta U(\vx) = \vgrad U\cdot \valpha(\vx)$ in the gradient approximation. Since the gradients are random they are generally not aligned, and hence both components of the deflection angle can be measured \emph{without} assuming that
the deflection field is a gradient. In this case, the curl-free property of the deflection angle field is then a consistency check.
The exceptions are the measure-zero cases when the two gradients are aligned, in which case the analysis is equivalent to the lensing of the temperature anisotropies described above.
On larger scales the story is more complicated. However, since we expect there to be no unlensed $B$-modes (except possibly on scales above a degree or so)
the situation is still much better than with the temperature: if we assume there are no unlensed $B$-modes, we only have to solve for the unlensed Gaussian $E$ field and the lensing potential.
Since we observe both the lensed $E$ and $B$ fields there should be enough information to solve for both if we assume a curl-free deflection angle. The number of modes that cannot be recovered have measure zero~\cite{Hirata:2003ka}, for example when the unlensed polarization happens to be zero over a finite area.
Furthermore the polarization $E$-mode power spectrum is skewed to smaller scales than the temperature, so there are effectively more background shape-correlated CMB blobs available to constrain the large-scale lensing modes. This reduces the noise on large-scale mode reconstruction from the lensed non-Gaussianity, though on small scales this effect is compensated by the fact that small-scale power acts as a source of correlated noise on top of the otherwise clean unlensed background gradients.

How well can the lensing $B$-mode signal actually be subtracted? Early work on subtraction methods applied quadratic estimators to reconstruct the lensing potential, and then used the estimated lensing potential to delens the sky. These methods indicated that the lensing power can be reduced by a factor $\sim 10$ in the zero-noise limit~\cite{Knox:2002pe,Kesden:2002ku,Seljak:2003pn}. Although nearly optimal for large noise levels, as discussed in the previous section the quadratic estimator is not optimal due to the unnecessarily large cosmic-variance error.

It was pointed out by Refs.~\cite{Hirata:2003ka,Seljak:2003pn} that in fact one should be able to do much better than the quadratic estimator. The reasoning is as follows. The primordial $B$-mode signal peaks at $l\sim 100$, and falls off on smaller scales (see Fig.~\ref{ClRange}). So we are really interested in subtracting the lensing signal only on large scales, $l\alt 200$ say. The $B$-modes on smaller scales, peaking at $l\sim 1000$, will be dominated by the lensing signal. There is therefore a very large amount of unambiguous information available in the $B$-modes at $l\agt 200$, which can be used to reconstruct the lensing potential. Since the lensing potential peaks at $l\sim 60$, fairly large-scale lensing modes will be responsible for most of lensing signal. The large-scale modes in the lensing potential can therefore be constructed extremely accurately using the large number of small-scale lensed $B$-mode observations.

If the low-$l$ $B$-mode lensing signal had no power from small-scale lensing modes, the reconstruction from lowest-order lensing might be exact in principle: there would be an essentially infinite number of small-scale $B$-modes available to reconstruct a small finite number of large scale lensing potential modes. In fact the low-$l$ $B$-modes are sensitive to small scale power in the lensing potential ($\sim 10\%$ of $C^B_l$ comes from $C^\psi_l$ at $l \agt 1000$ --- see Fig.~\ref{BBlmax}), so the actual limit in practice is not obvious. In reality the improvement over the quadratic estimator methods depends on the noise level. For low noise levels, Refs~\cite{Hirata:2003ka,Seljak:2003pn} claim that at least a factor 40 delensing reduction in the lensing power using their approximate iterative maximum-likelihood method. The ultimate limit for ideal low noise experiments may therefore approach $r\sim \clo(10^{-6})$, depending on the optical depth, corresponding to a potential $V_*^{1/4} \sim 10^{15}\,\GeV$~\cite{Seljak:2003pn}. In reality foreground and experimental issues are likely to be a significant obstacle~\cite{Hu:2002vu,Tucci:2004zy,Verde:2005ff}.

Technically the posterior distribution of the unlensed sky is given by Eq.~\eqref{PpsiX}, integrated over all possible lensing potentials. However the lensing matrix $\mLambda(\psi)$ is a non-linear function of $\psi$, and so cannot easily be integrated analytically. We must either resort to approximations like those discussed above, or use sampling methods. One potentially promising sampling method is Gibbs sampling, applied to the simpler unlensed case in Ref.~\cite{Wandelt:2003uk}.
The method works by iteratively sampling from the marginal distributions, here $P(\vX|\psi,\vX^\tot)$ and $P(\psi|\vX,\vX^\tot)$. Asymptotically it can be shown that a random sample from this iteration corresponds to a sample from the joint distribution $P(\vX,\psi|\vX^\tot)$ (except in pathological cases), though convergence can be arbitrarily slow. Gibbs sampling requires being able to sample from the marginal distributions; this is fine for $P(\vX|\psi,\vX^\tot)$ since this distribution is Gaussian, however $P(\psi|\vX,\vX^\tot)$ is non-Gaussian. One possible approximation is to use a series expansion in the deflection angle. However unfortunately the two marginal distributions $P(\vX|\psi,\vX^\tot)$ and $P(\psi|\vX,\vX^\tot)$ are in general both \emph{much} narrower than the marginalized distribution $P(\vX|\vX^\tot)$: given a particular lensing potential the delensed sky is given essentially by a delta function.  This means that naive Gibbs iterations will not converge within a reasonable time. At the time of writing there are no known practical methods for sampling from the full posterior distribution.

Other tracers may also be used to learn about the lensing potential and hence delens the CMB signal. In particular high-redshift 21-cm observations could be a powerful way to improve tensor-mode limits in principle~\cite{Sigurdson:2005cp}.

%% file: curv.tex
The flat-sky approximation is very useful in that it makes calculations simpler and is likely to be quite a good approximation for observations covering only a small fraction of the sky. However it may be inadequate for observations covering a significant fraction of the sky. Furthermore, even though the lensing deflections are tiny ($\clo(10^{-3})$), they are correlated over degree scales. This relatively large-scale correlation suggests that corrections due to sky-curvature may be non-negligible even on quite small scales~\cite{Hu:2000ee}. The $B$-modes induced by lensing of $E$-modes are of most interest on large scales where they confuse the tensor-mode signal, so it may be important to include the effects of sky curvature.

The temperature field on the sphere can be expanded in spherical harmonics $Y_{lm}$. The deflection angle and polarization fields can be expanded in terms of spin $\pm1 $ and spin $\pm 2$ spherical harmonics ${}_{\pm1}Y_{lm}$ and ${}_{\pm2}Y_{lm}$ respectively~\cite{Zaldarriaga:1996xe,Lewis:2005tp}, as defined in Section~\ref{spinYlm}. Equivalently the fields could be expanded in terms of vector and tensor spherical harmonics~\cite{Kamionkowski:1996ks}. The variance of the harmonic coefficients then defines power spectra for the fields in the usual way, with the $C_l$ being independent of $m$ by the assumption of statistical isotropy, e.g.
\begin{equation}
 \la \Theta_{lm} \Theta_{l'm'}^*\ra = \delta_{l l'}\delta_{m m'} C_l^\Theta.
\end{equation}
The temperature correlation function is defined by
\begin{equation}
\xi(\beta) \equiv \la \Theta(\vnhat_1) \Theta(\vnhat_2) \ra ,
\end{equation}
where $\beta$ is the angle between the two directions ($\vnhat_1 \cdot
\vnhat_2 = \cos\beta$).


The effect of sky curvature on the $\clo(C^\psi_l)$ series-expansion calculation of the lensed CMB power spectra was given in Ref.~\cite{Hu:2000ee} (see also~\cite{Challinor:2002cd}), and shown to be above cosmic variance\footnote{It is necessary to reduce theoretical
 errors that are systematic (i.e.\ of constant sign) over a range of scales
 to \emph{well below} cosmic variance on an individual $C_l$.  This is to avoid bias in parameters
 that are sensitive to a broad range of scales.
 By `cosmic variance' here we mean the error in a bin of some width in $l$ comparable to the range in which the effect is important. Generally $\Delta C_l / C_l \ll 1 / l$ is sufficient for bias on an amplitude parameter to be small compared to its posterior width (c.f. e.g.  Ref.~\cite{Seljak:2003th}).
}. However as discussed in Section~\ref{sec:temp} the perturbative calculation is not actually accurate, and the series-expansion approximation is in fact much worse than the neglect of curvature. An accurate result can however be calculated using the full-sky correlation functions~\cite{Challinor:2005jy}, which is the approach we follow here.

Since deflection angles are small we can treat these accurately
with flat-sky approximations, but we must use
curved-sky results for terms involving larger scale correlations. On the sphere it is simplest to describe the correlation of the deflection angles in terms of the spin-1 deflection field ${}_1 \alpha\equiv \valpha \cdot
(\ve_\theta + i \ve_\phi)$, where $\ve_\theta$ and $\ve_\phi$
are the unit basis vectors of a spherical-polar coordinate system.
Rotating to the basis defined by the geodesic connecting $\vnhat_1$ and
$\vnhat_2$, the spin-1 deflection (denoted with an overbar in the geodesic
basis) has real and imaginary components
\begin{equation}
\alpha_1 \cos\psi_1 = \Re {}_1 \bar{\alpha}(\vnhat_1), \quad
\alpha_1 \sin\psi_1 = \Im {}_1 \bar{\alpha}(\vnhat_1), \quad
\end{equation}
and similarly at $\vnhat_2$. Here,
$\alpha_1 = |\valpha(\vnhat_1)|$ is the length of the lensing displacement
at $\vnhat_1$ and $\psi$ is the angle it makes with the geodesic from
$\vnhat_1$ to $\vnhat_2$.
In terms of these angles the lensed correlation function is
\begin{eqnarray}
\xil(\beta) &=& \la \Theta(\vnhat_1') \Theta(\vnhat_2') \ra \nonumber\\& =&
 \sum_{lm} C_l^\Theta \langle Y_{lm}
(\vnhat_1') Y^*_{lm}(\vnhat_2') \rangle\nonumber\\&=&
\sum_{lmm'} C_l^\Theta d^l_{mm'}(\beta) \langle Y_{lm}
(\alpha_1,\psi_1) Y^*_{lm'}(\alpha_2,\psi_2) \rangle.
\label{curv_rot}
\end{eqnarray}
Here $d^l_{m m'}(\beta)\equiv D^l_{m m'}(0,\beta,0)$ are the reduced Wigner functions (see e.g. Refs.~\cite{Brink93,AngularMom}). The easiest way to see the last step is to put $\vnhat_1$ along the $z$-axis, and
$\vnhat_2$ in the $x$-$z$ plane so that $\vnhat_1'$ has polar coordinates
$(\alpha_1,\psi_1)$. The harmonic at the deflected position
$\vnhat_2'$ can be
evaluated by rotation: $Y_{lm}(\vnhat_2') = [\hat{D}^{-1}(0,\beta,0)
Y_{lm}](\alpha_2,\psi_2)$, where $[\hat{D}Y_{lm}](\vnhat)$ is a spherical
harmonic rotated by the indicated Euler angles.
As usual we have
neglected the small correlation between the deflection angle and
the temperature so that they may be treated as independent fields. The remaining average is over possible realizations of the lensing field.

The last line of Eq.~\eqref{curv_rot} gives the correlation function in terms of a long-range curved-sky term ($d^l_{mm'}(\beta)$) and a covariance that only depends on deflection-angle scales. This suggests that we can use a small-angle approximation to evaluate the latter term. For small $\alpha$ the spherical harmonics can be approximated to within $\clo(1/l^2)$ factors as~\cite{AngularMom}
\begin{eqnarray}
{}_s Y_{lm}(\alpha,\psi) &\approx& (-1)^{m} \sqrt\frac{2l+1}{4\pi} e^{i m \psi} J_{m+s}\left[(l+1/2)\alpha\right] \\
&=&  i^{m} (-i)^s \sqrt{\frac{2l+1}{4\pi}} \frac{e^{-i s\psi}}{2\pi}\int_0^{2\pi} d\phi_\vL\,
e^{i \vL \cdot \bar{\valpha}} e^{i(m+s)\phi_\vL},
\end{eqnarray}
where $\vL \equiv (l+1/2)(\cos\phi_\vL , \sin\phi_\vL)$ and
$\bar{\valpha} \equiv \alpha(\cos\psi , \sin\psi)$.
Hence
\begin{equation}
\xil(\beta) \approx \sum_l  \frac{2l+1}{4\pi} C_l^\Theta \sum_{m m'} i^{m-m'} d^l_{mm'}(\beta) \frac{1}{(2\pi)^2} \int d\phi_\vL\, d\phi_{\vL'}\,
 e^{i (m \phi_\vL -  m' \phi_{\vL'})}\left\la e^{i(\vL\cdot\bar{\valpha}_1-\vL'\cdot\bar{\valpha}_2)}  \right\ra.
\end{equation}
The expectation can easily be calculated as in the flat calculation using Eqs.~\eqref{gauss_avg} and~\eqref{flat_expect}:
\begin{equation}
\left\la e^{i(\vL\cdot\bar{\valpha}_1-\vL'\cdot\bar{\valpha}_2)}  \right\ra = e^{-\la (\vL\cdot\bar{\valpha}_1-\vL'\cdot\bar{\valpha}_2)^2\ra/2} = e^{-L^2[\Cgl(0) - \Cgl(\beta)\cos(\phi_\vL - \phi_{\vL'}) + \Cgltwo(\beta) \cos(\phi_\vL + \phi_{\vL'})]/2}.
\end{equation}

The angular integrals can now be performed to get give modified Bessel functions involving $\Cgl(0)$, $\Cgl(\beta)$ and $\Cgltwo(\beta)$, where on the full sky we now have
\begin{eqnarray}
 -\Cgltwo(\beta) \equiv \langle {}_1 \bar{\alpha}(\vnhat_1) {}_1 \bar{\alpha}(\vnhat_2) \rangle &=&
- \sum_l \frac{2l+1}{4\pi} l(l+1) C_l^\psi d^l_{-1 1}(\beta)  , \nonumber \\
\Cgl(\beta) \equiv
\langle {}_1 \bar{\alpha}^*(\vnhat_1) {}_1 \bar{\alpha}(\vnhat_2) \rangle &=&
\sum_l \frac{2l+1}{4\pi} l(l+1) C_l^\psi d^l_{1 1}(\beta).
\label{cgl_def_curv}
\end{eqnarray}
The fully non-perturbative result (new to this paper) is
\begin{multline}
\xil(\beta) \approx \sum_l  \frac{2l+1}{4\pi} C_l^\Theta e^{-L^2 \Cgl(0)/2} \sum_{m m'} d^l_{mm'}(\beta)  I_{\frac{m+m'}{2}}\left[L^2\Cgl(\beta)/2\right] I_{\frac{m-m'}{2}}\left[L^2 \Cgltwo(\beta)/2\right] \\
= \sum_l \frac{2l+1}{4\pi} C_l^\Theta e^{-L^2 \sigma^2(\beta)/2} \sum_{\mu\nu} d^l_{(\mu+\nu)(\mu-\nu)}(\beta)
  e^{-L^2 \Cgl(\beta)/2}I_\mu\left[L^2\Cgl(\beta)/2\right] I_\nu\left[L^2 \Cgltwo(\beta)/2\right]
\end{multline}
where $m+m'$ is even and we defined $\sigma^2(\beta) \equiv \Cgl(0) - \Cgl(\beta)$. Unlike the flat result, this result depends on $\Cgl(\beta)$ explicitly as well as via $\sigma^2(\beta)$. In the large $l$ limit it is consistent with the flat result of Eq.~\eqref{flatt_exact} because $d^l_{(\mu+\nu)(\mu-\nu)}(\beta) \rightarrow J_{2\nu}(l\beta)$ and we then have
\begin{equation}
e^{-L^2\Cgl(\beta)/2} \sum_{\mu=-\infty}^{\infty} I_\mu\left[L^2\Cgl(\beta)/2\right] = 1,
\end{equation}
which follows from setting $\phi=\pi$ in Eq.~\eqref{I_n_expand}.
This means that contributions from $\Cgl(\beta)$ can only be important at low $l$. At low $l$ the $I_\mu$ term is close to $\delta_{\mu 0}$ because $\Cgl(\beta)$ is small, suggesting that we can neglect the  $\Cgl(\beta)$ dependence entirely. This is indeed a good approximation. Another way to see this is by noting that deflections angles that do not change the unlensed separations must have no effect: equal deflections along the direction of the connecting geodesic cannot affect the correlation function, and similarly for non-parallel deflections leaving the separation invariant. In the first-order $\alpha$ approximation this means that the un-averaged correlation function can only be a function of $\bar{\valpha}_1 - \bar{\valpha}_2$ (for any $\beta$), though at higher order there will be non-zero terms in $\bar{\valpha}_1 + \bar{\valpha}_2$. Since our approximation for the ${}_s Y_{lm}$ is only first order in $\alpha$, we can consistently neglect any higher-order dependence in $\bar{\valpha}_1 + \bar{\valpha}_2$, as we also can at high $l$ where the flat-sky approximation applies. Since explicit $\Cgl(\beta)$ dependence only enters when there are $\bar{\valpha}_1 + \bar{\valpha}_2$ terms present, it is safe to neglect it (a zeroth-order expansion in $\Cgl(\beta)$ should be accurate).

\begin{figure}
\begin{center}
\psfig{figure=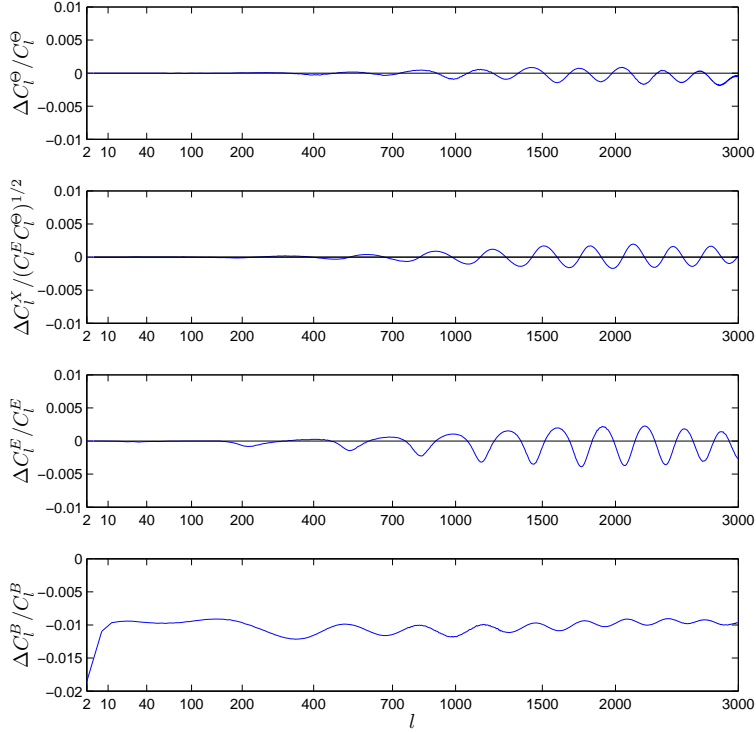,width=10cm}
\caption{The fractional difference between the lensed CMB power spectra using the flat-sky approximation compared to the full-sky method (using the second-order result given in Ref.~\cite{Challinor:2005jy}).
\label{curv_diff}}
\end{center}
\end{figure}

At our level of approximation we can take $L^2 = (l+1/2)^2 = l(l+1) + 1/4 \approx l(l+1)$, and the  lensed correlation function is then given to good accuracy by
\begin{eqnarray}
\xil & \approx &  \sum_l \frac{2l+1}{4\pi} C^\Theta_l
e^{-l(l+1)\sigma^2/2} \sum_{n=-l}^l  I_n\left[l(l+1) \Cgltwo/2\right] d^l_{n-n} \nonumber \\
&=& \sum_l \frac{2l+1}{4\pi} C^\Theta_l  e^{-l(l+1)\sigma^2/2} \biggl\{
 d^l_{00} + \frac{l(l+1)}{2} \Cgltwo d_{1-1}^l  + \dots \biggr\}.
\label{curvt}
\end{eqnarray}
Here and below we suppress the $\beta$ arguments for clarity. The first line gives a fully non-perturbative result, valid on all scales, that generalizes the flat-sky result of Eq.~\eqref{flatt_exact}. The last line gives the first terms in a perturbative expansion. The partial flat-sky approximation we have used here is very good: compared to the full second-order calculation of Ref.~\cite{Challinor:2005jy} the above result if correct to within $\clo(1/l^2)$ factors, giving lensed $C_l$ indistinguishable from the full result (differences certainly well below cosmic variance on scales of interest).


Results for the polarization correlation functions can be worked out using a similar method.
As for the temperature, we evaluate the polarization correlation functions by
taking $\vnhat_1$ along the $z$-axis and $\vnhat_2$ in the $x$-$z$ plane at
angle $\beta$ to the $z$-axis.
With this geometry, the polar-coordinate basis is
already the geodesic basis connecting $\vnhat_1$ and $\vnhat_2$ so
that the lensed correlation functions are~\cite{Challinor:2005jy}
\begin{eqnarray}
\xil_+ &\equiv & \langle \tilde{P}^*(\vnhat_1)
\tilde{P}(\vnhat_2) \rangle \nonumber\\
 &=& \sum_{lmm'} (C_l^E+C_l^B) d^l_{mm'}(\beta)
\langle e^{-2i\psi_1} {}_2 Y_{lm}^*(\alpha_1,\psi_1) {}_2 Y_{lm'}
(\alpha_2,\psi_2) e^{2i\psi_2} \rangle ,  \\
\xil_- &\equiv& \langle \tilde{P}(\vnhat_1)
\tilde{P}(\vnhat_2) \rangle \nonumber\\
 &=& \sum_{lmm'} (C_l^E-C_l^B) d^l_{mm'}(\beta)
\langle e^{2i\psi_1} {}_{-2} Y_{lm}^*(\alpha_1,\psi_1) {}_2 Y_{lm'}
(\alpha_2,\psi_2) e^{2i\psi_2} \rangle ,  \\
\xil_X &\equiv& \langle \tilde{\Theta}(\vnhat_1)
\tilde{P}(\vnhat_2)\rangle \nonumber\\
 &=& \sum_{lmm'} C_l^X d^l_{mm'}(\beta)
\langle Y_{lm}(\alpha_1,\psi_1) {}_2 Y_{lm'}
(\alpha_2,\psi_2) e^{2i\psi_2} \rangle. \label{pol_rot_curv}
\end{eqnarray}
The additional exponential factors here arise from rotating the spin harmonics
at the two points from the polar basis to the appropriate geodesic bases;
see Ref.~\cite{Challinor:2005jy} for details.
Applying the same approximations and methods as for the temperature gives the results
\begin{eqnarray}
\xil_+\angarg &\approx&
\sum_{l} \frac{2l+1}{4\pi}
(C_l^E + C_l^B)e^{-l(l+1)\sigma^2/2}\sum_{n=-l}^l  I_n\left[l(l+1) \Cgltwo/2\right] d^l_{(n+2)(-n+2)}
\nonumber\\
&=&
\sum_{l} \frac{2l+1}{4\pi}
(C_l^E + C_l^B)e^{-l(l+1)\sigma^2/2} \Big\{  d^l_{22}\angarg
+ \frac{1}{2} l(l+1)\Cgltwo\angarg
  d^l_{31}\angarg + \dots\Big\}\\
\xil_-\angarg &\approx& \sum_{l} \frac{2l+1}{4\pi}
(C_l^E - C_l^B)e^{-l(l+1)\sigma^2/2}
\sum_{n=-l}^l  I_n^{(2)}\left[l(l+1) \Cgltwo/2\right] d^l_{n-n}
\nonumber\\
&=& \sum_{l} \frac{2l+1}{4\pi}
(C_l^E - C_l^B)e^{-l(l+1)\sigma^2/2}\Big\{  d^l_{2\, -2}\angarg + \frac{1}{4} l(l+1)\Cgltwo\angarg  \left( d^l_{1\,-1}\angarg + d^l_{3\,-3}\angarg\right) + \dots\Big\}\\
\xil_X\angarg &\approx&
\sum_{l} \frac{2l+1}{4\pi}
C_l^X e^{-l(l+1)\sigma^2/2} \sum_{n=-l}^l  I_n'\left[l(l+1) \Cgltwo/2\right] d^l_{(n+1)(-n+1)}
\nonumber\\
&=&\sum_{l} \frac{2l+1}{4\pi}
C_l^X e^{-l(l+1)\sigma^2/2} \Big\{d^l_{02}\angarg + \frac{1}{4}l(l+1)\Cgltwo\angarg
 \left(d^l_{11}\angarg + d^l_{3\,-1}\angarg \right) + \dots
\Big\}
\end{eqnarray}
where $I^{(2)}_n \equiv 2I_n'' - I_n$ and dashes denote derivatives with respect to the argument. Note that the lowest-order curved-sky results can be less accurate than a higher-order flat-sky result; for numerical work, higher-order terms in $\Cgltwo$ must be included~\cite{Challinor:2005jy}.

The lensed correlation functions can be used to calculate the lensed full-sky $C_l$ using the relations
\begin{eqnarray}
\tilde{C}_l^\Theta &=& 2\pi\int_{-1}^1 \xil(\beta)
d^l_{00}(\beta) \ud \cos\beta , \\
\tilde{C}_l^E - \tilde{C}_l^B &=& 2\pi\int_{-1}^1 \xil_-(\beta)
d^l_{2-2}(\beta) \ud \cos\beta , \\
\tilde{C}_l^E + \tilde{C}_l^B &=& 2\pi\int_{-1}^1 \xil_+(\beta)
d^l_{22}(\beta) \ud \cos\beta , \\
\tilde{C}_l^X  &=& 2\pi\int_{-1}^1 \xil_X(\beta)
d^l_{20}(\beta) \ud \cos\beta .
\end{eqnarray}

 The full curvature corrections to the flat correlation function result are shown in Fig.~\ref{curv_diff}. The corrections are at the small fraction of a percent level for all but the $B$ spectrum, though above cosmic variance on small scales. Due to the mode mixing induced by the lensing the corrections appear at all scales in the lensed CMB power spectra, and in particular there is a percent level effect on the $\tC^{B}_l$ power spectrum at all $l$. Calculation of  $\tC^{B}_l$ to this precision is difficult as it depends on the small-scale non-linear shape of the lensing potential power spectrum, so other errors and uncertainties may dominate; however the full-sky result is numerically not significantly harder to calculate than the flat-sky approximation and hence should be used. It is included in the numerical CMB codes \CAMB\footnote{\url{http://camb.info}} and \CMBEASY\footnote{\url{http://cmbeasy.org}}. Note however that theoretical uncertainties in the recombination ionization history and various late-time secondaries are likely to be much more important than the sky-curvature correction.

Useful relations between full-sky and flat-sky harmonics and correlation functions are given in the appendices of Refs.~\cite{Hu:2001fa,Okamoto03}. Most of the flat-sky results we have given in previous sections can be generalized to the full sky, e.g. the  quadratic estimators for the potential~\cite{Okamoto03} and results for the trispectrum structure~\cite{Hu:2001fa}.

%% file: sim.tex
Simulating lensed CMB maps is straightforward in the lowest weak lensing approximation if both the lensing potential and unlensed CMB fields are Gaussian and statistically isotropic: make two maps, one of the gradient of the lensing potential and one of the CMB, and a lensed map can then be made by remapping points on the CMB map according to the deflection vectors $\valpha = \grad\psi$ from the lensing map using $\vXt(\vnhat) = \vX(\vnhat + \valpha)$~\cite{Zaldarriaga:1998te,Bernardeau:1998mw,Hu:2001tn}. The gradient of the deflection angle can be taken easily in harmonic space, then transformed to get maps of the $x$ and $y$ components.
In practice the CMB map needs to be interpolated or generated at high resolution in order not to suffer from pixelization issues when remapping the points.

On the full sky a pixel remapping scheme can also be used~\cite{Lewis:2005tp}. The deflection vector field can be constructed from vector spherical harmonics (or equivalently spin-1 harmonics), and spherical trigonometry used to calculate the pixel remapping from the geodesic in the direction of the deflection vector. A parallelized code based on \Healpix~\cite{Gorski:2004by} is publicly available\footnote{\url{http://cosmologist.info/lenspix/}}. Simulating lensed skies is significantly slower than unlensed skies because of the higher resolution required for the unlensed map, however it is efficiently paralellizable. Note that convolution with the observational beam takes place on the \emph{lensed} sky, so for an accurate simulation the unlensed sky must be simulated at high resolution independent of the beam.

On smaller scales we may need to model the effects of non-linear evolution, perform ray-tracing to test the weak lensing approximations, and include other effects (such as SZ) in order to test their correlation and confusion with the lensing signal. In general this requires doing detailed numerical simulations for the density field~\cite{Jain:1999ir,White:1999xa,Amblard:2004ih,Anton:2005ip,Vale:2004rh} (for a recent review on $N$-body codes and references see e.g. Ref.~\cite{Heitmann:2004gz}).
The photon path can be computed by ray tracing, though in many cases the Born approximation can be used along an undeflected path to good accuracy to obtain results which include just the non-linear physics.

%% file: apps.tex
\subsection{Observational status}

At the time of writing there is no detection of lensing of the CMB or of lensing-induced cross-correlation with large scale structure~\cite{Hirata:2004rp}. This is expected to change very shortly, either by direct detection of the non-Gaussian signature in the small-scale CMB, or by the improvement to model fitting by using the lensed as opposed to the unlensed CMB power spectra. Current data restrict the amplitude of the lensing effect to be within a factor of three of the expected result, but consistent with zero. However lensing is a robust prediction, so in the next subsections we discuss what new cosmological and/or astrophysical information we might be able to extract from future observations of CMB lensing.

\subsection{Cosmological model constraints}

Cosmological parameter analyses using the lensed CMB power spectra (and ignoring the non-Gaussian structure) is straightforward using standard Markov Chain Monte Carlo methods (e.g. using \COSMOMC\footnote{\url{http://cosmologist.info/cosmomc/}}~\cite{Lewis:2002ah}), and will probably be adequate for the immediate future. Although the lensing effect on the power spectrum at $l\alt 2000$ is significant, the effect on recovered constraints of standard flat-universe cosmological parameters is mild~\cite{Zaldarriaga:1997ch,Lewis:2005tp}. For non-flat models the lensing signal can however be used to break the geometrical degeneracy, allowing a better CMB constraint on the curvature and dark energy~\cite{Stompor:1998zj}.

With sensitive observations, information in the lensing potential can in principle be used to extract extra information about cosmological parameters. For example the unlensed CMB anisotropy is only weakly sensitive to neutrino masses sufficiently light that they are relativistic at recombination. However the lensing potential probes later times, and the small scale damping effect on the lensing potential power spectrum can be significant. Ref.~\cite{Kaplinghat:2003bh} finds that a neutrino with mass $0.1\eV$ only changes the unlensed spectra at the sub-percent level, but the effect on the potential power spectrum is $\sim 5\%$. If the lensing potential can be reconstructed to within this accuracy, this extra information can be used to improve the limits on the neutrino mass from CMB data alone. Future experiments may ultimately be able to probe neutrino masses with error of $\pm 0.035\eV$~\cite{Lesgourgues:2005yv}. Although probably not competitive with future Lyman-alpha and galaxy survey constraints, it is an independent check that is much less sensitive to non-linear evolution and biasing issues.

\begin{figure}
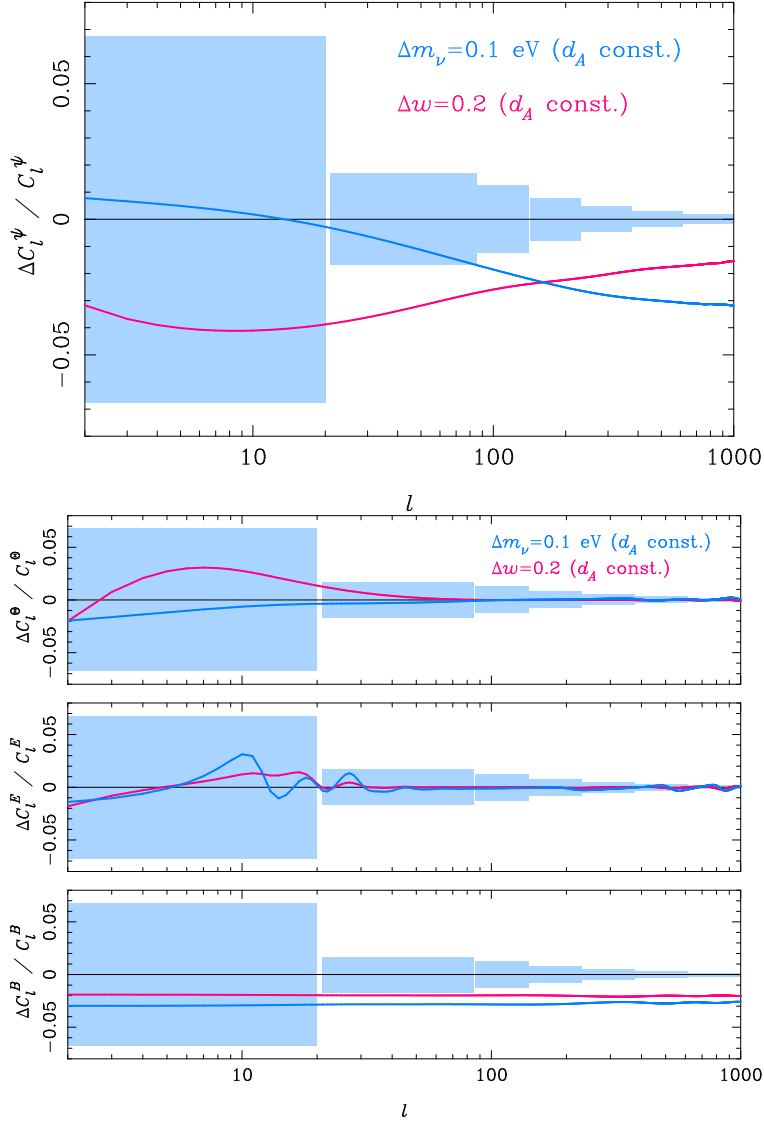

\begin{center}
\psfig{figure=lens_power_derivs_dA.ps,angle=-90,width=10cm}
\psfig{figure=CMB_power_derivs_dA.ps,angle=-90,width=10cm}
\caption{Effect on lensing potential power spectrum (top) and
lensed CMB spectra (bottom)
of varying the neutrino mass (assumed the same for all
three families) and equation of state of dark energy at fixed angular
diameter distance to last scattering in a flat model.
The error boxes are cosmic variance errors and ignore non-Gaussianity
for the lensed CMB spectra.
\label{fig:lens_param_dependence}}
\end{center}
\end{figure}

The lensing potential is also a probe of the late-time evolution, and can be used to help constrain the dark energy model.
Cross-correlation of the lensing potential with the temperature can probe curvature and dark energy via the effect on the large-scale integrated Sachs-Wolfe (ISW) effect~\cite{Seljak:1998nu,Goldberg:1999xm,Hu:2001fb,Giovi:2003ri,Gold:2004ee}. Further information can be extracted by using probes that are sensitive to somewhat different redshift ranges, for example galaxy weak lensing surveys~\cite{vanWaerbeke:1999jd,Benabed:2000jt}.
Since the growth of large-scale perturbations is sensitive to the dark energy sound speed, cross-correlation studies may also provide constraints on the dark energy clustering. Unfortunately cosmic variance dominates on the largest scales where the cross-correlations are significant, so the precision of the constraints is limited.

More direct constraints on the dark energy come from the angular and growth function effects on the lensing potential power spectrum~\cite{Acquaviva:2005xz}.
The effect of massive neutrinos and details of the dark energy model (such
as the equation of state) can be separated by differences in the
scale-dependence of their effect on the lensing potential power
spectrum~\cite{Kaplinghat:2003bh}. This is illustrated for a flat universe in
Fig.~\ref{fig:lens_param_dependence}, which shows the fractional change when the equation of state
parameter $w$ and neutrino mass are varied, but keeping the angular diameter distance to last scattering fixed. The fiducial model is a cosmological
constant model ($w=-1$) with massless neutrinos.
These effects are, however, very nearly degenerate if only the lensed CMB power spectra are used: there is no angular scale for which the
dominant contribution is from the large-angle lenses
for which the effects of massive neutrinos and changes in the dark energy model
are distinct~\cite{Smith:2005ue}; see Fig.~\ref{fig:lens_param_dependence}. Lensing potential reconstruction (or a full non-Gaussian likelihood analysis) is required to provide good non-degenerate constraints.

\subsection{Cluster and galaxy mass}
\label{subsec:clusters}

In Section~\ref{sec:temp} we discussed the effect of lensing on the statistics of the temperature power spectrum. On small scales the CMB has very little power, and what power there is comes mainly from the lensing signal from
large scale structure (and other second-order effects). With very dense clumps of matter it may be possible to detect them individually via their lensing effect, hence the idea of using CMB lensing to learn about cluster and galaxy mass.
Since lensing is sensitive to the potential gradient, lensing constrains the projected mass distribution directly rather than most other observational signatures such as thermal SZ which are only sensitive to the baryonic component.

\begin{figure}
\begin{center}
\psfig{figure=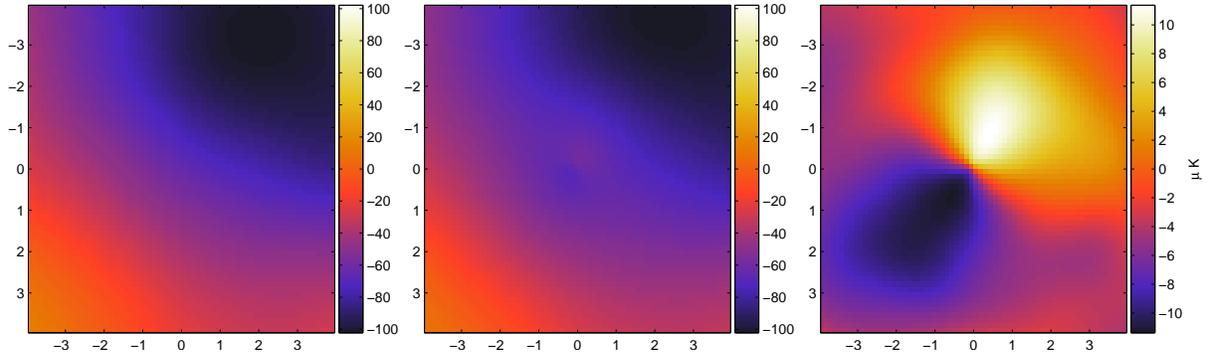,width=16cm}
\caption{Simulated effect of cluster lensing on the CMB temperature. Left: the unlensed CMB; middle: the lensed CMB; right: the difference due to the cluster lensing. The cluster is at redshift one, and has a spherically symmetric NFW profile with mass of $m_{200} = 10^{15} h^{-1} M_\odot$ and concentration parameter $c=5$ (see Refs.~\cite{Dodelson:2004as,Lewis:2005fq}). Distances are in arcminutes, and can be compared to the cluster virial radius of $3.3\,\arcmin$. In the middle figure note the direction of the gradient inverts inside the $\sim 1\,\arcmin$ Einstein radius where the lensing deflections cross the centre of the cluster. This is a rather clean realization, in general the dipole pattern can be weaker and/or more complicated.
\label{clustermassT}}
\end{center}
\end{figure}

The small-scale unlensed CMB is very smooth due to diffusion damping, and can be locally approximated by a gradient. Lensing makes photons appear to originate further from the centre of the cluster than they actually do, so the cold side of the gradient will look hotter after lensing, and the hot side will look colder. This gives a distinctive `wiggle' in the temperature in a direction aligned with the direction of the background CMB gradient.  The idea is therefore to model these lensing wiggles in the observed temperature (and polarization) in an otherwise smooth unlensed background~\cite{Seljak:1999zn,Zaldarriaga:2000ud,Dodelson:2004as,Holder:2004rp,Vale:2004rh}.
For a cluster deflection field $\valpha(\vr)$, the wiggle due to the lensing is given simply by $\valpha(\vr)\cdot\vgrad \Theta$ in the gradient approximation.
A particular example of the cluster lensing effect is shown in Fig.~\ref{clustermassT}.

The rms temperature gradient is given by Eq.~\eqref{RT_def} and is $\sim 14 \muK /\arcmin$ for standard models. Massive clusters can give deflections of the order of an arcminute, so the lensing signal can be at the $\sim 10\muK$ level. The scale of the dipole-like pattern induced by the cluster lensing is much smaller than the unlensed CMB, and so should in principle easily be observable for observations with high enough resolution and low enough noise. However the signal depends on the background gradient, so only clusters in front of a significant gradient can have their mass constrained this way.

Note that in the central region of the cluster deflection angles can cross the centre, where the lensing is usually called strong lensing. The radius at which the deflections meet at a point in the centre is called the Einstein radius.
However even in the strong lensing region the deflection angles are still small, and the Born approximation should still be valid for a single thin lens, so this region can be handled in just the same way as the weak lensing signal at larger radii.

If the background gradient could be measured cleanly away from the cluster, the cluster deflection angles could be reconstructed directly from observations, and hence used to solve for the cluster profile and mass given certain assumptions. Unfortunately the situation is more complicated because the unlensed CMB isn't \emph{exactly} a gradient: the background gradient cannot be measured reliably far from the cluster due to small scale power in the CMB (as well as the effect of lensing by the tails of the cluster). Fortunately the problem is statistically straightforward if we take the unlensed CMB field to be Gaussian: the correlation function of the observed temperature should just be given by the correlation function at the undeflected position on the last scattering surface. We can therefore work out the likelihood of any given cluster deflection field $\valpha(\theta)$ for some set of cluster parameters $\theta$ using
\begin{equation}
-2 \log P(\valpha(\theta)| \tilde{\vTheta}) = \tilde{\Theta}(x_i) C^{-1}(x_i+\alpha_i,x_j+\alpha_j) \tilde{\Theta}(x_j) + \log |C(x_i+\alpha_i,x_j+\alpha_j)|
\label{cluster_like}
\end{equation}
in the absence of noise~\cite{Lewis:2005fq}. Here $\mC$ is the covariance matrix given by the correlation function of the unlensed field, and noise can easily be included using the noise correlation function for the observed positions. The effect of lensing by other perturbations along the line of sight can be crudely modelled by using the lensed power spectrum, though a more careful analysis would require a study of the non-Gaussian distribution numerically using simulations.

This picture is complicated further by other secondary signals that cannot easily be distinguished from the lensing signal. The effect of inhomogeneous kinetic SZ has been investigated from numerical simulations, and found to be a major source of confusion in many cases~\cite{Cooray:2003wd,Holder:2004rp}. Even in idealized cases the kinetic SZ signal can be important. For example if the cluster is rotating there will be a dipole-like signature that looks very like that expected from CMB lensing, and hence acts as a source of confusion~\cite{Lewis:2005fq} (though not aligned with the background gradient direction). The Ostriker-Vishniac and kinetic SZ signals from inhomogeneous reionization can also greatly increase the small scale power in the unlensed CMB, adding an effective additional correlated noise.  The moving lens effect (see Section.~\ref{sec:movinglenses}) also introduces a dipole-like signal at the few $\muK$ level for massive clusters~\cite{Birkinshaw83,Aghanim:1998ux}. However unlike the static lensing signal this should be uncorrelated with the direction of the background CMB gradient, and is generally subdominant to fluctuations in the kinetic-SZ signal.
Taken together these issues make it very challenging to obtain useful cluster constraints from the CMB temperature, and certainly in the near future galaxy lensing is likely to be a much better way to constrain cluster masses~\cite{Lewis:2005fq}.

 If information about the mass of lensing clusters can be obtained by other means (e.g. using SZ observations with some assumptions), CMB lensing studies of many clusters can in principle be used to constrain cosmological parameters~\cite{Cooray:2003wd}. The complicating signals like  kinetic-SZ would have to be distinguished, for example by using their correlation to the frequency-dependent thermal-SZ effect and different morphology~\cite{Cooray:2003wd,Holder:2004rp}. Residual signals degrade the cluster lensing measurement and this is likely to be difficult in practice~\cite{Vale:2004rh}.

We might also want to attempt a general reconstruction of the projected density field of the cluster profile. Ref.~\cite{Maturi:2004zj} suggest a modification to the quadratic estimator potential reconstruction methods of Section~\ref{sec:recon} they claim is better on cluster scales, and obtained constraints on cluster masses by averaging over many clusters.

If there were sufficient small-scale structure in the CMB on cluster scales, e.g. due to secondary effects behind the cluster, these could also be used to measure the shear induced by the lens and hence used to constrain cluster mass profiles~\cite{Zaldarriaga:1998te}. However this is likely to be difficult in practice as it requires very high resolution observations.

Cluster lensing techniques can in principle also be extended to galaxy CMB lensing~\cite{Dodelson:2003gv}, though the signal is much smaller (tenths of a micro-Kelvin) and arcsecond resolution is required. Cross-correlation of the CMB lensing signal with other sources can help to constrain cosmological parameters or learn about the source distributions, for example correlation with the far-infrared background~\cite{Song:2002sg}.

\subsubsection{Cluster lensing of polarization}
\label{subsec:cluster_pol}

Cluster lensing of the CMB polarization~\cite{Lewis:2005fq} is conceptually very similar to the temperature lensing, with potentially two new gradients in $Q$ and $U$ (only correlated at the 10\% level with the temperature gradient). The rms gradient of the Stokes' parameters is $\sim 1\muK/\arcmin$, so a signal $\sim 1\muK$ is expected from massive clusters. Detection therefore requires noise levels about ten times lower than for the temperature. If the temperature signal could be used at this sensitivity level the temperature constraints would always be much better. However, as mentioned above, the temperature signal is complicated by confusion due to kinetic SZ (and other signals). By contrast the polarization signal is expected to be relatively clean~\cite{Valageas:2000ke,Gibilisco:1997wr,Hu:1999vq,Challinor:1999yz,Shimon:2006hn}. Since there are both $Q$ and $U$ Stoke's parameters there are also effectively two random gradients to play with, so polarization cluster constraints suffer less from random variations in the gradient being lensed. Furthermore lensing of a temperature gradient cannot be used to detect deflections orthogonal to the gradient, but with the two Stokes' parameter gradients this degeneracy can in general be broken: polarization cluster lensing should therefore be useful to improve general profile reconstruction.

In a fixed flat sky basis, using the gradient approximation,  some deflection field $\valpha(\vr)$ gives the lensed polarization field
\begin{equation}
\tP(\vr) = P(\vr) +  \valpha(\vr) \cdot \vgrad P,
\end{equation}
where as usual $P\equiv Q+iU$. A constant gradient is neither $E$ nor $B$, because making the $E$/$B$ decomposition locally requires taking two gradients (see Section.~\ref{sec:pol}). Using the definition of the harmonic components in Eqs.~\eqref{pol_transform}, and assuming a suitably well behaved deflection field, the lensed components are
\begin{eqnarray}
\tE(\vl) &=& -\valpha(\vl) \cdot \left( \cos (2\phi_\vl)\vgrad Q + \sin (2\phi_\vl) \vgrad U\right) \nonumber\\
\tB(\vl) &=& -\valpha(\vl) \cdot \left( -\sin (2\phi_\vl)\vgrad Q + \cos (2\phi_\vl) \vgrad U\right).
\label{cluster_EB}
\end{eqnarray}
At a point the gradients of $Q$ and $U$ are uncorrelated, and averaging over (assumed Gaussian) background polarization gradients we get
\begin{eqnarray}
\la |\tE(\vl)|^2\ra_{\vgrad P} &=& \frac{1}{2} R^E |\valpha(\vl)|^2 \nonumber\\
\la |\tB(\vl)|^2\ra_{\vgrad P}&=& \frac{1}{2} R^E|\valpha(\vl)|^2,
\end{eqnarray}
where $R^E = \la |\vgrad Q|^2\ra = \la |\vgrad U|^2\ra$ as in Eq.~\eqref{RE}. An arbitrary deflection field therefore gives identical $E$ and $B$ modes on average when lensing a pure gradient field. Taking the average over a Gaussian deflection field gives the result we found previously in Eqs.~\eqref{EE_highl_approx} and~\eqref{BB_highl_approx}. If the deflection field has circular symmetry (as from a spherical cluster), the angular average of the $E$ and $B$ mode power are equal for any fixed polarization gradient:
\begin{eqnarray}
\frac{1}{2\pi}\int \ud\phi_\vl |\tE(\vl)|^2 &=& \frac{\alpha^2(l)}{4}\left(|\vgrad Q|^2 + |\vgrad U|^2\right)\nonumber\\
\frac{1}{2\pi}\int \ud\phi_\vl |\tB(\vl)|^2 &=& \frac{\alpha^2(l)}{4}\left(|\vgrad Q|^2 + |\vgrad U|^2\right).
\end{eqnarray}

Cluster lensing therefore generates equal amplitude $E$ and $B$ in the gradient approximation~\cite{Lewis:2005fq}.
However there is very little power in the unlensed $E$ or $B$ polarization on cluster scales,
so the lensed $E$ contains almost as much information as the lensed $B$: the $E$/$B$ decomposition is not as useful as on larger scales.
If the full likelihood function is used there is in fact no need to use $E$ and $B$, the correlation function can be calculated directly for the $Q$ and $U$ Stoke's parameters, and the likelihood calculated using the analogue of Eq.~\eqref{cluster_like}. However for nearby large clusters the $B$ mode signal may allow the cluster signal to be distinguished from unlensed CMB `noise', so unlike in the temperature case cluster lensing may not be CMB noise limited for large cluster sizes.

Observations of CMB lensing by clusters can be used to constrain the masses of clusters at any redshift. For the bulk of clusters at redshift $z \alt 1$ mass measurements using lensed galaxy shears are however likely to be much better~\cite{Lewis:2005fq}, even allowing for very optimistic future CMB observations. For high-redshift clusters, for which there are very few source galaxies that can be used for shear measurements, CMB lensing may however be a better way to constrain the mass if high sensitivity measurements of a clean polarization signal are available. The largest identified sources of confusion are likely to be re-scattering from anisotropic thermal-SZ giving a frequency-dependent cluster polarization signal $\alt 0.7\muK$~\cite{Lavaux:2003qf,Shimon:2006hn}, and cluster-scattering of the CMB quadrupole giving a frequency-independent signal $\alt 0.1\muK$~\cite{Kamionkowski:1997na,Cooray:2002cb,Amblard:2004yp}. However only dipole-like components should confuse the lensing signal of the cluster mass.

\subsection{Moving lenses and dipole lensing}
\label{sec:movinglenses}
If a lensing mass is moving with respect to the CMB it corresponds to an evolving potential well that induces anisotropies in a manner similar to the ISW and Rees-Sciama effects. If a lens is moving transverse to the line of sight, a photon passing the front side of the lens (with respect to the direction of motion) will see a weaker potential on its way into the lens than on the way out, and hence receives a net redshift. Similarly a photon on the other side will receive a net blue shift. The moving lens therefore induces a dipole-like temperature anisotropy in the photons crossing the lens' path~\cite{Birkinshaw83,Gurvits86}. For small transverse 3-velocity $\vv_\perp$ (natural units), the temperature anisotropy induced by a small moving lens is at lowest order
\begin{equation}
\frac{\Delta \Theta}{\Theta} = -2 \int \ud\chi\, \vv_\perp \cdot \vgrad_\perp \Psi = \vv_\perp \cdot \vdelta\beta,
\end{equation}
where the integral is along the photon path, $\vv_\perp$ is transverse to the line of sight and $\vdelta\beta$ is the deflection angle of the photon at the lens (related to the observed deflection angle $\valpha$ by the ratio of the relevant angular diameter distances). The signature from a cluster is therefore rather similar to that coming from the lensing of the background CMB temperature gradient, with the difference that here the dipole pattern is aligned with the transverse velocity of the cluster rather than the background gradient. For a $10^{15} h^{-1} M_\odot$ mass cluster moving transverse to the line of sight at $600 \,\rm{km}\, s^{-1}$ the amplitude of the signal is $\sim 5\times 10^{-7}$, corresponding to a signal of $\sim 1\muK$. It is therefore generally subdominant to the effect of lensing of the background CMB gradient.

As described above in the rest frame of the CMB the moving lens contribution is not really a lensing effect, but a contribution to the Rees-Sciama effect~\cite{ReesSciama} (the other contributions being due to the rest-frame change in the lensing mass' potential with time). However in the rest frame of the lens the moving lens effect can be interpreted as CMB dipole lensing, as follows. The velocity of the lens relative to the rest frame of the CMB induces a dipole
in the lens' frame so that the temperature coming from direction $\hat{\vr}$
is $\Theta(1+\vv\cdot\hat{\vr})$ at lowest order. In its rest frame,
the static lens deflects the photon by an angle $\vdelta \beta$ 
perpendicular to the line of sight $\vnhat$, so the observed temperature
along this direction is $\Theta[1+\vv\cdot (\vnhat + \vdelta\beta)]$.
Transforming back to the original frame, we have a temperature anisotropy
$\Delta\Theta = \Theta \vv_\perp \cdot \vdelta\beta$.
On the back side the photons are deflected from the hot side of the dipole so they are hotter that they would have been without the deflection. On the other side they are colder. This temperature anisotropy is exactly the same
pattern we calculated above, so dipole lensing is an equally valid way to interpret the signal.

What happens if we change \emph{our} velocity to be comoving with the lens? In this frame the lens is at rest, but our observed CMB has a dipole. We might therefore think that we could compute $\vgrad \Theta \cdot \vgrad \psi$ to get the contribution from CMB lensing by the cluster.
This however gives a \emph{different} answer in general: the lensing potential $\psi$ depends on the distance to the cluster and source via the angular $1-\chi/\chi_*$ factor, whereas the effect we computed above is independent of distance. This is essentially because the relevant dipole is the one seen by the lens, not the one seen by us. It is therefore not valid to consider the dipole due to our local motion as a dipole source at last scattering that is then deflected by the lens~\cite{Cooray:2005my}. The velocity dipole source must be taken to be at infinite distance to get a consistent answer. The primordial
dipole at last scattering is however lensed, and is  physically distinct from the local motion-induced dipole (which is also associated with other higher order signals arising from the transformation from the CMB frame~\cite{Challinor:2002zh}; see Section~\ref{sec:gauges}).

Typical peculiar velocities are $\clo(10^{-3}c)$, and the moving-lens signature could in principle be used to measure the transverse component. However this may well be impossible in practice due to confusion with other larger signals~\cite{Tuluie:1995ut,Aghanim:1998ux,Cooray:2002ee}. The moving-lens signal itself is also a potential source of confusion when trying to measure cluster properties from CMB lensing as it gives a similar dipole-like pattern, albeit uncorrelated with the background gradient.

There will also be some local effect due to our local mass distribution moving relative to the CMB, giving a local contribution to the large-scale CMB anisotropies~\cite{Vale:2005mt}. This effect is however probably very small~\cite{Cooray:2005my}.

An interesting example of a moving-lens effect is the signal generated
by any cosmic strings that may be present within our
horizon volume (for a review see e.g.\ Ref.~\cite{Vilenkin2000}). A cosmic string with tension
(energy per length) $\mu$ alters the surrounding geometry by a deficit angle $8\pi G \mu$. This gives an effective rest-frame deflection angle $\delta\beta = 4\pi G \mu$ for lines of sight perpendicular to the string. The deflection angle is small for models consistent with current observations ($\delta\beta \alt 10^{-4}$) and likely to be significantly smaller, so the signal from lensing of CMB anisotropies is small. However for strings moving close to the speed of light, this should induce temperature fluctuations $\Delta \Theta/\Theta \sim 4 \pi G \mu$ with a step-like discontinuity across the string. This is called the (Gott-)Kaiser-Stebbins effect~\cite{Kaiser:1984iv,Gott:1984ef} and acts in addition to the signal generated from string-induced perturbations
at last scattering. The Kaiser-Stebbins effect can also be understood in the
CMB frame in terms of integrated anisotropy from the scalar, vector
and tensor perturbations that the moving string generates.
It may be possible to detect this cosmic-string imprint in future
high-resolution data using its distinct non-Gaussian character; see
Ref.~\cite{Lo:2005xt} for a targeted search and constraints with WMAP data.

Finally, if a lens is moving along the line of sight the deflection angle calculated in the rest frame acquires a $\clo(\vv\cdot \vnhat/c)$ correction when transformed back to the CMB frame. This is generally totally negligible~\cite{Schaefer:2005up}, and can be consistently neglected in any lowest-order weak lensing calculation for $\clo(10^{-3}c)$ velocities.